\documentclass[traditabstract]{aa}
\usepackage{graphicx}
\usepackage{txfonts}
\usepackage{array}
\usepackage{amssymb}
\usepackage{natbib}
\usepackage{lscape}
\usepackage{rotating}
\usepackage{longtable}
\usepackage{supertabular}
\usepackage{color}

\newcommand{\lxlb}{$\log(L_{\rm X}/L_{\rm BOL})$}
\newcommand{\xmm}{{\sc{XMM}}\emph{-{\it Newton}}}
\newcommand{\ch}{{\it Chandra}}
\newcommand{\sw}{{\it Swift}}

\newcommand{\ergs}{\,erg\,s$^{-1}$}

\newcommand{\Lx}{$\rm L_{\rm X}$}
\newcommand{\loglxlb}{$\log(\rm L_{\rm X}/\rm L_{\rm BOL})$}
\usepackage{amstext}

\begin{document}

\title{Hot stars observed by \xmm\ II. A survey of Oe and Be stars\thanks{Based on observations collected with NASA missions \ch\ and \sw\ as well as the ESA observatory \xmm , an ESA Science Mission with instruments and contributions directly funded by ESA Member States and the USA (NASA). Tables are also available in electronic form at the CDS via anonymous ftp to cdsarc.u-strasbg.fr (130.79.128.5) or via http://cdsweb.u-strasbg.fr/cgi-bin/qcat?J/A+A/}}

\author{Ya\"el~Naz\'e\inst{1}\fnmsep\thanks{F.R.S.-FNRS Research Associate.} and Christian Motch\inst{2}}

\institute{Groupe d'Astrophysique des Hautes Energies, STAR, Universit\'e de Li\`ege, Quartier Agora (B5c, Institut d'Astrophysique et de G\'eophysique), All\'ee du 6 Ao\^ut 19c, B-4000 Sart Tilman, Li\`ege, Belgium\\
\email{ynaze@uliege.be}
\and Universit\'e de Strasbourg, CNRS, Observatoire Astronomique de Strasbourg, 11 rue de l'Universit\'e, F-67000 Strasbourg, France
}

\authorrunning{Naz\'e et al.}
\titlerunning{Hot stars observed by \xmm\ II.}
\abstract{We perform a survey of Oe and Be stars in the X-ray range. To this aim, we cross-correlated \xmm\ and \ch\ catalogs of X-ray sources with a list of Be stars, finding 84 matches in total. Of these, 51 objects had enough counts for a spectral analysis. This paper provides the derived X-ray properties (X-ray luminosities, and whenever possible, hardness ratios, plasma temperatures, and variability assessment) of this largest ever sample of Oe and Be stars. The targets display a wide range in luminosity and hardness. In particular, the significant presence of very bright and hard sources is atypical for X-ray surveys of OB stars. Several types of sources are identified. A subset of stars display the typical characteristics of O-stars, magnetic OB stars, or pre-main-sequence (PMS) objects: their Be nature does not seem to play an important role. However, another subset comprises $\gamma$\,Cas analogs, which are responsible for the luminous and hard detections. Our sample contains seven known $\gamma$\,Cas analogs, but we also identify eight new $\gamma$\,Cas analogs and one $\gamma$\,Cas candidate. This nearly doubles the sample of such stars. }
\maketitle

\section{Introduction}
Since the first sky surveys have been made in the high-energy range, it has been clear that stars of all types emit X-rays. In low-mass and young stars, this emission is directly associated with stellar activity driven by convection and hence appears linked to rotation and age \citep{pal1981}. The situation is different for massive stars, where stellar winds play a key role. These line-driven winds are indeed unstable and therefore shocks naturally occur within them, ultimately leading to soft ($kT\sim 0.2-0.6$\,keV) and mild (\lxlb$\sim-7$) X-ray emission \citep{pal1981,ber97}. Additional wind-related phenomena also occur, such as wind-wind collisions in binaries or wind magnetic confinement in magnetic objects \citep[see][for recent reviews]{rau16,udd16}. They lead to an increased X-ray luminosity, generally in the hard X-ray range. 

In this context, O stars have generally attracted more attention than B-type stars. The reason probably is that O stars have naturally stronger winds, therefore their X-ray emission is brighter and easier to study. However, surveys and cluster observations have brought some information on B stars as well. In contrast to O stars, B stars have a lower detection rate, show large scatter in \lxlb, and have hotter plasma \citep{ber97,coh97,naz09,naz11,rau15}. Because of these properties and of their more tenuous winds, it has often been hypothesized that except for the earliest B stars, X-ray emission with luminosities below \Lx $\,\sim\,10^{31}$\ergs\ arises from a young, solar-type, undetected companion, the intrinsic emission of the B star being negligible. 

In massive stars, however, a subset of Oe and early Be stars called the $\gamma$\,Cas analogs were found to strongly differ from this general picture. They display a hard ($kT\sim 10$\,keV) and relatively bright (\Lx\,= 10$^{32-33}$\ergs) X-ray emission \cite[see][for a recent review]{smith2016}. 

The origin of their peculiar X-rays is debated. Models have considered either a relatively low mass-accretion rate onto a compact companion or magnetic star-disk interactions. Evidence of a correlation between UV/optical and X-ray variations on short timescales \citep{smith1999,robinson2002} as well as marked dependencies of the X-ray luminosities on decretion disk density with basically zero delay \citep{mot15,rau18} appear to favor the latter scenario. The $\gamma$\,Cas nature of $\pi$\,Aqr (a system with a ``normal'', not compact, companion near the disk, \citealt{naz17}) further support an X-ray source located close to the Be star. It has been suggested by \citet{mot15} that a high stellar rotation rate close to critical plays a key role in the phenomenon, but much remains to be done to reach a full understanding of this peculiar X-ray emission. 

In this paper, we perform a census of the X-ray emission of Oe-Be stars using \xmm, \ch, and \sw\ data, with two main objectives: (1) examine whether the $\gamma$\,Cas analog list is complete considering all currently available observationsand (2) study which physical parameters are responsible for the X-ray characteristics of Oe-Be stars. Our search for $\gamma$\,Cas analogs among identified Be stars is a useful complement to the search for Be stars as counterparts of X-ray sources detected in all-sky X-ray surveys \citep{nebot2013,nebot2015}. The definition of the sample and the finding of counterparts are presented in Section 2, while details on data reduction appear in Section 3. The results are discussed in Section 4, and a summary concludes the paper in Section 5.

\section{Sample}
For this project, a list of Be stars is required first of all. Many catalogs of such stars exist, but some are too old to account for recent discoveries and their source content can be found in more recent catalogs, while others are incomplete in terms of sky coverage (they are specific to a region of the sky, e.g., \citealt{men02}) or contain only candidates, with some contamination by non-Be stars (e.g., following \citealt{gko16}, only 70\% of candidates in \citealt{wit08} truly are Be stars). Although some recent papers provide a few newly discovered objects \citep[e.g.,][]{li18}\footnote{For completeness, we did cross-correlate the new Oe stars that were identified in that paper with the X-ray catalogs mentioned below, but no match was found.}, we restricted our analysis to a single global catalog to preserve the homogeneity of our sample, and we finally decided to use the Be Star Spectra (BeSS) catalog, which is recent and ``as complete as possible'' \citep{nei11}. It also has the additional advantage of offering optical spectra datasets of the targets, facilitating follow-up studies. 

We limited ourselves to the Galaxy, because studying individual stars in the Magellanic Clouds remains a challenge even with dedicated \xmm\ or \ch\ observations \citep[e.g.,][]{osk13,naz14n11}. The low spatial resolution as well as the limited sensitivity of current X-ray facilities renders the identification of individual stellar sources with \Lx\ $<10^{32-33}$\ergs\ difficult if not impossible. In addition, only ``classical'' Be stars were kept: very young objects (e.g., HAeBe) and Be stars belonging to high-mass X-ray binaries (HMXBs) were excluded, as their high-energy emission is of a different nature than the emission we wished to study. Finally, we had to exclude NGC 884 2079 from the list since it is a late-A star, not a Be star \citep{bragg2002}, and we merged the two entries Cl*\,NGC\,884\,LAV\,1703 and [KW97]\,11-2, which correspond to the same coordinates.

\subsection{\xmm\ serendipitous source catalog}
We cross-correlated the resulting Be star list with the latest release of the \xmm\ serendipitous source catalog (3XMM-DR7, see also \citealt{ros16}) using the HEASARC archives\footnote{https://heasarc.gsfc.nasa.gov/cgi-bin/W3Browse/w3browse.pl}. This catalog contains 499\,266 unique X-ray sources, detected in more than 9\,700 EPIC observations made between 2000 February 3 and 2016 December 15. The correlation radius was set to 5'', which is the point spread function (PSF) size of \xmm\ and a typical value for such correlations (e.g., \citealt{ant08,cla11}). This led to 53 matches. The \xmm\ exposures associated with these Be stars were downloaded and processed locally, leading us to discard two matches (see section 3.1 for more details). The stars and their associated 3XMM matches are presented in Table \ref{journal}. This table also provides the distance between the star and its X-ray counterpart, as well as a variability flag set during the 3XMM processing. Whenever an X-ray source was bright enough, its EPIC time series were automatically extracted and analysed by a $\chi^2$ variability test in the processing. Sources were considered variable within the time span of the specific \xmm\ exposure if the null hypothesis (constancy) was rejected with a significance level of $10^{-5}$ or lower. This flag was therefore only set for short-term variability in bright X-ray emitters; it does not evaluate long-term variations.

In addition, we also correlated our Be star list with the recent \xmm\ archives: we searched whether any other Be star was observed in exposures that were not used for the DR7 release but are publicly available as of March 2018. Ten additional Be stars lie within the EPIC field of view of such recent \xmm\ observations. We retrieved and processed these data, running a source-detection algorithm onto each of them: of the ten stars, only two (Menkhib and HD90563) are detected and were thus also added to the detection list. In total, there are thus 53 Be stars detected in \xmm\ pointed observations (Table \ref{journal}). 

\subsection{CXOGSG}
The observations made by \ch\ constitute another large set of X-ray archives. \xmm\ and \ch\ were both launched in 1999, ensuring two decades of observations. However, only the first version of the \ch\ Source Catalog (CSC) was fully available when we began this project, and this version is limited to data that were public before the end of 2009. Therefore, we preferred to use the Chandra ACIS GSG Point-Like X-Ray Source Catalog \citep[CXOGSG,][]{wan16}. It contains 217\,828 distinct X-ray sources (twice more than CSCv1.1) that were found in 10\,029 ACIS observations archived before 2014 December 4. As the PSF of \ch\ is narrower than that of \xmm, we reduced the radius to 2'' for the cross-correlation with the Be star list, allowing for small astrometric errors in both X-ray and optical catalogs. This led to 31 matches, which are reported in Table \ref{journal} with the distances of their X-ray counterparts as wall as their variability flags (they were set when the source had more than ten counts and were found to be variable at a significance level of 1\% in a Kolmogorov-Smirnov test).

As for \xmm, we also correlated the Be star list with the recently archived \ch\ exposures. Four Be stars lie in the field of view of \ch\ observations, but only two (HD\,215227 and Cl*\,NGC\,3293\,FEAS\,32) were actually detected. Furthermore, one additional star that has been detected in the 3XMM-DR7, HD\,42054, was observed with grating data; it is thus not included in the CXOGSG, but we downloaded and processed these observations as well. In total, there are thus 34 Be stars detected in the X-ray range by \ch, 16 of which were also detected by \xmm.

\subsection{\xmm\ slew survey}
Finally, we cross-correlated the Be star list with the ``clean'' version of the second catalog of X-ray sources found in \xmm\ slew data (XMMSL2, see also \citealt{sax08}). This catalog contains 29\,393 bright X-ray emitters detected in data taken between 2001 August 26 and 2014 December 31. The astrometric precision of slew data is smaller than that of pointed observations, hence the correlation radius was here enlarged to 18''. We found 26 matches, half of them having been observed and detected by \xmm\ or \ch\ in pointed observations (Table \ref{journal}). We requested \sw\ exposures for 6 of the remaining sources to assess the identifications with the Be stars and secure X-ray spectra (see next sections for details). 

\subsection{Stellar properties}
In total, 84 Be stars are associated with X-ray sources in current X-ray archives (Table \ref{journal}). About two-thirds of them have usable spectra and are thus studied in more detail in Sect. 4. Table \ref{journal2} lists them along with their main physical parameters, which are used throughout this paper: spectral types, magnitudes and color excesses, distances, bolometric luminosities, effective temperatures, and projected rotational velocities. 

Most spectral types were extracted from the BeSS database. In the absence of a given luminosity class, we assumed a class V. Simbad provided us with V-band magnitudes, and distances were derived from Hipparcos and Gaia (DR1 and DR2) parallaxes or from photometric estimates given in the literature. In all cases, we retained the most accurate measurement. Following \cite{luri2018}, we quadratically added a 100\,$\mu$arcsec systematic uncertainty to all Gaia DR2 parallaxes. Some of the brightest stars have no parallaxes in Gaia DR2 \citep{gaiadr2g} or have parallaxes that appear to be less accurate than those provided by DR1 or by the new Hipparcos reduction of \cite{vanl2007}. 

The vast majority of the E(B-V) excesses were extracted from the Stilism database\footnote{http://stilism.obspm.fr/} \citep{capitanio2017} using the distance estimates. For the most distant targets that are located beyond the limit of the  Stilism database, we used the infrared 3D map of \cite{marshall2006} or took values from the literature. We assumed Av = 3.1$\times$E(B-V).

To derive bolometric luminosities, we used the bolometric corrections from \cite{nieva2013}, $\log(T_{eff})$ given by \cite{dejager1987} and absolute V magnitudes from \cite{wegner2007}. For binaries, this implies that the bolometric luminosities correspond to the pair, not to the Oe-Be star alone. Moreover, the quoted errors on bolometric luminosities only reflect distance uncertainties. In general, our photometric bolometric luminosities match those expected for the stellar spectral types quite well \cite[using the calibration of][e.g.]{dejager1987}. In addition, projected rotational velocities were extracted from various catalogs queried in Vizier; references are provided in the table footnote.

Finally, flags are provided in the last column of the table to indicate if a star is a known (close) binary, pulsating star, $\gamma$\,Cas analog \citep{smith2016}, or magnetic object \citep{gru17,sch2017}. Regarding the latter characteristics, a comment must be made. Magnetic fields are not detected in ``true'' Be stars \citep{gru12} and theoretical models indeed demonstrate that the presence of a disk is incompatible with the strong dipolar fields detectable in spectropolarimetric surveys \citep{udd17}. However, magnetic stars may display emission lines, hence some of them may have been classified as ``Be'' in the past and appear in the BeSS. We thus kept these few cases as a comparison point (for more details on the high-energy emission of magnetic massive stars, see \citealt{naz14}).

\begin{sidewaystable*}
\centering
\caption{List of Be stars detected in the X-ray range, ordered by RA.}
\label{journal}
\tiny 
\setlength{\tabcolsep}{2pt}
\begin{tabular}{cl|cccc|cccc|cc|c}
\hline\hline
\# & Star name & 3XMM & d('') & var.? & \xmm\ ObsID & CXOGSG & d('') & var.?& \ch\ ObsID & XMMSL2 & d('') & \sw\ ObsID \\
\hline
1 & $\gamma$\,Cas &J005642.4+604300     &0.0&y&0651670201/301/401/501,0743600101 & & & & & J005642.6+604301 & 1.0 \\
2 & Achernar      &J013742.7-571413     &1.7&n&0402120101 & & & & & & & \\
3 & V782\,Cas     &J020845.6+650215     &1.5&n&0102580801 & & & & & & & \\
4 & Cl*\,NGC\,869\,LAV\,1039& & & & &J021908.5+570348 & 1.2 &y & 5407,9912/3,12021 & & \\
5 & NGC\,869\,1164$:$&J021914.1+571105 &3.4& &0201160201 & & & & & & & \\
6 & HD\,14162     & & & & &J021927.5+570817     &0.7&n&5407,9912/3,12021 & & \\
7 & Cl*\,NGC\,884\,LAV\,1703$:$& J022125.1+571148 &4.7& &0201160301 & & & & & & & \\
8 & HD\,17505     & & & & & & & & &J025107.7+602503     &1.9 \\
9 & HD\,19818     & & & & & & & & &J030827.0-592231     &9.4 & 00010532001/2\\
10& Merope        & & & & &J034619.5+235653     &0.3&n&13,366,17250/1/2 & &\\
11& Menkhib       &none             & 1.3   & &0770990101/201 & & & & & J035858.0+354729        &2.5\\
12& $\lambda$\,Eri&J050908.8-084514 &0.6&n&0402120301 & & & & & & &\\
13& 25\,Ori       &J052444.8+015047     &0.5&n&0554610101&J052444.8+015047 &0.1& &8571& & \\
14& V1230\,Ori    &J053520.8-052144     &1.0&n&0212480301,0403200101&J053520.7-052144&0.2&y&$^c$ & & \\
15& 43\,Ori       &J053522.8-052457     &0.7&n&0212480301,0403200101&J053522.8-052457&0.5&y&$^c$ & & \\
16& HD\,42054     &J060703.6-341843     &0.0&y&0402121401&none & & & 11021,12226 & & \\
17& PZ\,Gem       &J062715.8+145321     &0.5& &0670080301,0760220601 & & & & & & &\\
18& Cl*\,NGC\,2244\,PS\,26 & & & & &J063129.7+045449&0.4&n&1874,3750& &\\
19& Cl*\,NGC\,2244\,JOHN\,33 & & & & & J063215.4+045520&0.3&y&1874,3750& &\\
20& 15\,Mon       &J064058.6+095345     &0.7&n&0011420101&J064058.6+095344&0.2&n&5401,6247/8&J064058.2+095348&6.6 \\
21& 19\,Mon       &J070254.7-041421     &0.3& &0761090901 & & & & & & &\\
22& HD\,57682     &J072202.1-085845     &0.8&n&0650320201 & & & & & & &\\
23& BN\,Gem       &J073705.7+165415 &0.3& &0670080201 & & & & & & &\\
24& V392\,Pup     &J074610.5-375601     &0.2&n&0694730301/401 & & & & & & &\\
25& V374\,Car     &J075850.5-604928     &0.2&n&0113890601/1001/1101,&J075850.5-604928 &0.1&n&65/6,1229,1232,1458& & \\
&     & &&&0126511201,0134531201/301/501  & & & & & & &\\
26& HD\,90563     &none             & 2.3   & &0780070701 & & & & & & &\\
27& Cl*\,NGC\,3293\,FEAS\,32& & & & &none & 0.3 & & 16648& &\\
28& HD\,93190     &J104419.5-591700     &1.8& &0109530101/201/301/401&J104419.5-591659 &0.3&y&9484& & \\
29& HD\,93843     & & & & &J104837.7-601325     &0.2&n&9508,9857& &\\
30& HD\,305891$:$&J110230.1-604854      &2.0&y&0111210201,0152570101 & & & & & & &\\
31& Phecda        & & & & & & & & &J115350.6+534143     &6.1\\
32& V863\,Cen     &J120805.2-503941     &0.5&n&0742340101 & & & & & & &\\
33& $\delta$\,Cen &J120821.5-504320     &0.4&n&0742340101 & & & & & & &\\
34& BZ\,Cru       &J124250.3-630331     &0.1&y&0504730101 & & & & & J124250.7-630328    &2.9 \\
35& BQ\,Cru$:$   &J124332.8-630607      &4.5& &0109480101/201/401 & & & & & & &\\
36& HD\,117357    &J133115.1-614357     &2.8& &0720300201 & & & & & & &\\
37& HD\,119682    &J134632.5-625524     &0.2&y&0087940201,0551000201 &J134632.5-625523 &0.2&n&8929,10834/5/6&J134631.8-625529&7.7 \\
38& $\mu$\,Cen$:$&J134936.7-422822 &3.5&n&0402121701 & & & & & & &\\
39& V767\,Cen     &J135357.1-470741     &1.2&y&0402121801 & & & & & J135357.1-470739&2.5\\
40& CQ\,Cir       & & & & & & & & &J145050.6-601659     &5.6 & 00042457001/2, 00042464001, 00010536001\\
41& V1040\,Sco    & & & & &J155355.8-235841     &0.4&n&13624 & &\\
42& $\delta$\,Sco &J160019.9-223718     &0.4&n&0743660101 & & & & & & &\\
43& $\zeta$\,Oph  & & & & &J163709.5-103401     &0.4&n&2571,4367&J163709.4-103402       &1.3 \\
44& HD\,153295    & & & & &J170025.2-421900     &0.1& &8234,8237& & \\
45& V1075\,Sco    &J171519.2-333254     &0.1&n&0554440101 & & & & & J171518.7-333251&7.8\\
46& $\gamma$\,Ara &J172523.7-562239     &0.6&n&0201550101 & & & & & J172523.8-562240&0.7\\
47& V750\,Ara     &J172754.7-470134     &0.5&y&0551020101 & & & & & J172754.3-470132&5.7\\
48& $\alpha$\,Ara & & & & & & & & &J173150.8-495233     &2.8\\
49& V864\,Ara     & & & & & & & & &J173311.8-583324     &5.2\\
\hline
\end{tabular}
\end{sidewaystable*}
\setcounter{table}{0}
\begin{sidewaystable*}
\centering
\caption{Continued.  }
\tiny 
\setlength{\tabcolsep}{2pt}
\begin{tabular}{cl|cccc|cccc|cc|c}
\hline\hline
\#& Star name & 3XMM & d('') & var.? & \xmm\ ObsID & CXOGSG & d('') & var.?& \ch\ ObsID & XMMSL2 & d('') & \sw\ ObsID \\
\hline
50& Cl*\,NGC\,6383\,FJL\,24 &J173448.0-323520 &0.8&n&0001730201 & & & & & & &\\
51& V3892\,Sgr    &J174445.7-271344     &0.7&y&0201200101,0691760101& J174445.7-271344 &0.2&n&8647&J174445.6-271351&6.9\\
52& HD\,316341    & & & & &J174835.5-295729     &0.1&y&8743& & \\
53& V771\,Sgr     & & & & & & & & &J175328.3-244630     &2.3 & 00043742001/2002/9002, 00010533001/2\\
54& HD\,316568    &J175442.6-294346     &0.9&n&0206590201,0402280101&J175442.7-294347 &0.9&n&4547,5303& & \\
55& ALS\,4570$:$ & & & & &J180223.0-230158      &1.9&y&2566 & &\\
56& Cl*\,NGC\,6530\,ZCW\,175 &J180411.1-242145 &0.6&n&0008820101,0720540401/501/601 & J180411.1-242145 &0.3&y &977,3754,4397,4444& & \\
57& HD\,164906               &J180425.8-242309 &1.1&n&0008820101,0720540401/501/601     & J180425.8-242308 &0.1&n &977& & \\
58& Cl*\,NGC\,6530\,ZCW\,221 &J180429.3-242527 &3.2&n&0008820101,0720540401/501/601     & J180428.9-242525 &0.6&n &977& & \\
59& Cl*\,NGC\,6530\,ZCW\,228 &J180432.9-241844 &0.4&n&0008820101,0720540401/501/601     & J180432.9-241844 &0.2&n&977,3754,4397,4444& & \\
60& V4379\,Sgr    & & & & & & & & &J180533.3-194514     &0.9 & 00010535001\\
61& HD\,165783    &J180827.1-195207     &1.1& & 0152833001,0672320201 & & & & & & &\\
62& Cl*\,NGC\,6611\,PPM\,38& & & & & J181839.7-134656 &0.2&&978& & \\
63& BD-13\,4928   &J181842.8-134649     &1.7& & 0605130101& J181842.7-134650 &0.2& y&978& & \\
64& Cl*\,NGC\,6611\,BKP\,29783$:$& J181846.1-135438&3.2&n& 0605130101 & & & & & & &\\
65& BD-13\,4933   & & & & &J181904.8-134820     &0.4&n&978,8932,9864/5,9872& & \\
66& EM*\,AS\,315  &J184059.9-052749     &2.6& & 0604820301 & & & & & & &\\
67& CX\,Dra       & & & & & & & & &J184643.3+525916     &2.2 \\
68& HD\,344783    & & & & & J194306.7+231612 &0.3&n&10502,10517& & \\
69& HD\,190864    &J200539.7+353628     &0.2&n&0556260301 & & & & & & &\\
70& HD\,228438    &J201350.1+363721     &1.4& &0720600101 & & & & & & &\\
71& HD\,228860$:$&J201851.5+365740 &0.5& &0510011401,0552350101  & & & & & & &\\
72& V2188\,Cyg    & & & & & J203318.5+411535 & 0.2 & & 4511,109556/7/8& & \\
73& W\,Del        &J203740.0+181703     &0.4& &0742500601 & & & & & & &\\
74& HD\,198931    &J205209.7+442604     &0.5&n&0679580201 & & & & & & &\\
75& V2156\,Cyg    & & & & & & & & &J212501.9+442707&3.1 & 00010534001/2/3\\
76& Alfirk        &J212839.5+703338     &0.1&n&0300490201/301/401/501 & & & & & J212839.7+703339 &0.8\\
77& $\epsilon$\,Cap     & & & & & & & & &J213704.0-192755&12.1\\
78& EM\,Cep       & & & & & J215348.1+623651 &0.2&n&8938,10818/19/20&& \\
79& $\pi$\,Aqr    &J222516.6+012238     &0.2&y&0720390701& & & & & J222517.0+012226& 14.4\\
80& HD\,215227    &J224257.2+444315     &3.2& &0723610201&none & 0.9 & &16753& &\\
81& EM*\,MWC\,659 & & & & & & & & &J224743.4+571652     &17.0 & 00010537001\\
82& BD+61\,2355$:$&J225229.0+624112&3.5& &0743980301 & & & & & & &\\
83& V810\,Cas     & & & & &J232019.0+554828     &0.1&n&3453& & \\
84& $\beta$\,Scl  & & & & & & & & & J233258.5-374906    &3.3\\
\hline
\end{tabular}
\\
\tablefoot{$:  $ A colon indicates a dubious identification (see section 3 for details) ; $^c$=3,4,18,1522,2567/8,3498,3744,4373/4,4395/6,4473/4,7407/8/9/10/11/12,8568,8589,8895/6/7,13637,14334/5,15546,17735. ALS\,4570 is used for the BeSS entry HD164492D since the two objects have the same coordinates. In case of multiple observations (hence multiple detections of a star under the same X-ray source name), the closest separation is provided. }
\end{sidewaystable*}

\begin{table*}
\centering
\setlength{\tabcolsep}{3pt}
\caption{Physical properties of the targets.}
\label{journal2} 
\small
\begin{tabular}{cllrllccclc}
\hline
\hline
\#& Object                 & Spectral             & V band & E(B-V)$^{b}$    & distance                  & Origin of    & $\log(L_{\rm BOL}/L_{\odot})$     & $\log(T_{eff})^{d}$ & $v_{rot} sini ^{e}$ & Flags\\  
&                       & type$^{a}$                       &        &                 &   (pc)                    & distance$^{c}$           &                   &            & (km\,s$^{-1}$)              & \\
\hline
1 & $\gamma$\,Cas           & B0IVpe                     & 2.39   & 0.044           & 190   $\pm$ 20         & DR1                & 4.73$\pm$ 0.09    & 4.486      & 295 $^{17}$          & $^\gamma$ \\
2 & Achernar               & B6Vpe                      & 0.46   & 0.001           & 43    $\pm$ 1          & DR1                & 3.42$\pm$ 0.02    & 4.138      & 218 $^{18,19}$       \\
3 & V782\,Cas               & B2.5III:[n]e+$^{s}$        & 7.62   & 0.483           & 955   $\pm$ 96         & DR2                & 4.15$\pm$ 0.09    & 4.284      & 188 $^{20,21}$        \\
4 & Cl*\,NGC\,869\,LAV\,1039   & B3e                        & 12.37  & 0.609           & 2293  $\pm$ 85    & $^{10}$         & 3.20$\pm$ 0.03    & 4.297      &                  \\
5 & NGC\,869\,1164           & A0e                        & 14.72  & 0.630           & 2293  $\pm$ 85         & $^{10}$            & 1.63$\pm$ 0.03    & 3.991      &                       \\
6 & HD\,14162               & Be                         & 9.29   & 0.642           & 2549  $\pm$ 680        & DR2                & 4.92$\pm$ 0.24    & 4.477      &                       \\
7 & Cl*\,NGC\,884\,LAV\,1703   & B2IVe                      & 12.24  & 0.662           & 2346  $\pm$ 87    & $^{10}$         & 3.43$\pm$ 0.03    & 4.339      &                  \\
8 & HD\,17505               & O6.5III((f))n+O8V $^{s}$   & 7.07   & 0.763           & 3032  $\pm$ 886        & DR2                & 6.36$\pm$ 0.26    & 4.596      & 126 $^{22}$         & $^+$  \\
9 & HD\,19818               & B9.5Vne                    & 9.06   & 0.017           & 309   $\pm$ 10         & DR2                & 1.47$\pm$ 0.03    & 4.023      &                       \\
10& Merope                 & B6IVe                      & 4.18   & 0.021           & 106   $\pm$ 5          & DR2                & 2.80$\pm$ 0.04    & 4.166      & 229 $^{17,18,19}$    \\
11& Menkhib                & O7.5IIIe         & 4.06   & 0.201           & 397   $\pm$ 78           & DR1                & 5.03$\pm$ 0.17    & 4.553      & 230 $^{23}$            & $^+$  \\
12& $\lambda$\,Eri          & B2IVne                     & 4.27   & 0.021           & 249   $\pm$ 11         & DR1                & 3.87$\pm$ 0.04    & 4.339      & 295 $^{18,19}$      \\
13& 25\,Ori                 & B1Vpe                      & 4.96   & 0.026           & 260   $\pm$ 25         & DR2                & 3.78$\pm$ 0.08    & 4.405      & 278 $^{17,18,19}$    \\
14& V1230\,Ori              & B8IVve                     & 9.74   & 0.038           & 436   $\pm$ 22         & DR2                & 1.66$\pm$ 0.04    & 4.088      &                       \\
15& 43\,Ori                 & O9.5Vpe          & 6.39   & 0.052           & 459   $\pm$ 61          & DR2                & 3.95$\pm$ 0.12    & 4.508      & 133 $^{23}$            & $^+$  \\
16& HD\,42054               & B4IVe                      & 5.84   & 0.010           & 337   $\pm$ 15         & DR2                & 3.31$\pm$ 0.04    & 4.252      & 220 $^{18}$           \\
17& PZ\,Gem                 & O9pe                       & 6.64   & 0.139           & 834   $\pm$ 76         & DR2                & 4.54$\pm$ 0.08    & 4.536      & 265  $^{18,19}$     & $^\gamma$ \\
18& Cl*\,NGC\,2244\,PS\,26     & B7Ve                       & 11.41  & 0.266           & 759   $\pm$ 83    & $^{11}$         & 1.77$\pm$ 0.10    & 4.096      &                  \\
19& Cl*\,NGC\,2244\,JOHN\,33   & B6Vne                      & 12.01  & 0.382           & 1390  $\pm$ 100   & $^{12}$         & 2.29$\pm$ 0.06    & 4.138      &                  \\
20& 15\,Mon                 & B1Ve             & 4.64   & 0.011           & 288   $\pm$ 41          & DR1                & 3.97$\pm$ 0.12    & 4.405      & { }55 $^{19,23}$    & $^{+,m?}$    \\
21& 19\,Mon                 & B1Ve                       & 5.00   & 0.032           & 376   $\pm$ 31         & H07                & 4.09$\pm$ 0.07    & 4.405      & 272 $^{17,18,24}$    & $^*$ \\
22& HD\,57682               & O9Ve                       & 6.43   & 0.143           & 1266  $\pm$ 177        & DR2                & 4.99$\pm$ 0.12    & 4.536      & { }16 $^{25}$       & $^m$ \\
23& BN\,Gem                 & O8Vpev                     & 6.87   & 0.031           & 650   $\pm$ 176        & DR1                & 4.11$\pm$ 0.24    & 4.554      & 247 $^{18,19}$       \\
24& V392\,Pup               & B7Ve                       & 5.87   & 0.010           & 204   $\pm$ 6          & DR2                & 2.53$\pm$ 0.03    & 4.096      &                       \\
25& V374\,Car               & B2IVnpe                    & 5.81   & 0.088           & 325   $\pm$ 26         & H07                & 3.57$\pm$ 0.07    & 4.339      & 224 $^{18}$           \\
26& HD\,90563               & B2Ve                       & 9.86   & 0.685 $^{1}$     & 3743  $\pm$ 1277      & DR2                & 4.72$\pm$ 0.31    & 4.320      &                       \\
27& Cl*\,NGC\,3293\,FEAS\,32   & B8Ve                       & 12.87  & 0.290 $^{2}$    & 2750  $\pm$ 250   & $^{2}$          & 2.25$\pm$ 0.08    & 4.058      &                  \\
28& HD\,93190               & O9.5e                      & 8.84   & 1.500 $^{3}$     & 3276  $\pm$ 1024      & DR2                & 6.42$\pm$ 0.28    & 4.500      &                       \\
29& HD\,93843               & O6IIIe                     & 7.33   & 0.250 $^{4}$     & 2846  $\pm$ 791       & DR2                & 5.53$\pm$ 0.25    & 4.578      &                     & $^{m?}$  \\
30& HD\,305891              & B3Ve                       & 10.37  & 1.500 $^{3}$     & 2737  $\pm$ 767       & DR2                & 3.74$\pm$ 0.25    & 4.274      &                       \\
31& Phecda                 & A0Ve+K2V                   & 2.44   & 0.001           & 26    $\pm$ 1          & H07                & 1.86$\pm$ 0.02    & 3.991      & 168 $^{26,27,28}$   & $^+$ \\
32& V863\,Cen               & B6IIIe                     & 4.47   & 0.009           & 102   $\pm$ 3          & DR2                & 2.58$\pm$ 0.03    & 4.140      & 111 $^{18}$          & $^m$ \\
33& $\delta$\,Cen           & B2Vne                      & 2.52   & 0.011           & 128   $\pm$ 8          & H07                & 3.94$\pm$ 0.05    & 4.320      & 228 $^{18,19}$       \\
34& BZ\,Cru                 & B0.5IVpe                   & 5.31   & 0.331           & 421   $\pm$ 28         & DR2                & 4.55$\pm$ 0.06    & 4.455      & 338 $^{18,19}$      & $^\gamma$ \\
35& BQ\,Cru                 & Be                         & 12.05  & 0.950 $^{5}$     & 2600  $\pm$ 80        & $^{5}$             & 3.08$\pm$ 0.03    &             &                      \\
36& HD\,117357              & O9.5Ve                     & 9.58   & 0.365           & 1605  $\pm$ 615        & DR1                & 4.09$\pm$ 0.35    & 4.508      & { }78 $^{18}$      & $^*$ \\
37& HD\,119682              & B0Ve                       & 7.90   & 0.332           & 1752  $\pm$ 325        & DR2                & 4.86$\pm$ 0.16    & 4.512      & 200 $^{19}$         & $^\gamma$  \\
38& $\mu$\,Cen              & B2Vnpe                     & 3.43   & 0.013           & 155   $\pm$ 4          & H07                & 3.75$\pm$ 0.02    & 4.320      & 150 $^{18,19}$       \\
39& V767\,Cen               & B2Ve                       & 6.10   & 0.070           & 830   $\pm$ 100        & $^{13}$            & 4.20$\pm$ 0.11    & 4.320      & 100 $^{19}$           \\
40& CQ\,Cir                 & B1Ve                       & 10.04  & 0.462           & 1803  $\pm$ 333        & DR2                & 3.96$\pm$ 0.16    & 4.405      & 335 $^{19}$          \\
41& V1040\,Sco              & B2Ve                       & 5.40   & 0.110 $^{6}$     & 141   $\pm$ 4         & DR2                & 3.00$\pm$ 0.02    & 4.320      & 312 $^{17,18,19}$   & $^m$ \\
42& $\delta$\,Sco           & B0.2IVe          & 2.32   & 0.142           & 154   $\pm$ 21          & H07                & 4.69$\pm$ 0.12    & 4.486      & 165 $^{17}$            & $^+$  \\
43& $\zeta$\,Oph            & O9Ve                       & 2.56   & 0.203           & 113   $\pm$ 3          & H07                & 4.51$\pm$ 0.02    & 4.536      & 335 $^{18,19,23}$   & $^{m?}$ \\
44& HD\,153295              & B2e                        & 9.10   & 0.559           & 1684  $\pm$ 322        & DR2                & 4.17$\pm$ 0.17    & 4.300      &                       \\
45& V1075\,Sco              & O8Ve                       & 5.53   & 0.269           & 1022  $\pm$ 167        & DR2                & 5.35$\pm$ 0.14    & 4.554      & { }81 $^{18,25}$   & $^{m?}$     \\
46& $\gamma$\,Ara           & B1IIe                      & 3.34   & 0.100           & 343   $\pm$ 19         & H07                & 4.63$\pm$ 0.05    & 4.343      & 281 $^{29}$           \\
47& V750\,Ara               & B2Vne                      & 6.66   & 0.142           & 1124  $\pm$ 152        & DR2                & 4.33$\pm$ 0.12    & 4.320      & 277 $^{18}$         & $^\gamma$  \\
48& $\alpha$\,Ara           & B2Vne                      & 2.95   & 0.006           & 83    $\pm$ 6          & H07                & 3.38$\pm$ 0.06    & 4.320      & 279 $^{18,30}$       \\
49& V864\,Ara               & B7Vnnpe                    & 7.03   & 0.058           & 242   $\pm$ 6          & DR2                & 2.27$\pm$ 0.02    & 4.096      &                       \\
50& Cl*\,NGC\,6383\,FJL\,24    & B9Ve                       & 11.38  & 0.460           & 1518  $\pm$ 266   & DR2             & 2.46$\pm$ 0.15    & 4.023      &                  \\
51& V3892\,Sgr              & Oe                         & 9.13   & 0.783           & 1233  $\pm$ 175        & DR2                & 4.68$\pm$ 0.12    & 4.536      & 260 $^{18}$        & $^\gamma$   \\
52& HD\,316341              & B0Ve                       & 9.88   & 0.773           & 2022  $\pm$ 466        & DR2                & 4.65$\pm$ 0.20    & 4.477      & 120 $^{19}$          \\
53& V771\,Sgr               & B3/5ne $^{s}$              & 9.16   & 0.311           & 780   $\pm$ 97         & DR2                & 3.45$\pm$ 0.11    & 4.428      &                       \\
54& HD\,316568              & B2IVpe                     & 9.66   & 0.540 $^{3}$     & 2328  $\pm$ 597       & DR2                & 4.27$\pm$ 0.23    & 4.339      &                       \\
55& ALS \,4570               &                            & 6.80   & 1.630           & 2950  $\pm$ 450        & $^{14}$            & 7.48$\pm$ 0.13    & 4.564      & { }33 $^{23}$           \\
56& Cl*\,NGC\,6530\,ZCW\,175   & B3Ve                       & 10.37  & 0.388           & 1074  $\pm$ 131   & DR2             & 3.01$\pm$ 0.11    & 4.274      &                  \\
57& HD\,164906              & B0Ve                       & 7.38   & 0.431           & 1260  $\pm$ 177        & DR2                & 4.83$\pm$ 0.12    & 4.477      & 255 $^{19}$           \\
58& Cl*\,NGC\,6530\,ZCW\,221   & B2.5Ve                     & 10.70  & 0.350           & 1329  $\pm$ 191   & DR2             & 3.07$\pm$ 0.13    & 4.297      &                  \\
59& Cl*\,NGC\,6530\,ZCW\,228   & B2.5Vne                    & 10.45  & 0.375           & 1076  $\pm$ 131   & DR2             & 3.02$\pm$ 0.11    & 4.297      &                  \\
60& V4379\,Sgr              & B9Ve                       & 7.05   & 0.070           & 122   $\pm$ 45         & DR2                & 1.46$\pm$ 0.33   & 4.023      &                        \\
61& HD\,165783              & B4IIe                      & 8.37   & 0.634           & 1593  $\pm$ 278        & DR2                & 4.24$\pm$ 0.15    & 4.176      &                       \\
62& Cl*\,NGC\,6611\,PPM\,38    & B8Ve                       & 13.71  & 0.893           & 1750  $\pm$ 50    & $^{15}$         & 2.27$\pm$ 0.02    & 4.058      &                  \\
63& BD-13\,4928             & O9.5Vne                    & 10.04  & 0.330 $^{7}$     & 2220  $\pm$ 210       &$^{7}$              & 4.21$\pm$ 0.08    & 4.508      &                       \\
64& Cl*\,NGC\,6611\,BKP\,29783 & Be                         & 13.59  & 0.963           & 1750  $\pm$ 50    & $^{15}$         & 3.30$\pm$ 0.02    & 4.477      &                  \\
65& BD-13\,4933             & B0.5Ve                     & 10.63  & 1.100           & 2264  $\pm$ 631        & DR2                & 4.77$\pm$ 0.25    & 4.443      &                       \\
\hline
\end{tabular}
\end{table*}
\setcounter{table}{1}
\begin{table*}
\centering
\setlength{\tabcolsep}{3pt}
\caption{Continued}
\small
\begin{tabular}{cllrllccclc}
\hline
\hline
\#& Object                 & Spectral             & V band & E(B-V)$^{b}$    & distance                  & Origin of    & $\log(L_{\rm BOL}/L_{\odot})$     & $\log(T_{eff})^{d}$ & $v_{rot} sini ^{e}$ & Flags\\  
&                       & type$^{a}$                       &        &                 &   (pc)                    & distance$^{c}$           &                   &            & (km\,s$^{-1}$)              & \\
\hline
66& EM*\,AS\,315             & Be                         & 11.30  & 1.310 $^{3}$     & 4653  $\pm$ 1979      & DR2                & 5.41$\pm$ 0.39    & 4.477      &                       \\
67& CX\,Dra                 & B2.5Ve           & 5.90   & 0.024           & 357   $\pm$ 15          & DR2                & 3.45$\pm$ 0.04    & 4.297      & 159 $^{17,19,30}$  & $^+$  \\
68& HD\,344783              & B0IVe                      & 9.80   & 0.604           & 1785  $\pm$ 857        & DR2                & 4.30$\pm$ 0.45    & 4.486      &                       \\
69& HD\,190864              & O7IIIe                     & 7.78   & 0.600           & 2143  $\pm$ 458        & DR2                & 5.51$\pm$ 0.19    & 4.562      & { }55 $^{23,25}$        \\
70& HD\,228438              & B0.5IIIe                   & 8.37   & 0.660           & 2702  $\pm$ 716        & DR2                & 5.26$\pm$ 0.24    & 4.429      & 219 $^{18}$           \\
71& HD\,228860              & B0.5IVe                    & 9.72   & 1.070           & 2083  $\pm$ 440        & DR2                & 5.07$\pm$ 0.19    & 4.455      &                       \\
72& V2188\,Cyg              & B2Ve$^{*}$                 & 14.88  & 2.570 $^{8}$     & 1500  $\pm$ 200       & $^{16}$            & 4.34$\pm$ 0.12    & 4.339      &                       \\
73& W\,Del                  & A0Ve             & 9.81   & 0.050           & 820   $\pm$ 73          & DR2                & 1.98$\pm$ 0.08    & 3.991      &                & $^+$  \\
74& HD\,198931              & B1Vnne                     & 8.72   & 0.407           & 825   $\pm$ 172        & DR2                & 3.74$\pm$ 0.18    & 4.405      & 322 $^{18}$           \\
75& V2156\,Cyg              & B1.5Vnnpe                  & 8.91   & 0.312           & 943   $\pm$ 94         & DR2                & 3.59$\pm$ 0.09    & 4.364      &                       \\
76& Alfirk                 & B2IIIev           & 3.23   & 0.007           & 105   $\pm$ 11          & DR2                & 3.44$\pm$ 0.09    & 4.305      & { }26 $^{17,25}$    & $^+$    \\
77& $\epsilon$\,Cap         & B3Vpe                      & 4.55   & 0.006           & 155   $\pm$ 17         & DR2                & 3.19$\pm$ 0.09    & 4.274      & 237 $^{17,19,30}$    \\
78& EM\,Cep                 & B0.5Ve                     & 7.03   & 0.210           & 530   $\pm$ 141        & H07                & 3.85$\pm$ 0.24    & 4.443      &                       \\
79& $\pi$\,Aqr              & B1Ve                       & 4.64   & 0.059           & 241   $\pm$ 16         & H07                & 3.88$\pm$ 0.06    & 4.405      & 243 $^{17,18,19}$   & $^{+,\gamma}$ \\
80& HD\,215227              & B5ne                       & 8.81   & 0.250 $^{9}$     & 2575  $\pm$ 654       & DR2                & 4.05$\pm$ 0.23    & 4.200      & 262 $^{18}$           \\
81& EM*\,MWC\,659            & B0IIIpe                    & 10.15  & 0.420           & 4290  $\pm$ 1632       & DR2                & 4.69$\pm$ 0.35    & 4.465      &                       \\
82& BD+61\,2355             & B7IVe                      & 9.63   & 0.268           & 576   $\pm$ 50         & DR2                & 2.30$\pm$ 0.08    & 4.125      &                       \\
83& V810\,Cas               & B1npe                      & 8.59   & 0.288           & 1625  $\pm$ 271        & DR2                & 4.24$\pm$ 0.15    & 4.405      & 422 $^{18}$           \\
84& $\beta$\,Scl            & B9.5IVmnpe                 & 4.37   & 0.001           & 54    $\pm$ 1          & H07                & 1.86$\pm$ 0.01    & 4.053      & { }26 $^{28}$         \\
\hline
\end{tabular}
\tablefoot{ 
$^{a)}$ All spectral types are from the BeSS database except for those marked with $^{s}$ , which are from Simbad. In this context, we note that Simbad mentions O7V, F5III, and B0.5III as spectral types for the companions of 15\,Mon, CX\,Dra, and Alfirk, respectively.
$^{b)}$ all E(B-V) extracted from the Stilism interface at http://stilism.obspm.fr \citep{capitanio2017} unless specified otherwise. 
$^{c)}$ DR1; Gaia DR1 \citep{gaiadr1}, DR2: \citep{gaiadr2g}, H07: \citep{vanl2007}
$^{d)}$ after \cite{dejager1987}.
$^{e)}$ Mean of values from all quoted references. \\
Flags: An asterisk indicates a pulsating object; a cross identifies binaries; $^\gamma$ is set for known $\gamma$\,Cas analogs, and $^m$ indicates magnetic stars ($^{m?}$ is for claimed magnetic objects). We note that Alfirk's primary is a pulsating magnetic star, while the secondary is of Be type; the temperature and rotational velocity values provided here are those of the secondary. \\
References:
$^{1)}$ \cite{kal2012}.
$^{2)}$ \cite{baume2003}.
$^{3)}$ E(B-V) from \citep{marshall2006}.
$^{4)}$ \cite{wegner1993}.
$^{5)}$ \cite{piatti2001}.
$^{6)}$ \cite{gon2018}.
$^{7)}$ \cite{selim2016}. 
$^{8)}$ \cite{wrihgt2010}.
$^{9)}$ this paper. 
$^{10)}$ \cite{currie2010}.
$^{11)}$ \cite{sung1997}.
$^{12)}$ \cite{hens2000}.
$^{13)}$ \cite{sch2017}.
$^{14)}$ \cite{{savage1985}}.
$^{15)}$ \cite{guar2007}.
$^{16)}$ \cite{mary2017}.
$^{17)}$ \cite{abt2002}.
$^{18)}$ \cite{yudin2001}.
$^{19)}$ \cite{zorec2016}.
$^{20)}$ \cite{gle05}
$^{21)}$ \cite{jas82}
$^{22)}$ \cite{uesugi1970}.
$^{23)}$ \cite{sim2014}.
$^{24)}$ \cite{huang2010}.
$^{25)}$ \cite{sim2017}.
$^{26)}$ \cite{evans1967}.
$^{27)}$ \cite{david2015}.
$^{28)}$ \cite{royer2007}.
$^{29)}$ \cite{ochs1987}.
$^{30)}$ \cite{chauville2001}. 
}
\end{table*}
\normalsize

\section{Data reduction} 

\subsection{\xmm}
We downloaded from the archives all \xmm\ datasets associated with the Be stars detected in the 3XMM. In addition, we downloaded the recent datasets that are not included in the 3XMM DR7, with the Be stars Menkhib, HD\,55135, CD$-$44\,4392, HD\,90599 and HD\,90563, HD\,305627, HD\,306111, HD\,117172, V807\,Cen, and HD\,225985 in their fields of view. These \xmm\ data were locally processed with the Science Analysis Software (SAS) v16.0.0 using calibration files available in Oct. 2017 and following the recommendations of the \xmm\ team\footnote{SAS threads, see \\ http://xmm.esac.esa.int/sas/current/documentation/threads/ }. 

These observations were taken in various mode and filter combinations. We checked in each case that the filter was adapted to the target optical/UV luminosity. This was the case for all but three stars: (1) all observations of 48\,Lib were taken with a thin filter, while the target magnitude requires a thick filter and the X-ray data therefore suffer from heavy optical loading and cannot be used (this target was thus discarded from the list of matches reported in Table \ref{journal}); (2) One observation of V374\,Car (ObsID=0113891201, Rev.164) was taken with a medium filter rather than a thick one like the other exposures and this dataset was therefore discarded from further analyses; (3) The observations of V807\,Cen were taken with the thin filter rather than the medium one but since the star is not detected (see below), this does not change the analyses. We also checked modes, as they allow for different levels of pile-up. Considering the 3XMM count rates, three potential problems exist: (1) from its count rates, $\pi$\,Aqr is at the limit of pile-up, but a previous, dedicated study showed that there is no pile-up problem for the \xmm\ data \citep{naz17}; (2) BZ\,Cru is much too bright in X-rays for the full-frame mode used in ObsIDs 0109480101/201/401, hence those datasets were discarded; (3) $\gamma$\,Cas was observed in small-window mode, but its EPIC count rates still are above the pile-up limits, although a run of {\it epatplot} does not reveal the typical signatures of pile-up and we therefore kept the circular extraction as in \citet{smi12b}.

After the initial pipeline processing, the European Photon Imaging Camera (EPIC) observations were filtered to keep only the best-quality data ({\sc{pattern}} 0--12 for MOS and 0--4 for pn). To assess the crowding near the targets in order to choose the best extraction region, a source detection was performed on each EPIC dataset using the task {\it edetect\_chain}, which uses first sliding box algorithms and then performs a PSF fitting, on the 0.3--10.0\,keV energy band and for a log-likelihood of 10. In this context, we note that, regarding the recent datasets, only Menkhib and HD\,90563 are detected, and they were thus added to the detection list (Table \ref{journal}). Furthermore, the results on the older datasets indicated no formal detection for NGC\,869\,1164, HD\,93190, BQ\,Cru, or  HD\,228860: these sources thus are faint, at the limit of detection (indeed, the DR7 processing considers a log-likelihood detection level of 6, which is lower than our local processing). Except for HD\,93190, which was clearly detected by \ch\ as well, some doubt may thus be cast on the detection of the other three stars. In particular, it should be noted that BQ\,Cru is also detected to be extended in the 3XMM catalog, unlike all other matches that are considered as point-like sources. In this context, doubts may also arise when considering the case of HD\,305891, as the source appears to lie within extended emission, although the 3XMM detection officially has zero extension.   

Before we extracted the spectra, light curves for events beyond 10\,keV were calculated for the full cameras, and when background flares were detected, the corresponding time intervals were discarded. In some cases, the light curves indicated continuous flaring during the observation, and such exposures were discarded to avoid problems or strong noise in data. Other exposures of the same target usually existed, hence no information was lost. There is one exception, however: only one exposure exists for V782 Cas, and it is contaminated by strong flares. To check whether this caused any problem, we extracted the spectra of two other sources that were also situated at large off-axis angles in this (sparsely populated) field of view. One is brighter and one fainter than V782\,Cas. These spectra are all different, as could be expected if background is well corrected, hence the spectrum of V782\,Cas can be trusted. 

We extracted EPIC spectra using the task {\it{especget}} in circular regions centered on the Simbad positions of the targets and with radii between 12.5 and 30\arcsec, depending on the crowding. Only for V1230\,Ori, 43\,Ori, and V750\,Ara were elliptical regions used as the sources are far off-axis and their PSF is strongly distorted. Background was derived in nearby circular regions devoid of sources. Dedicated ancillary response file (ARF) and redistribution matrix file (RMF) response matrices, which are used to calibrate the flux and energy axes, respectively, were also calculated by this task. EPIC spectra were grouped with {\it{specgroup}} to obtain an oversampling factor of five and to ensure that a minimum signal-to-noise ratio of 3 (i.e., a minimum of ten counts) was reached in each spectral bin of the background-corrected spectra; unreliable bins below 0.25\,keV were discarded. 

Reflection grating spectrometer (RGS) data were also processed using the initial pipeline, but only for sources that are bright and centered in the field of view: $\gamma$\,Cas, V1075\,Sco, $\delta$\,Sco, Alfirk, and Menkhib. As for EPIC data, a flare filtering was also applied (using a threshold of 0.12\,cts\,s$^{-1}$) when needed. Dedicated response files were calculated for both orders and both RGS instruments, and were subsequently attached to the source spectra for analysis. These high-resolution data are not reported here in detail, but they were used to check the results of the EPIC spectral fits.

Finally, it is also important to examine the distance between the detected DR7 counterpart and the Be star. Most (80\%) of the X-ray counterparts appear within 2'' of the optical position of the star, including in the case of the new detections, but there are a few exceptions with separations of 3--5'': NGC\,869\,1164 (3XMM J021914.1+571105 at 3.4''), Cl*\,NGC\,884\,LAV\,1703 (3XMM J022125.1+571148 at 4.7''), BQ\,Cru (3XMM J124332.8-630607 at 4.5''), $\mu$\,Cen (3XMM J134936.7-422822 at 3.5''), Cl*\,NGC\,6530\,ZCW\,221 (3XMM J180429.3-242527 at 3.2''), HD\,215227 (3XMM J224257.2+444315 at 3.2''), Cl*\,NGC\,6611\,BKP\,29783 (3XMM J181846.1-135438 at 3.2''), and BD+61\,2355 (3XMM J225229.0+624112 at 3.5''). We examined each case in turn. First, we just registered doubts about X-rays associated with BQ\,Cru, and this decentering only added to the problem. To investigate the issue for other sources, we then examined the distance between other X-ray sources detected in the field of view and their known optical counterparts. For $\mu$\,Cen, only the high proper motion star UCAC3 96-143035 seems to be detected in addition to $\mu$\,Cen, but its 3XMM counterpart, 3XMM J134901.2-422841, lies at 0.7'' of the optical source. For BD+61\,2355, the field of view is centered on 2MASS J22535512+6243368, which is at 0.2'' from its counterpart 3XMM J225355.1+624336, and other sources seem well centered (e.g., 2MASS J22531578+6235262, which is at a similar off-axis angle as BD+61\,2355, is at 0.4'' of its counterpart 3XMM J225315.8+623526). In the field of Cl*\,NGC\,884\,LAV\,1703, the late-A star NGC\,884\,2079 lies at 0.5'' of 3XMM J022142.9+571830. This seems to indicate the absence of astrometric problems in these datasets, hence the association of these three Be stars with the 3XMM sources are doubtful. Finally, \ch\ data also exist for the remaining sources, and help settle the association question. For Cl*\,NGC\,6530\,ZCW\,221, there is an X-ray source at the position of the star, but it is very close to a brighter source, and this neighbor may have led to some confusion in the \xmm\ data, explaining the decentering. For HD\,215227, the \ch\ data clearly indicate an X-ray source at 0.8'' of the position of the star, so there is no doubt about the association. Conversely, the \ch\ data covering the positions of NGC\,869\,1164 and Cl*\,NGC\,6611\,BKP\,29783 reveal no source at the positions of the stars, but sources nearby, which may have led to an incorrect match in the \xmm\ data. For example, NGC\,869\,1171 is twice closer to 3XMM J021914.1+571105 than NGC\,869\,1164 and corresponds to an X-ray source in \ch\ data. All dubious identifications are clearly marked by a colon in Table \ref{journal}. In this context, it is also interesting to mention the case of Cl*\,NGC\,6530\,SCB\,790-HD\,164947\footnote{There are two separate entries in the BeSS for these stars, but Cl*\,NGC\,6530\,SCB\,790 is really HD\,164947A.}, which is at 2-2.6'' of 3XMM J180441.6-242056. The field is crowded, and in the \ch\ data, it is clear that the emitter is a neighboring source (CXOGSG J180441.6-242055 lies at 2'' of HD\,164947). We therefore discarded it from the list of matches. As a last step, the separation between X-ray and optical sources also needed to be compared to the uncertainty on the X-ray position. For the vast majority of the sources, both the separations and the errors are smaller than 2''. The separation is smaller than three times the positional error for all sources except for 15\,Mon (for which the error is small because of its X-ray brightness, but the separation is only 0.7'') and Cl*\,NGC\,884\,LAV\,1703 (as discussed above; the separation nearly reaches 5'' in this case). 

\subsection{\ch}
We downloaded from the archives all \ch\ datasets associated with the Be stars detected in the CXOGSG. In addition, we downloaded the grating data associated with HD\,42054 as well as recent datasets, not included in the CXOGSG, with the Be stars V402\,Car, CW\,Cep, Cl*\,NGC\,3293\,FEAS\,32, and HD\,215227 in their field of view. These \ch\ data were locally reprocessed with CIAO 4.9 and CALDB 4.7.3. 

A source detection using wavelets ({\it wavdetect}) was run on these new datasets, considering the full energy band, PSF maps, and two scale sizes. Only Cl*\,NGC\,3293\,FEAS\,32 and HD\,215227 are detected, providing for the latter a confirmation of the \xmm\ detection. 

Regarding CXOGSG counterparts, we may note that most matches occur within 1'', including for the new detections. Only Cl*\,NGC\,869\,LAV\,1039 and ALS\,4570 show larger distances (1.2 and 1.9'', respectively) from their CXOGSG counterpart. In the latter case, this separation reaches nearly twice the minimum positional error (set to 1'' in CXOGSG), which makes this optical-X-ray match clearly an outlier compared to others. The associated data reveal that ALS\,4570 appears at a moderate off-axis angle. Its PSF is elongated, but this is also the case of neighboring sources such as HD164492B, which lies at 0.1'' of its  CXOGSG counterpart, CXOGSG J180223.6-230145, or V2282\,Sgr, which is at 0.4'' of CXOGSG J180216.7-230346, hence the Be counterpart identification may be somewhat doubtful. 

\begin{figure*}
\includegraphics[width=4.5cm]{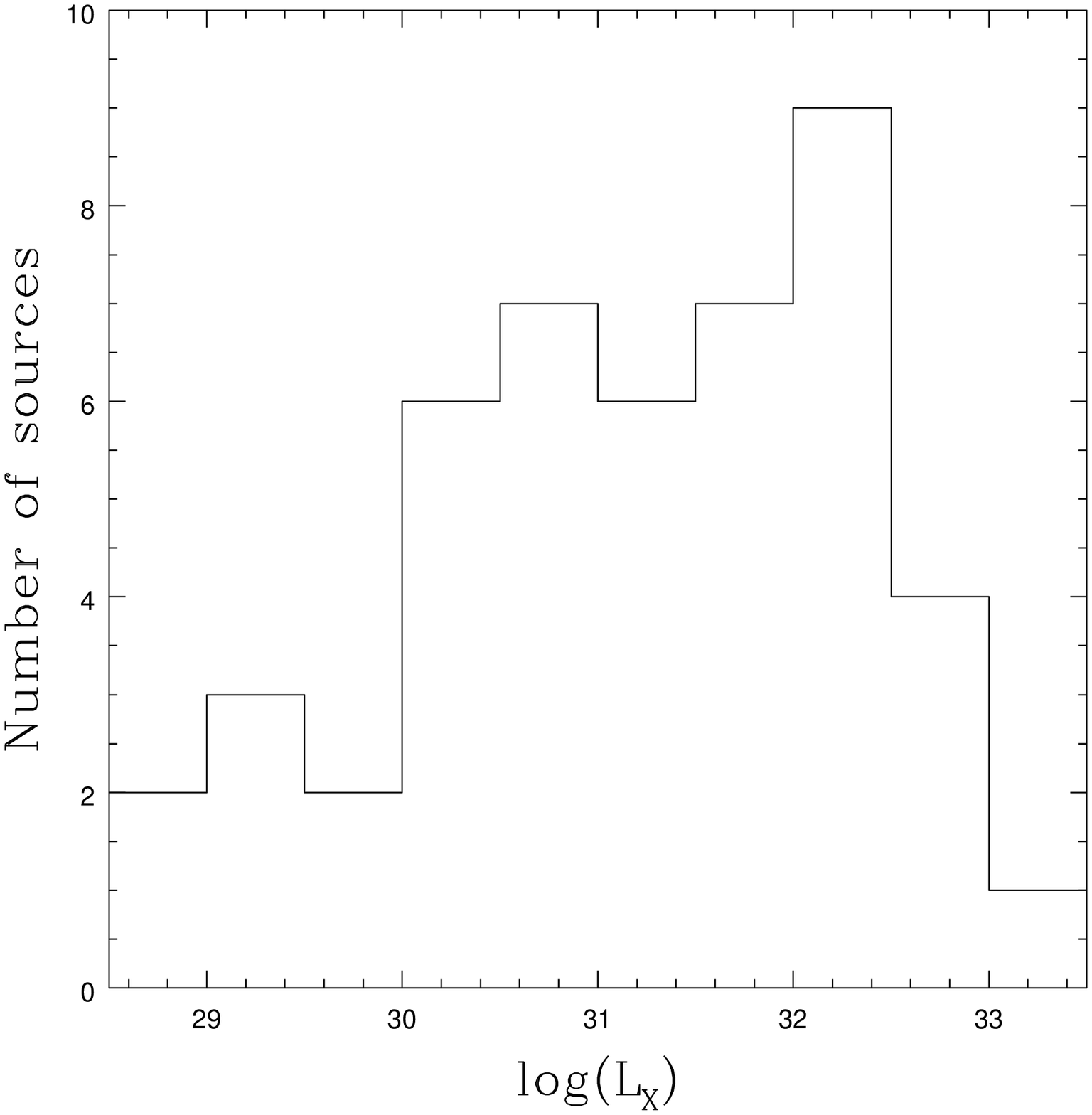}
\includegraphics[width=4.5cm]{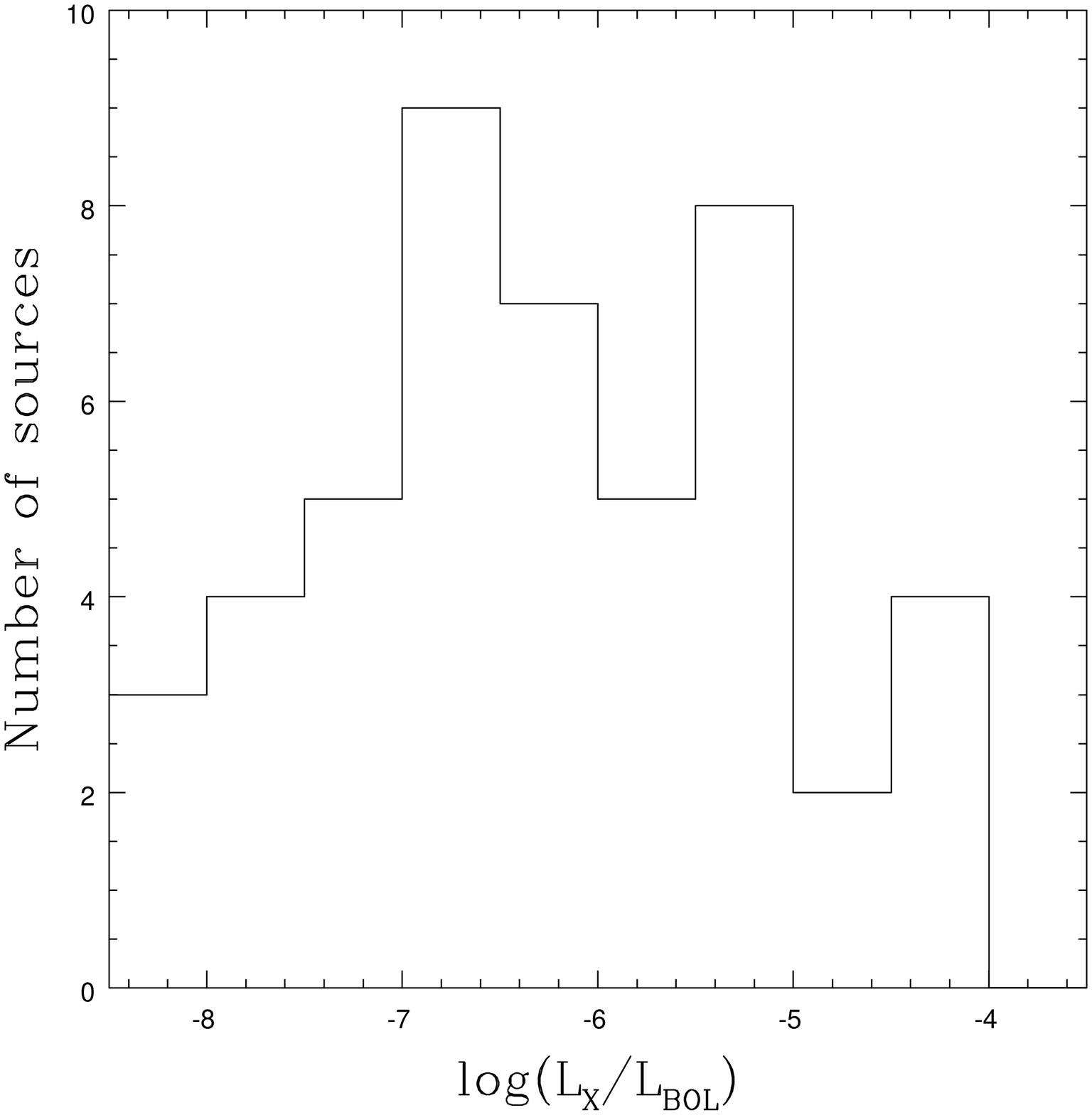}
\includegraphics[width=4.5cm]{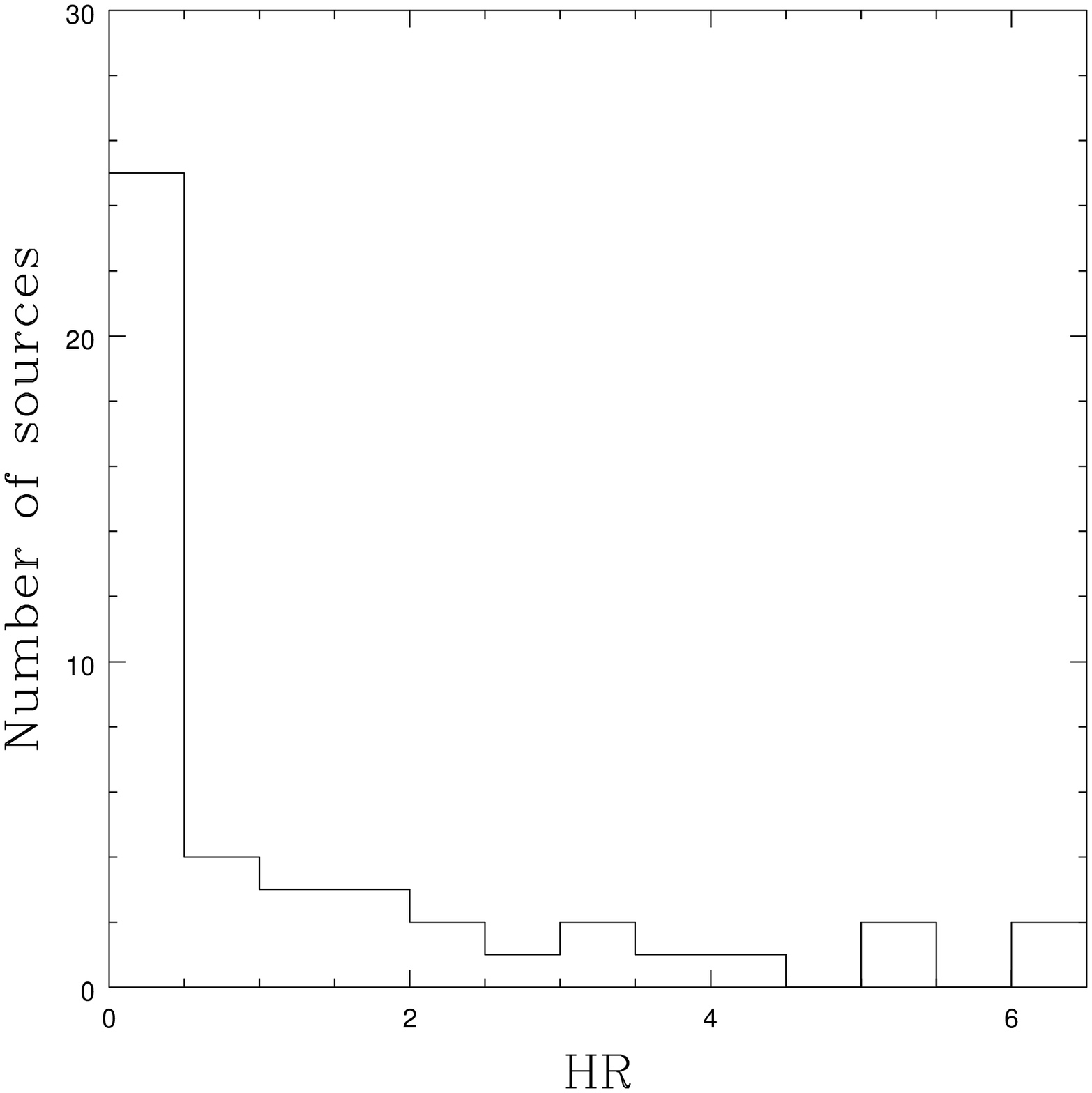}
\includegraphics[width=4.5cm]{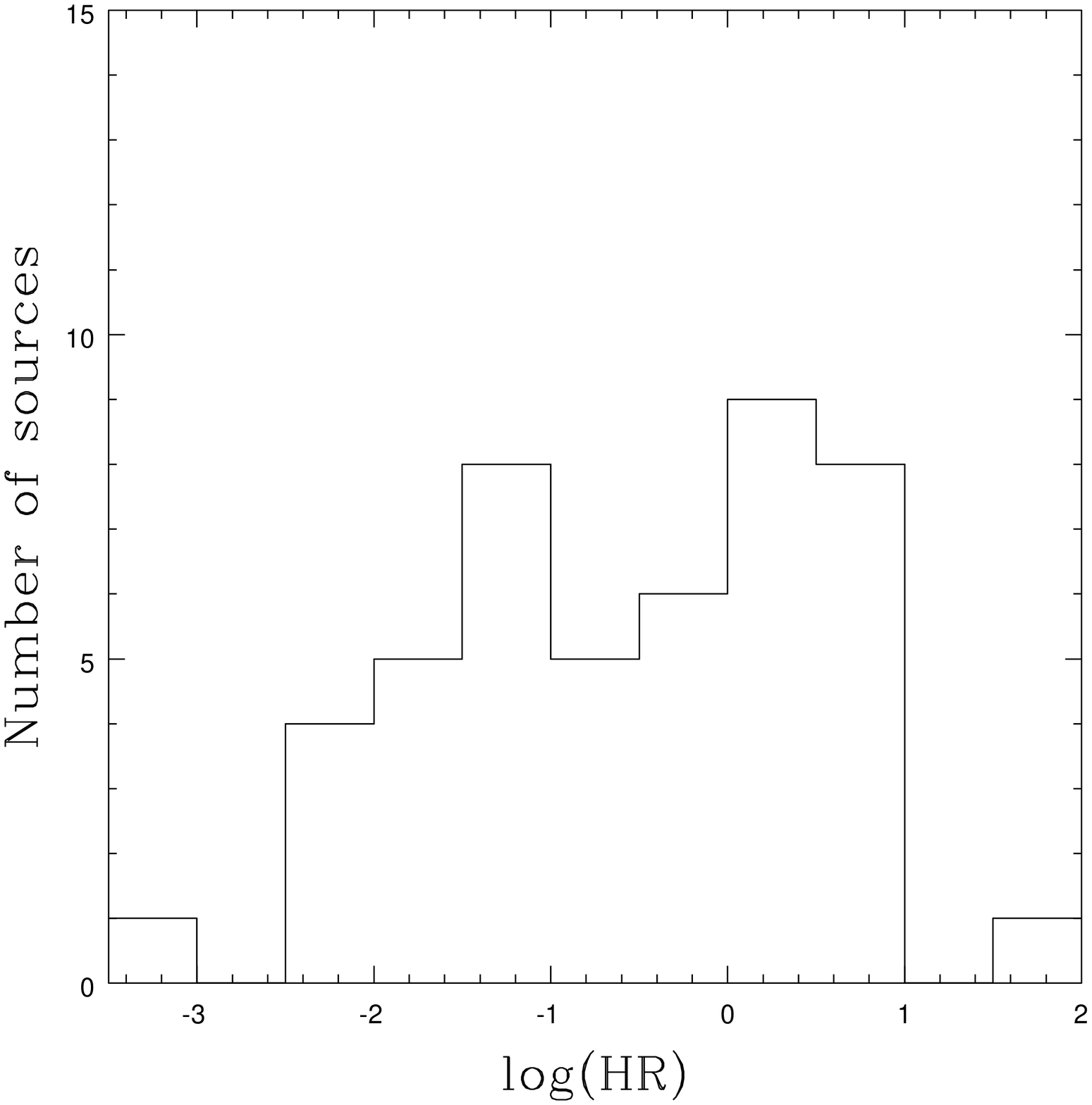}
\caption{Histograms of the X-ray luminosities, X-ray to bolometric luminosity ratios, and hardness ratios. When several observations of the same object existed, the values used to derive the histograms correspond to averages. The dubious cases ($\text{colons}$ in Table \ref{journal}) are not considered in these figures and that the high-luminosity parameters of PZ\,Gem were used. }
\label{histo}
\end{figure*}

ACIS spectra were then extracted using the task {\it specextract}, which also calculated the adequate {\it weighted} response matrices. Unweighted ARF and RMF matrices provide very similar results, although often with slightly higher fluxes, in all cases except for V374\,Car. Source regions were generally circles of 2.5'' radii centered on the Simbad positions of the sources, while the surrounding annuli, with radii 2.5'' and 7.5'', were used for background extraction. In some cases, however, the source region had to be elliptical rather than circular because of the PSF distorsion far off-axis; in other cases, depending on crowding, nearby circles also replaced the annuli for background definition. In addition, for bright sources, an annular source extraction was also performed to be able to check for pile-up. In case of grating data, zeroth-order spectra of the targets were always extracted. Moreover, if the target was observed on-axis and was X-ray bright (HD\,119682, HD\,42054, $\zeta$\,Oph, and 15\,Mon), spectra from orders +1 and --1 were combined using {\it combine\_grating\_spectra} to obtain the final HEG and MEG spectra. As for RGS, those high-resolution data are not reported here, but they were used to check the spectral fits. All \ch\ spectra were grouped in the same way as the \xmm\ spectra. 

Finally, it is important to note that we detected some pile-up in the ACIS-I, low-resolution spectra of 43\,Ori (most ObsID), 15\,Mon (ObsID 2550), HD\,119682 (ObsID 4554), and $\zeta$\,Oph (ObsID 14540). These spectra clearly appear harder than the available, non piled-up \xmm\ spectra, the zeroth-order spectra of the same targets in other \ch\ observations, or spectra corresponding to annular extraction regions. They were therefore discarded. Only the zeroth-order \ch\ spectra that were not affected were thus kept for these stars.

\subsection{\sw}
For a subset of stars detected in the XMM slew survey (Table \ref{journal}), XRT data from the Neil Gehrels \sw\ Observatory were requested. All but V4379\,Sgr are optically faint targets ($V>8.5$) and can therefore be observed with the XRT in PC mode. To avoid optical loading, the WT mode was used instead for V4379\,Sgr. The other XMMSL2 sources are too optically bright for the PC mode but also too X-ray faint for the WT mode and therefore cannot be observed by \sw. Data were processed locally using the XRT pipeline of HEASOFT v6.22.1 with calibrations v20170501. For V771\,Sgr, snapshots with ObsIDs 00043749001, 00043749003, and 00043749004 were discarded because very few counts were recorded for the source and/or a high background is present.

The source spectra were extracted within Xselect using circular regions of 20\,px (47.1\arcsec) radius. The surrounding annuli (with radii 20-60\,px, or 47.1-141.4\arcsec) were used for background for all but a few cases in which the source lies too close to the field-of-view edge; in these cases, a nearby circle devoid of sources was considered instead. The RMF matrix from the calibration database was used, but specific ARF response matrices were calculated for each dataset using {\it xrtmkarf}, considering the associated exposure map. Because of their small number of counts, we combined the spectra of the same source taken in different \sw\ exposures using the {\sc ftools} {\it addpha} and {\it addarf}. The weights for ARF combinations were in proportion to the number of counts collected in the individual spectra. Combinations of individual spectra were also provided by the online tool\footnote{http://www.swift.ac.uk/user\_objects/}. Their fitting formally yields slightly higher fluxes than the fitting of our locally processed and combined data, but the results remain similar within errors; the results provided below correspond to a simultaneous fitting of the online and local combined spectra. The spectra were binned using {\it grppha} in a similar manner as the \xmm\ spectra.  

\section{Results}

\subsection{Derivation of X-ray properties}
\subsubsection{Spectral fits}
We examined each individual spectrum and fit it only if there were enough counts (or bins) to do so, which was the case for 51 stars. Table \ref{fits} provides the fitting results. Fitting was made within Xspec v12.9.1p using absorbed optically thin thermal plasma models typical of massive stars X-ray emission (i.e., $tbabs\times phabs \times \sum apec$), with solar abundances of \citet{asp09}. In these models, the first absorption component represents the interstellar column, which was fixed to the value derived from the known color excess using the formula of \citet[$N_{\rm H}^{ISM}=6.12 \times 10^{21}\times E(B-V)$\,cm$^{-2}$]{gud12}. The second absorption allows for possible additional (local) absorption, for example, due to the stellar winds or the decretion disks. For the emission components, we used up to three temperatures: we added thermal components only if a single component did not provide a satisactory fit. For \xmm\ data, all available EPIC spectra were fit simultaneously. 

For the few sources in common, we checked our results with those reported in the Chandra Carina complex project \citep{naz11}, in the X-ray survey of magnetic stars \citep{naz14}, and in $\gamma$\,Cas analog papers \citep{lop07,lop10}. A good agreement was found (considering that slightly different distances and/or interstellar extinctions were sometimes used in those studies). 

For the brightest (and a few faintest) objects in the sample, the reduced $\chi^2$ of the best fits are sometimes larger than two, which rendered these fits formally inacceptable. These results were kept, however, notably because even in these cases, the fitting was good in the 0.5--10.0 keV energy band where the flux was estimated (some deviations may occur in 0.2--0.5\,keV). The problems mostly come from the impact of nonsolar abundances and/or unrealistically small error bars in the case of very bright sources (that do not account for the calibration systematics). For $\gamma$\,Cas objects, larger $\chi^2$ may also occur because their fluorescence Fe line is not fit (this choice was made to keep the spectral fitting homogeneous among all targets, and it did not affect the values of global fluxes or hardness ratios examined in this paper). In some cases, one spectral parameter (usually absorption) was fixed because unrealistic errors yielded erratic results that prevented us from deriving correct errors on the other parameters and on the fluxes. Finally, we calculated the X-ray luminosities, using the X-ray fluxes corrected for interstellar absorptions and the known distances (Table \ref{journal2}). The \loglxlb\ ratios further take into account the derived bolometric luminosities (see Section 2.4). Hardness ratios were calculated as the ratios between the interstellar medium (ISM) corrected fluxes in the hard (2.0--10.0\,keV) and soft (0.5--2.0\,keV) energy bands. 

\subsubsection{Simple detections}\label{sd}

When there were not enough counts for a meaningful spectral analysis, we derived the source X-ray luminosities, corrected for interstellar absorption, from the count rates. The \xmm\ catalogs provide such count rates in 0.2--12.\,keV (band 8) for MOS1, MOS2, and pn in the case of the XMM-DR7 and for pn only in the case of the XMM-SL2. The optical blocking filter that was used (thin, medium, or thick) is known for the DR7 cases, but not for the slew survey: for this survey, we considered both thin and thick filters to obtain extreme values. In one case, the detection is new (HD\,90563), so that the count rates of each instrument were estimated by a local run of the detection routine. For \sw, we derived the count rate in the 0.3--10.\,keV energy band from a run of the online tool\footnote{http://www.swift.ac.uk/user\_objects/}. For \ch, the CXOGSG catalog does not provide count rates; however, fluxes are available for most of the detections in the CSC (v1.1 or v2, although the latter is incomplete at the time of writing). These fluxes do not rely on any model, they simply correspond to the sum of the energies of each incident source photon, scaled by the local value of the ARF at the location of the incident photon.

We converted these count rates or observed photon fluxes using WebPIMMS\footnote{https://heasarc.gsfc.nasa.gov/cgi-bin/Tools/w3pimms/w3pimms.pl} for absorbed apec models, as suitable for massive stars. In these models, the absorbing column was fixed to the interstellar one (estimated using the color excess of Table \ref{journal2} and the formula of \citealt{gud12}). Furthermore, since the actual spectral properties are unknown and a range of temperatures was found for the thermal components in the spectral fits (see previous section), we decided to convert count rates for two temperatures in order to obtain upper and lower limits: values of 0.3\,keV and 15.\,keV were chosen as they were typically found as extremes in our spectral fits. Table \ref{det} provides the resulting ranges in X-ray luminosities (corrected for interstellar absorption) with the associated \loglxlb\ values. 

\begin{figure*}
\includegraphics[width=6cm]{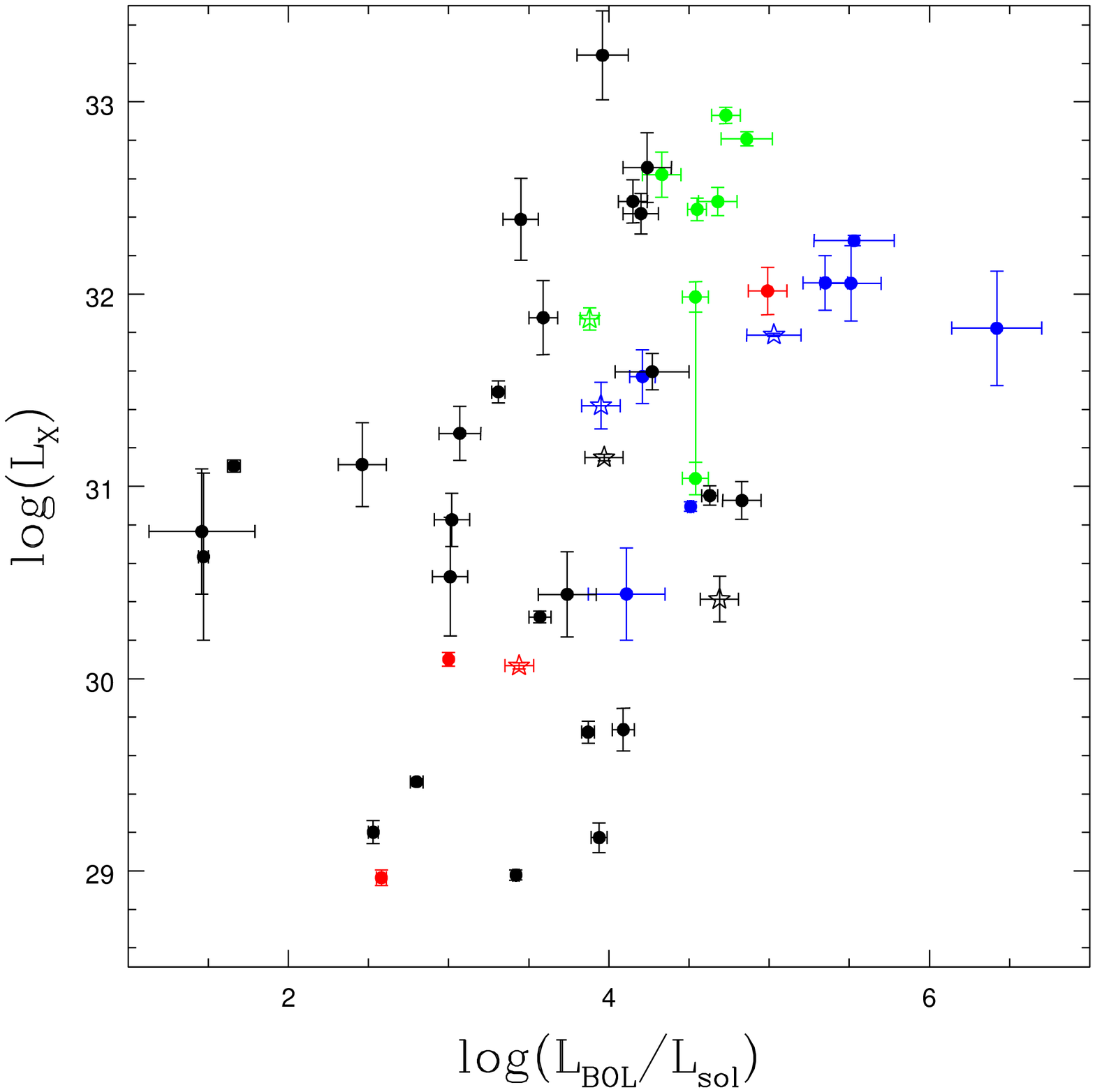}
\includegraphics[width=6cm]{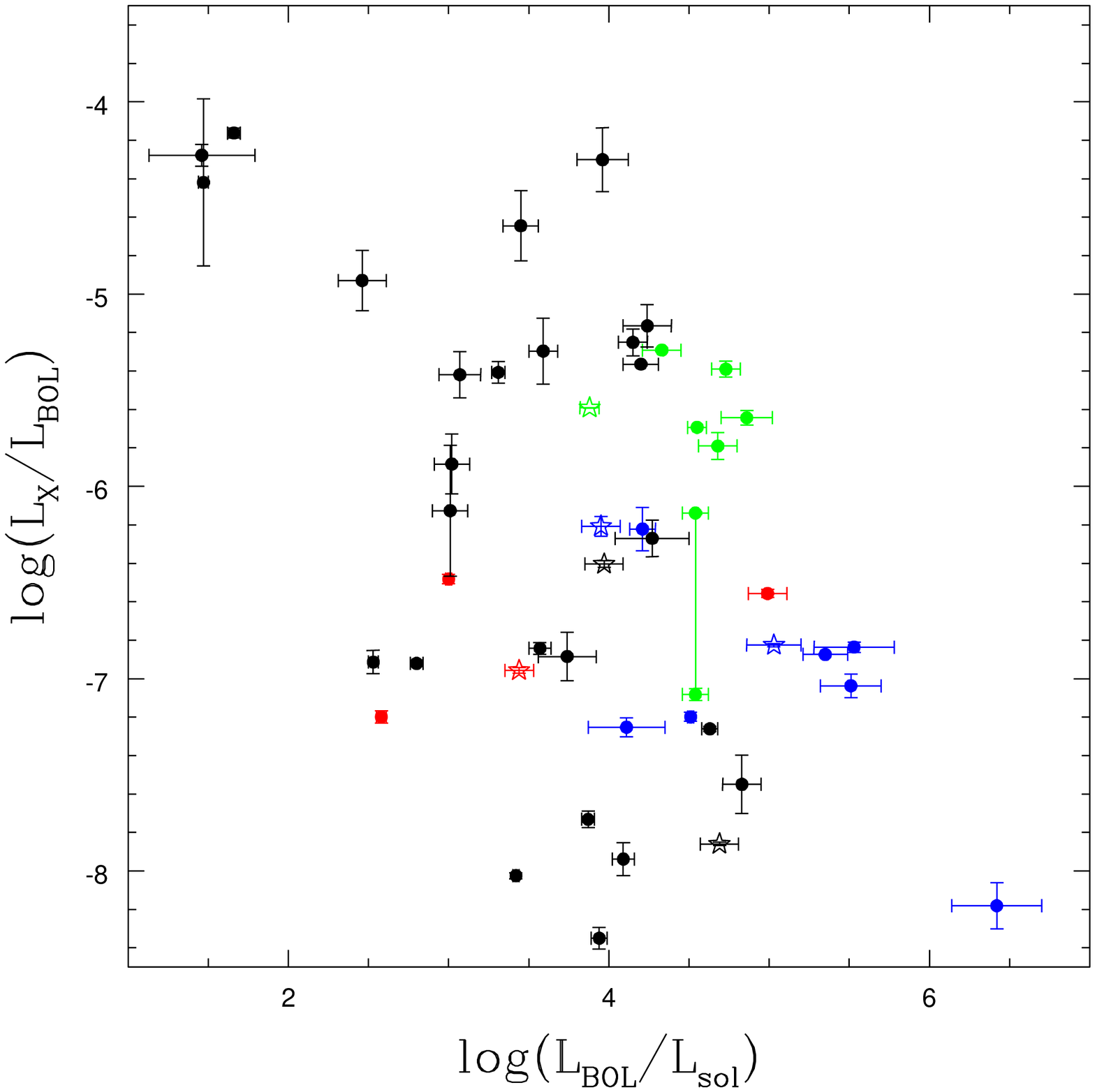}
\includegraphics[width=6cm]{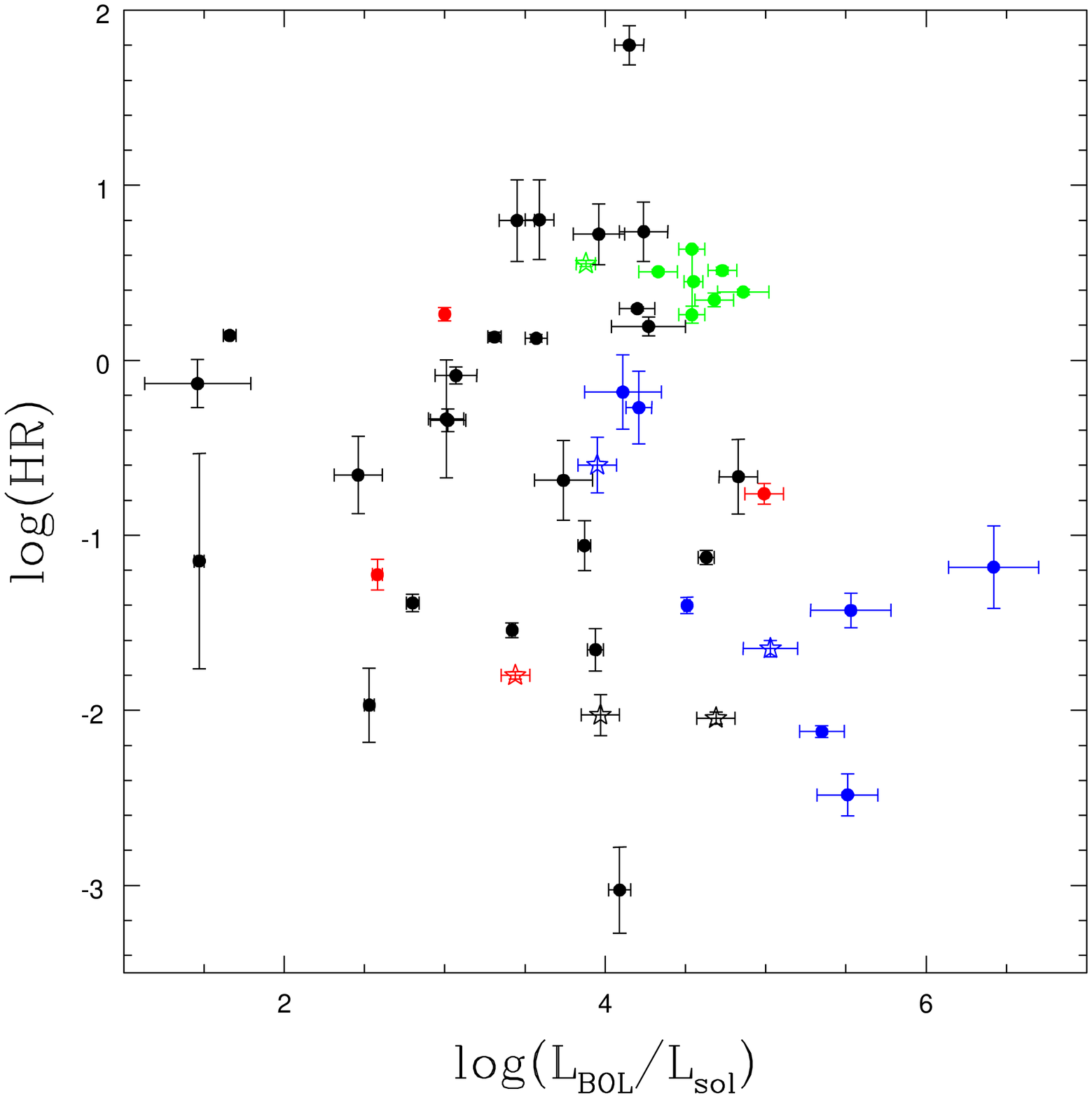}
\includegraphics[width=6cm]{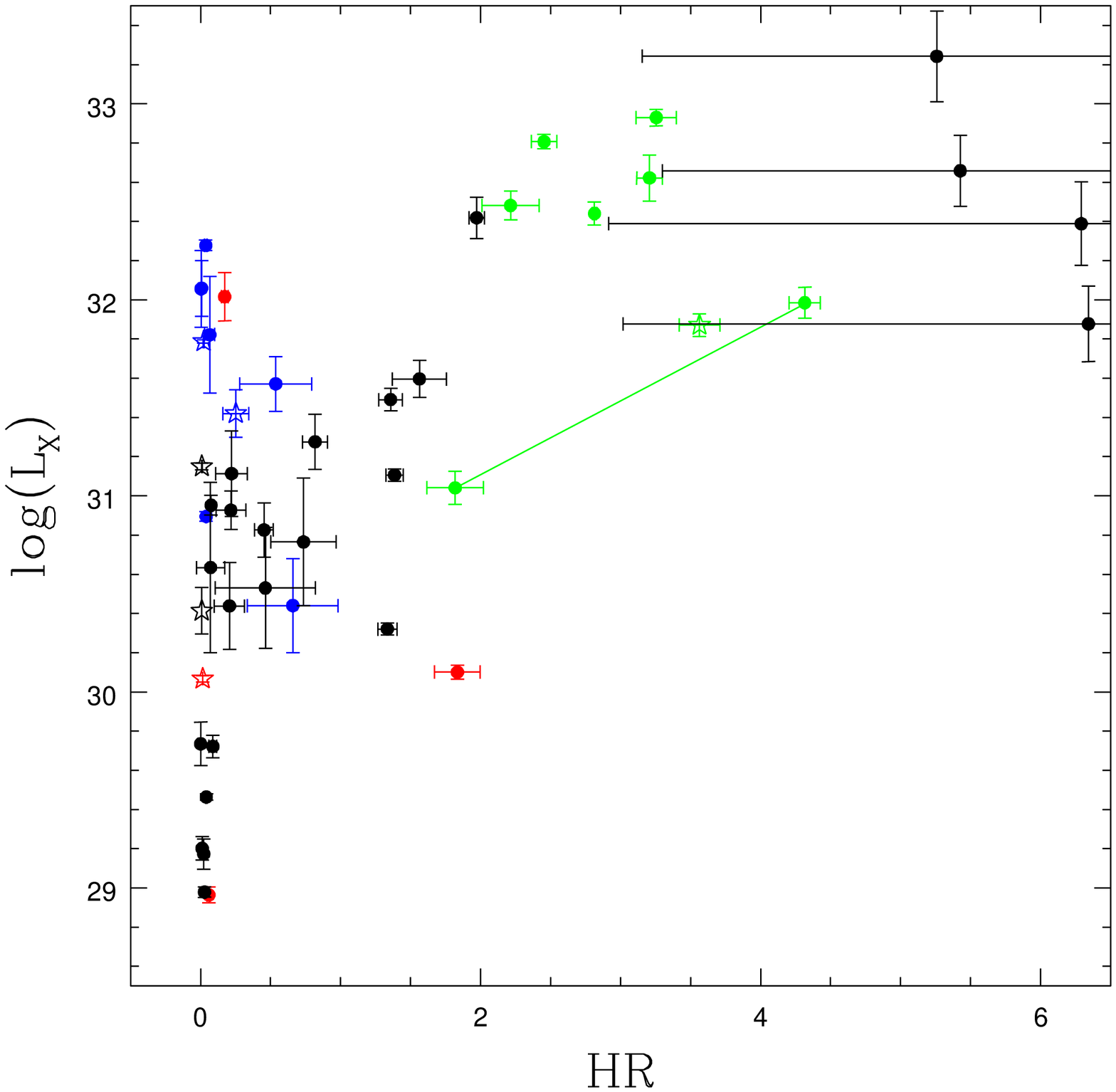}
\includegraphics[width=6cm]{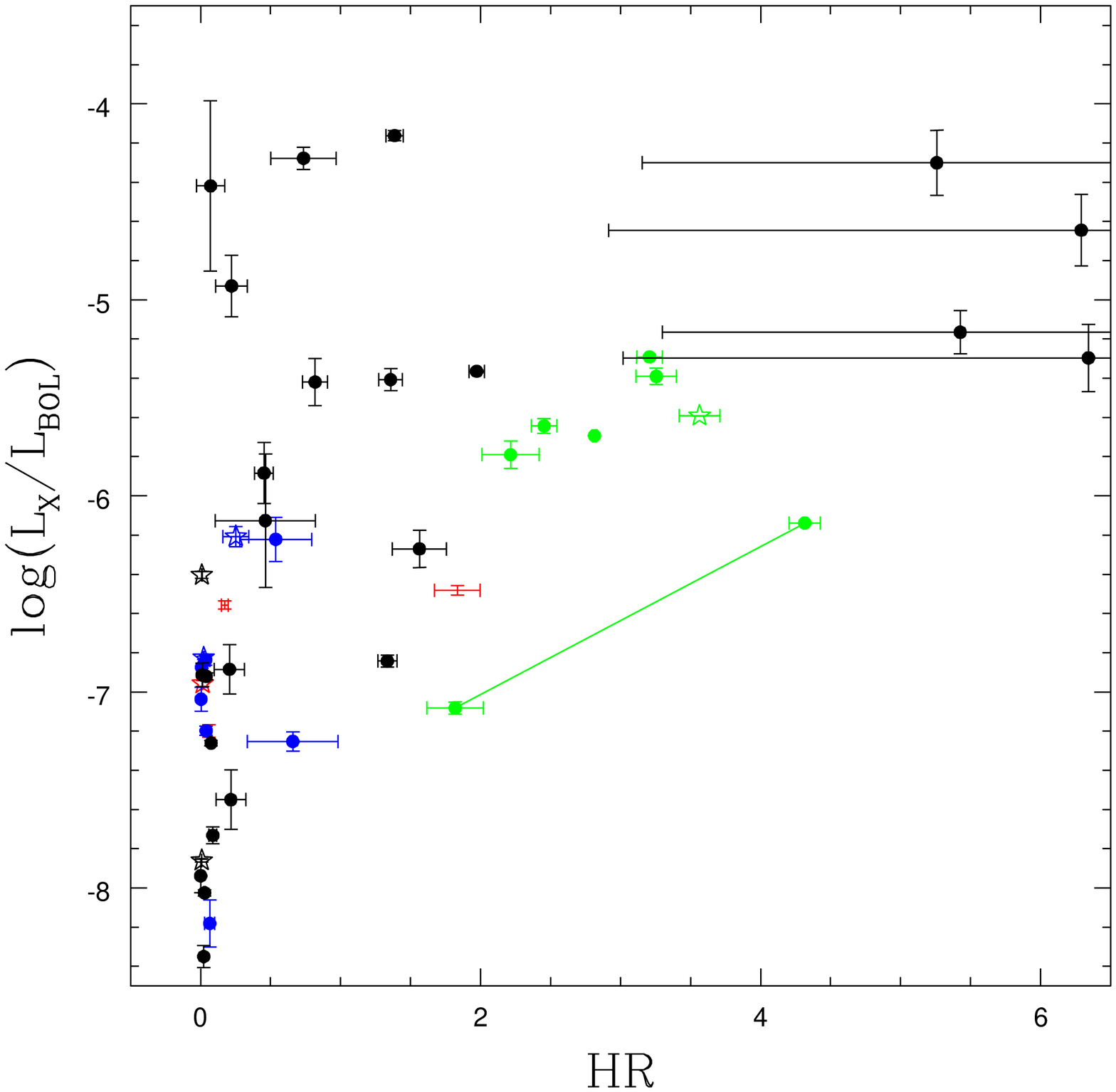}
\includegraphics[width=6cm]{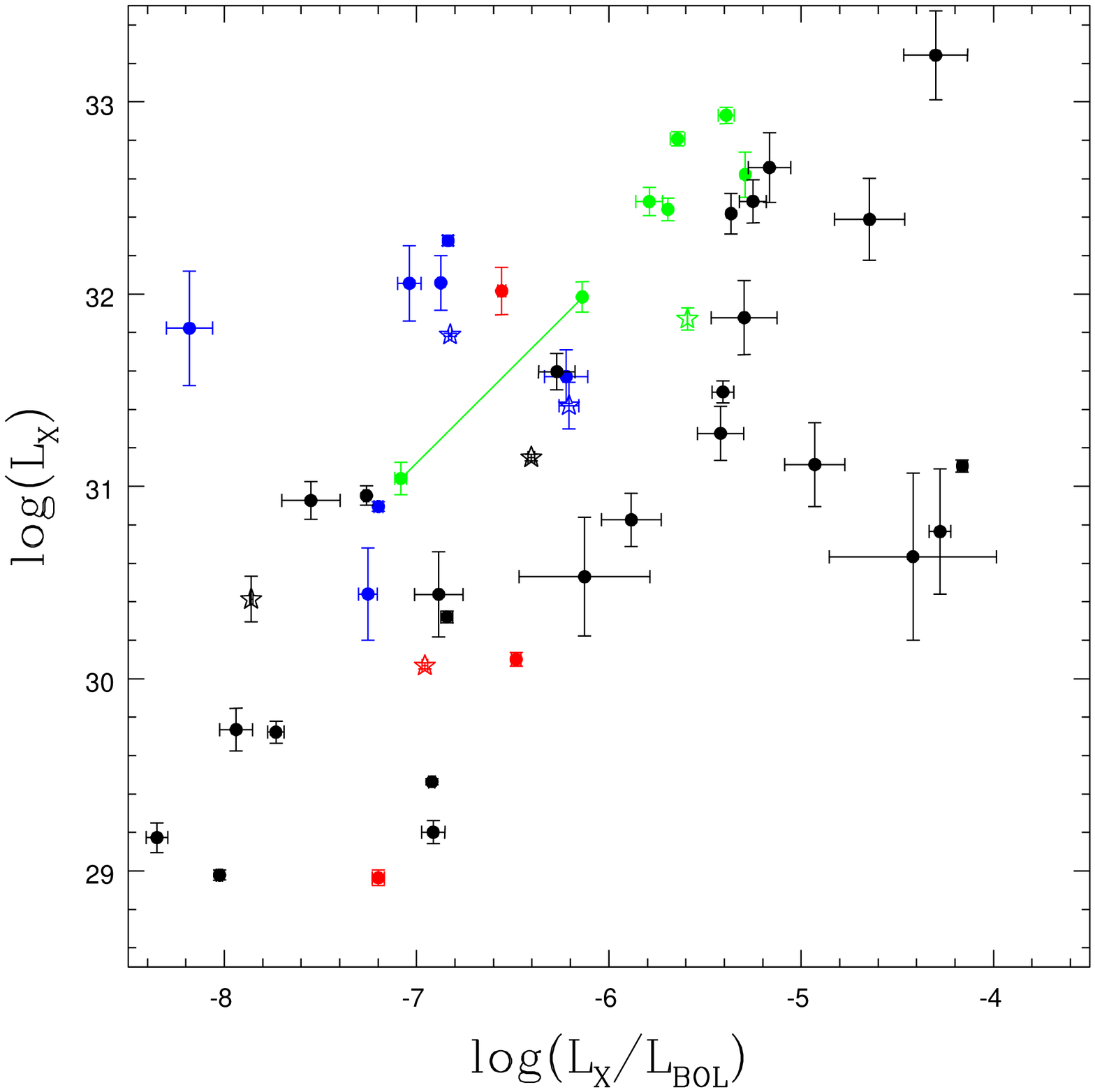}
\includegraphics[width=6cm]{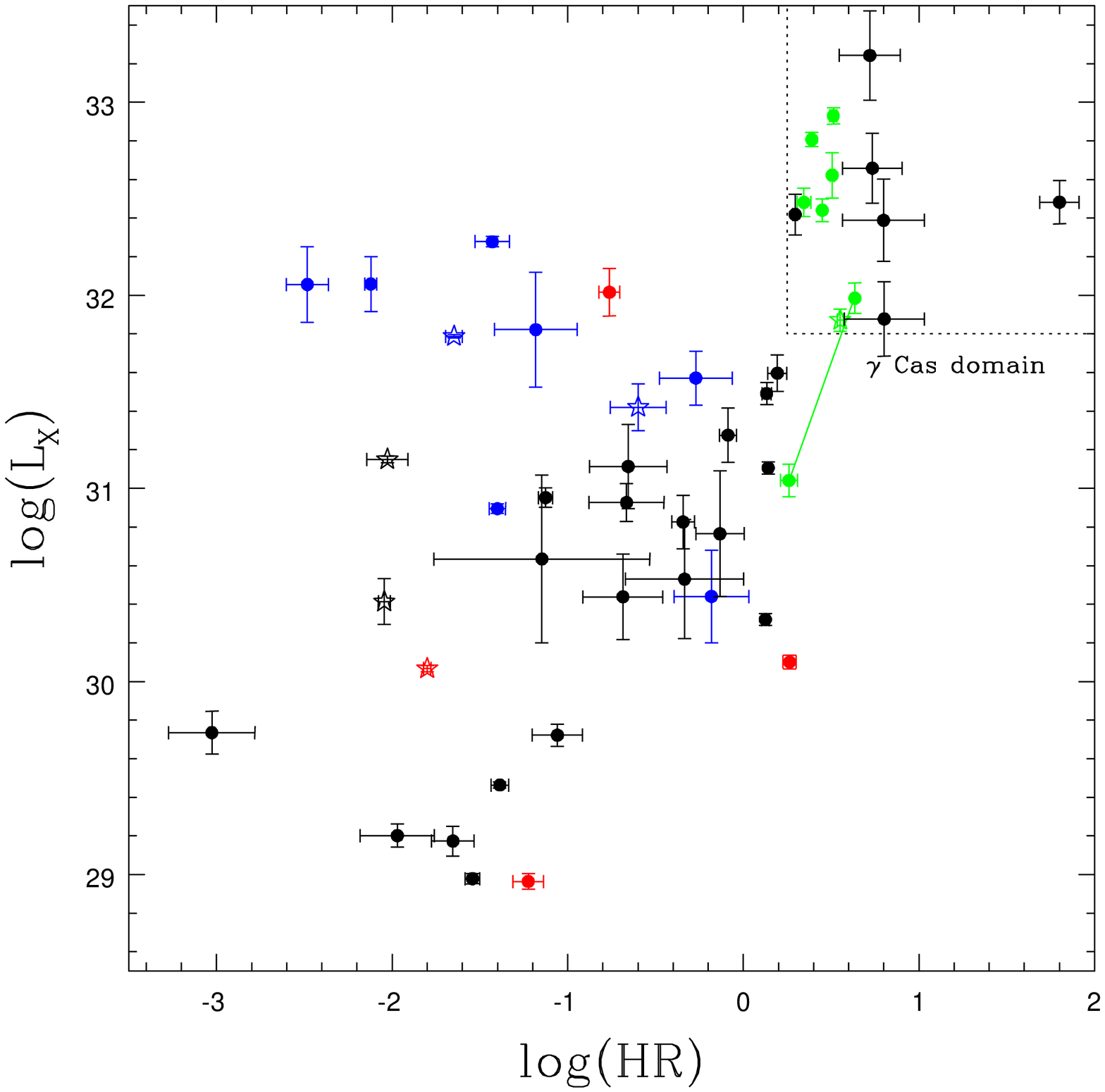}
\includegraphics[width=6cm]{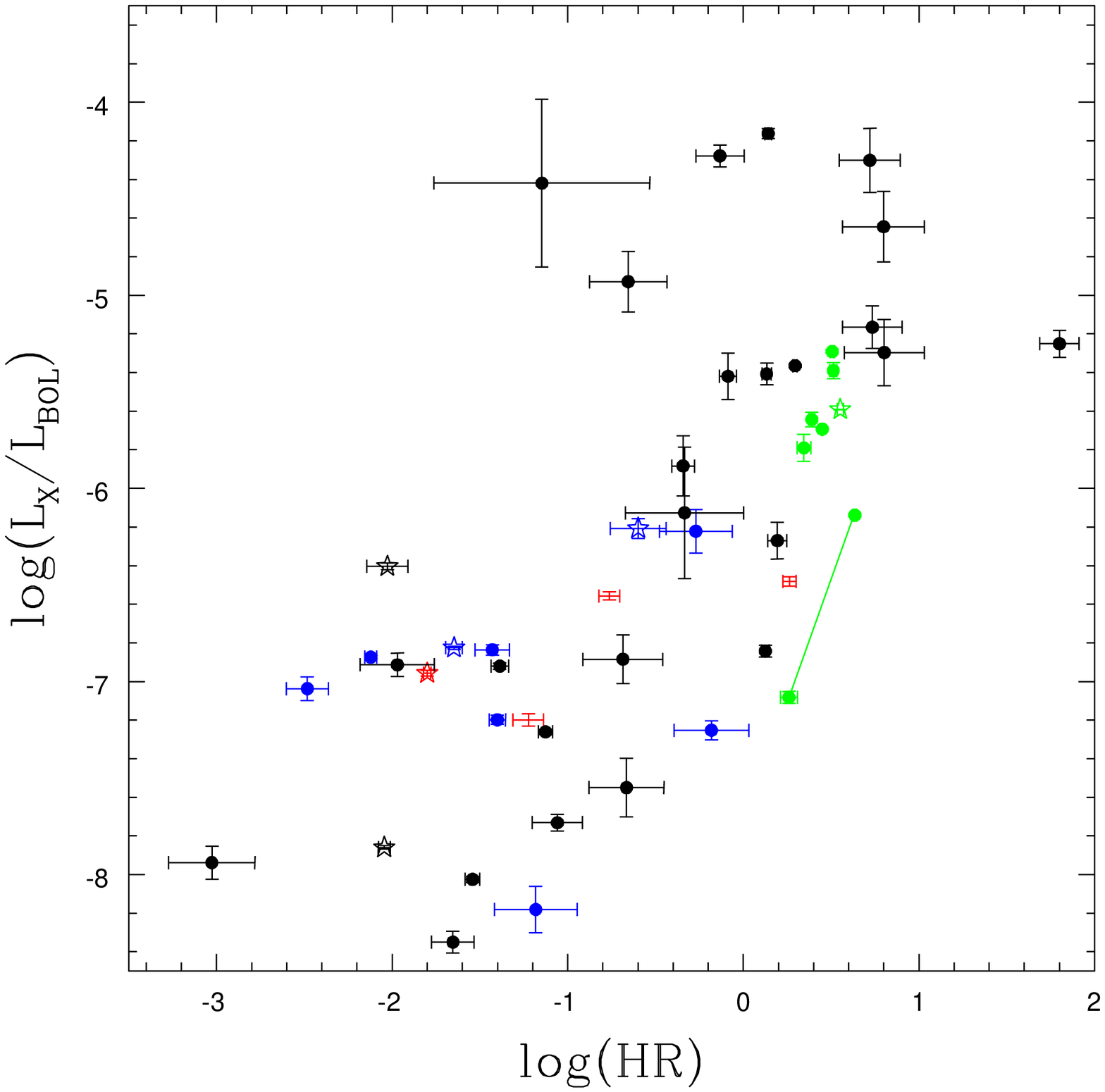}
\caption{Comparisons of X-ray luminosities, bolometric luminosities, X-ray to bolometric luminosity ratios, and hardness ratios, taken by pairs. Green, red, blue, and black symbols correspond to known $\gamma$\,Cas objects, known magnetic objects (including Alfirk because of its magnetic secondary), non-$\gamma$\,Cas and non-magnetic O-type stars, and all other stars, respectively. Stars are used for known binaries (see Table \ref{journal2}), simple dots otherwise. As in Fig. \ref{histo}, dubious cases are not shown and averages are used in case of multiple observations, the errors corresponding to the scatters around the means. For PZ\,Gem, both low-flux and high-flux cases are shown, as they differ significantly; they are connected so they can be easily spotted. }
\label{plot2d}
\end{figure*}

\subsection{Discussion}

While dedicated studies of a few Oe or Be stars exist, no large survey has been performed up to now. Only smaller studies have been published, with mixed results. \citet{meu92} reported the detection of ten Oe-Be stars in the {\it ROSAT all-sky-survey}. They found similar luminosities but a slightly lower detection rate for Oe-Be stars compared to OB stars. Also using {\it ROSAT}, \citet{coh97} in contrast reported a higher detection rate for Be stars than for B stars (all with types B1.5 or later). They also derived similar distributions in \loglxlb\ for both categories, although with a higher median value for their seven Be stars. These previous works possess two drawbacks: small statistics (only seven to ten stars in their samples) and a low-energy range (because of the use of {\it ROSAT}). Fortunately, this can now be easily corrected with the current generation of X-ray facilities, although the lack of all-sky surveys prohibits deriving global detection rates.

\begin{figure}
\includegraphics[width=9cm]{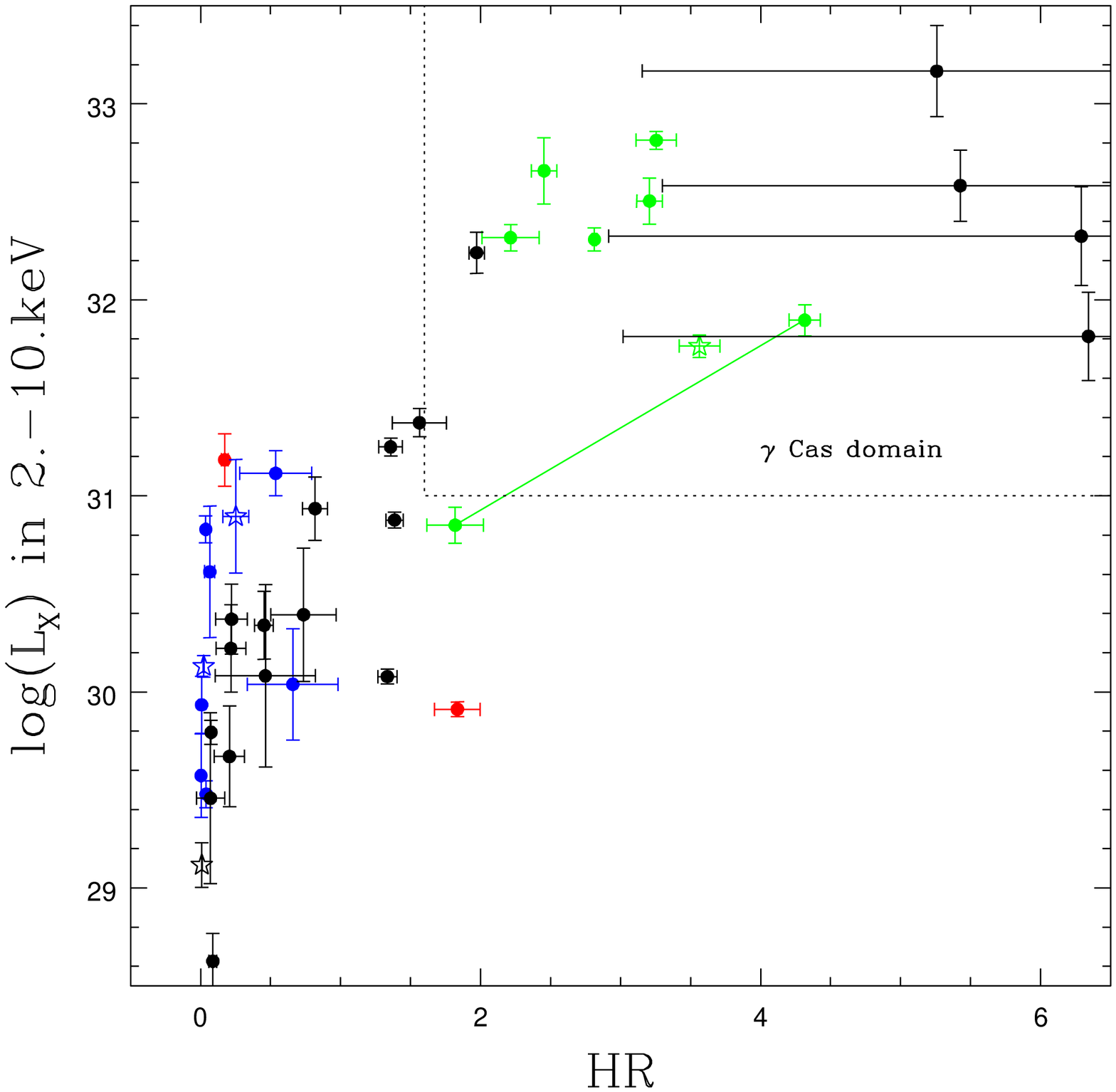}
\includegraphics[width=9cm]{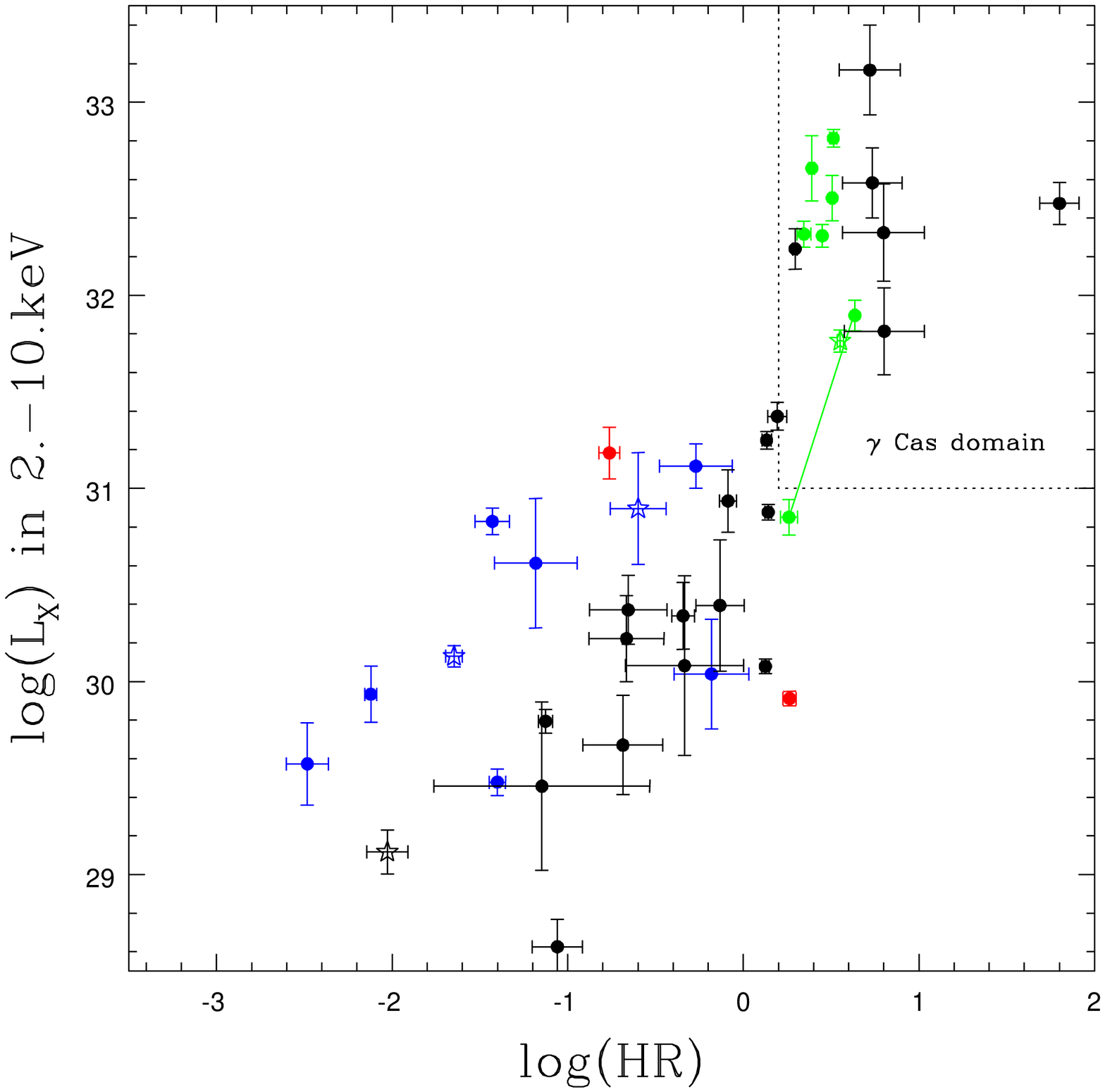}
\caption{Comparison of the X-ray luminosities in the hard (2.0--10.0\,keV) energy range and the hardness ratios. Symbols are the same as in Fig. \ref{plot2d}.}
\label{selection.fig}
\end{figure}

Figure \ref{histo} provides the histograms of the X-ray properties derived from our spectral fits (47 sources after discarding potentially dubious cases). Our targets cover a wide range of X-ray luminosities and \loglxlb\ ratios, larger than in general surveys of OB stars, but similar to what is seen for magnetic objects (see Fig. 2 of \citealt{naz14}). The presence of a second peak at \loglxlb $\sim-5.5..-5$ is unprecedented, however: in surveys, only few objects display such high ratios. In parallel, half of the stars display hardness ratios below 0.5: they are thus relatively soft sources as is usual for OB stars, magnetic or not (see the right panels of Fig. 8 of \citealt{naz14}). Again, there is a significant difference, however: our Be sample clearly displays a substantial fraction of much higher hardness values, which is not typical for OB stars. 

\begin{figure*}
\includegraphics[width=6cm]{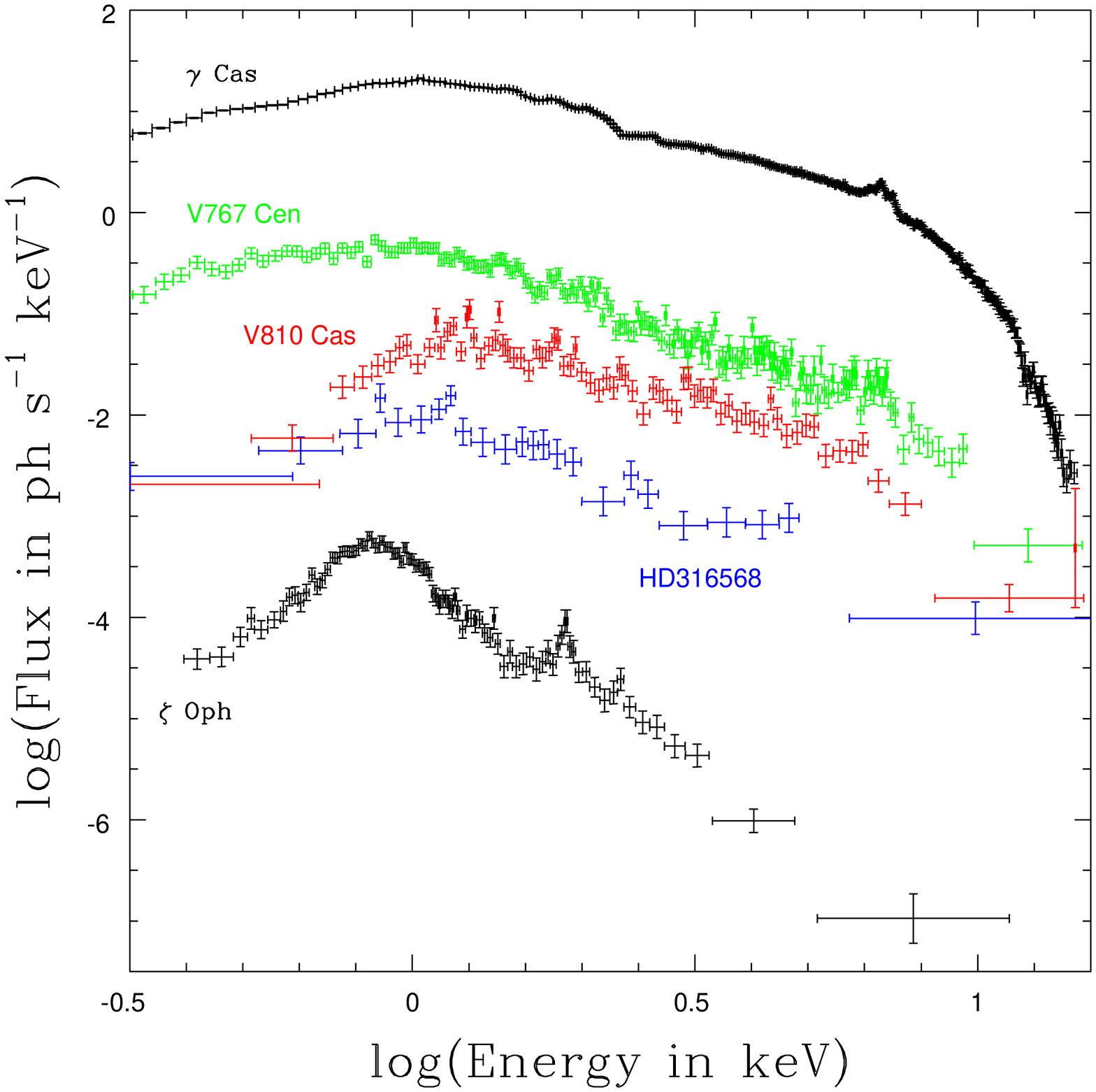}
\includegraphics[width=6cm]{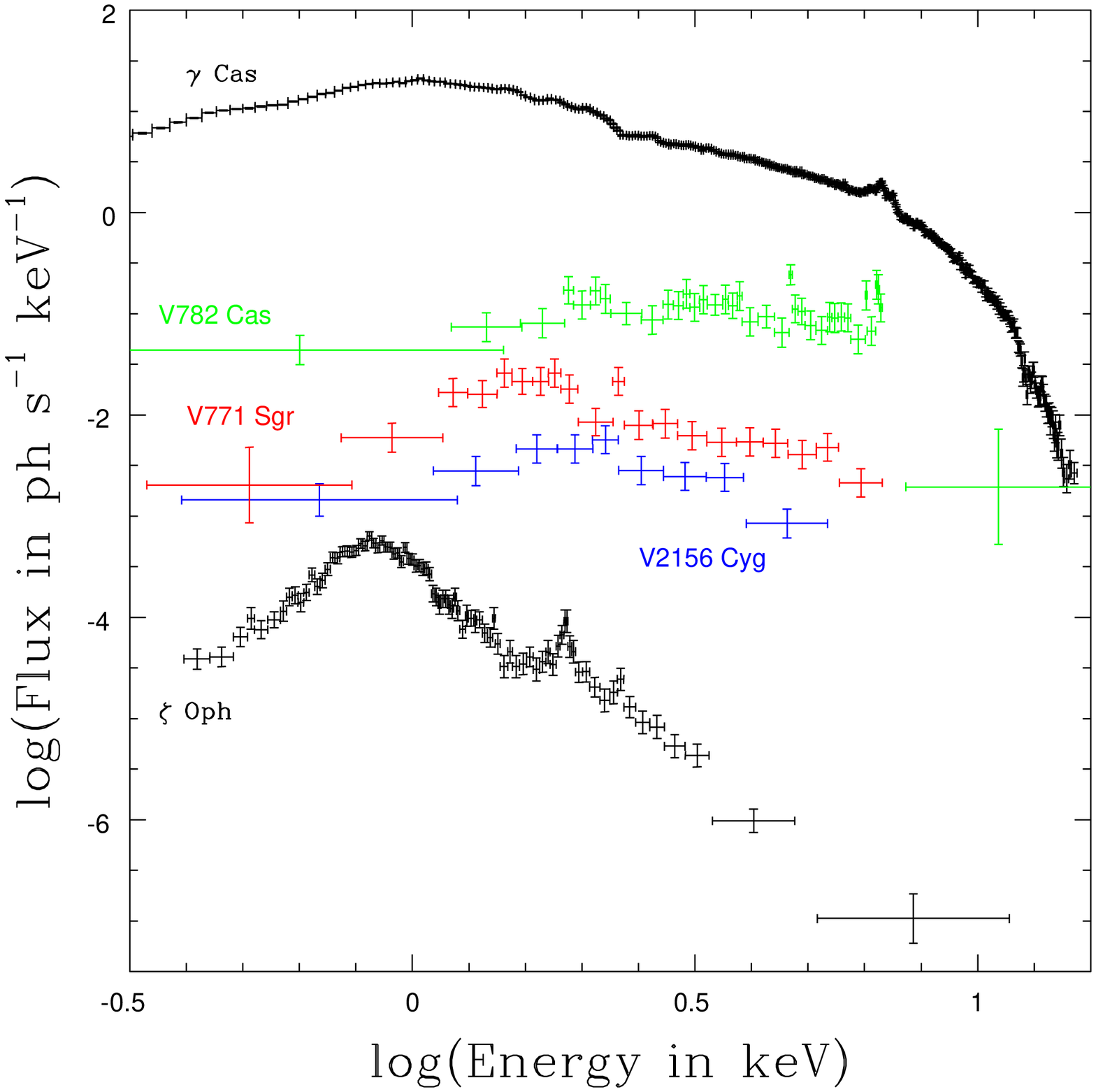}
\includegraphics[width=6cm]{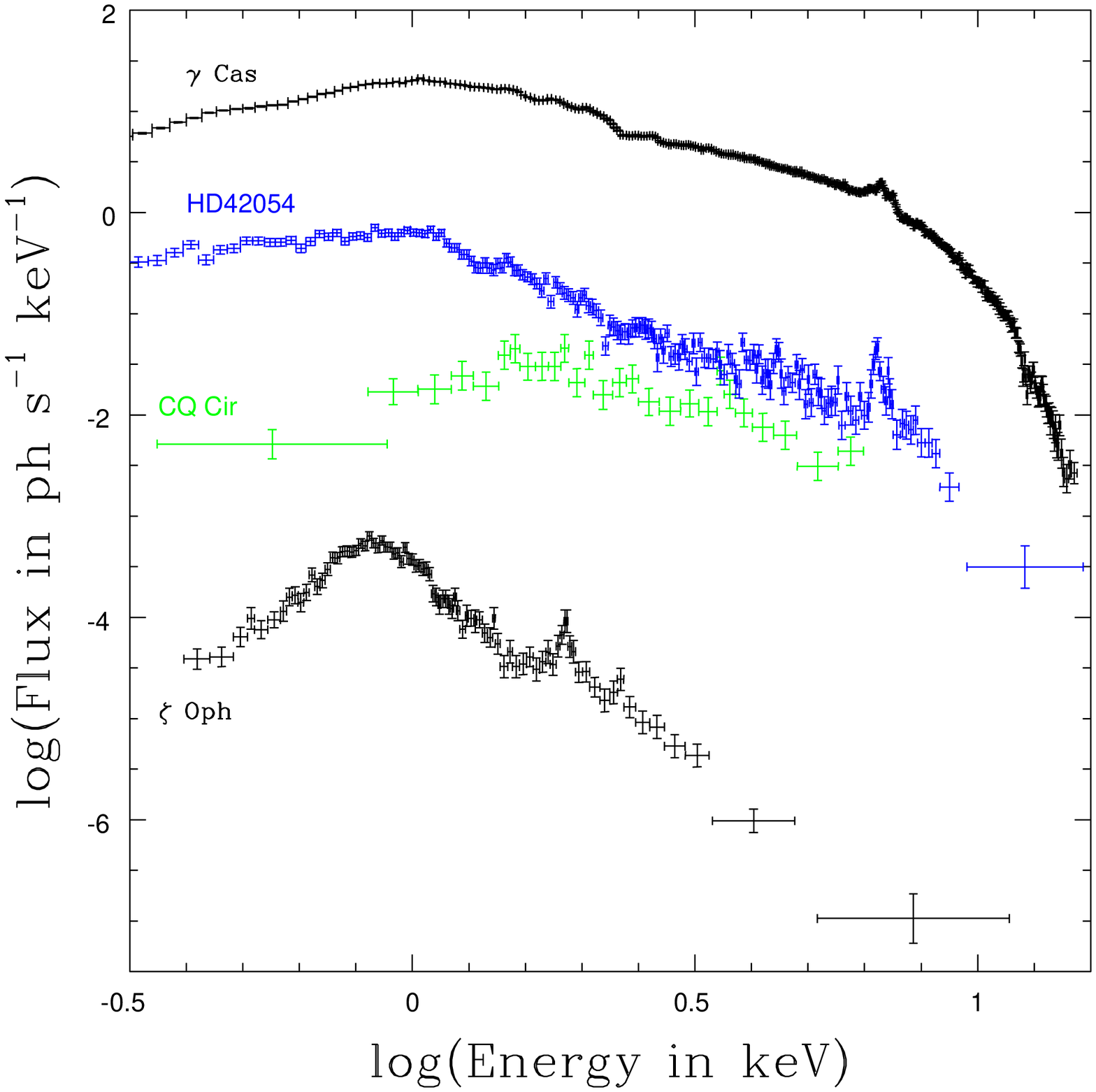}
\caption{Spectra of seven new $\gamma$\,Cas analogs and of one candidate, compared to those of $\gamma$\,Cas and $\zeta$\,Oph. The spectra slopes at high energies clearly are much flatter than for ``normal'' sources such as $\zeta$\,Oph. For clarity, the spectra of  $\zeta$\,Oph and V782\,Cas have been shifted by --2.5\,dex and +1\,dex, respectively.}
\label{spec}
\end{figure*}

More detailed plots (Fig. \ref{plot2d}) show a general trend between hardness and brightness, that is, the brightest sources (both in terms of \Lx\ and \loglxlb) always display a harder emission. The associated correlation coefficient is not very high, however: it reaches 69\% when considering \loglxlb\ and $\log(HR)$, and only 19\% when considering \Lx\ and $HR$. It may also be interesting to note that Fig. \ref{plot2d} does not show a clear-cut separation between stars or groups of stars. In particular, while the known $\gamma$\,Cas analogs appear mixed with a few objects presenting the same properties (see Section 4.2.2 below), the whole category (known cases + new detections) does not appear to be fully isolated from the other sources of the sample. 

\subsubsection{Oe and Be that are not $\gamma$\,Cas analogs}

Fig. \ref{plot2d} (for spectral fit cases) and Table \ref{det} (for HD\,117357) show that Oe stars display a soft emission with the typical \loglxlb\ of O-stars (\loglxlb $\sim-7$). The only exceptions are $\gamma$\,Cas analogs (see next subsection), BD--13$^{\circ}$4928, and the known binary 43\,Ori. In the latter case, colliding wind emission, magnetic activity, and/or contamination by companions have been discussed in the literature as possibly affecting its X-ray emission (see \citealt{sch06,gag08}). In particular, small-amplitude and short-term flares were attributed to a pre-main-sequence (PMS) neighbor \citep{gag08}, but we detect in our dataset a much larger variation in X-ray luminosity than reported before. In a \ch\ exposure (ObsID 4474), 43\,Ori brightens by one order of magnitude, reaching $\gamma$\,Cas characteristics (see below). However, some pile-up may artificially harden the spectrum in this case, therefore we do not count it as a secure $\gamma$\,Cas candidate. A more thorough investigation with more observations clearly needs to be performed to understand the nature of the X-ray emission of this system. Up to now, BD--13$^{\circ}$4928 was unknown to display a particularly bright X-ray emission. However, with only a slightly harder character than common O-stars, it certainly does not show a rather flat spectrum at high energies, typical of the very hot plasma detected in $\gamma$\,Cas objects. The most likely candidates for explaining its X-ray properties are thus colliding winds or magnetic confinement, hence this target should be monitored, both optically and in X-rays.

Magnetic objects display a wide spread in X-ray luminosities, wider than for the Oe stars in our sample and $\gamma$\,Cas objects. However, they remain less hard and less luminous than the latter ones, and their properties are in line with those reported in an X-ray survey of magnetic stars \citep{naz14}. 

PMS companions of massive stars are difficult to detect in the optical range, therefore it is possible that some of our Oe and Be targets possess such a companion. This is not without consequence on the X-ray properties.  PMS stars can indeed reach $10^{31}$\,erg\,s$^{-1}$ when flaring, and they also appear to be harder ($kT\sim1-2$\,keV) than typical O-stars at the time. The top panels of Fig. \ref{plot2d} show that the latest stars in our sample, that is, those with the lowest bolometric luminosities ($\log(L_{\rm BOL}/L_{\odot})<2$), have $L_{\rm X} \sim 10^{31}$\,erg\,s$^{-1}$ and moderate hardness ratios, reflecting such confusion with a nearby active star. The situation appears more mixed for $\log(L_{\rm BOL}/L_{\odot})=$ 2 to 4, as lower X-ray luminosities and hardness ratios also exist. The large scatter observed in this region suggests that it contains both contamination and true emission from the Be stars.

\subsubsection{$\gamma$\,Cas analogs: new detections}

One of our objectives was to determine whether the census of $\gamma$\,Cas analogs in current data was complete. Before we address this point, the characteristics of such stars must be recalled. $\gamma$\,Cas analogs were essentially defined on the basis of the X-ray and optical properties of the two brightest objects, namely, $\gamma$\,Cas itself and BZ\,Cru. The overall characteristics of these two stars and of other $\gamma$\,Cas analogs are summarized in \cite{smith2016}. Their most outstanding X-ray feature is the presence of a strong thermal component with kT $\geq$ 5-6\,keV (although lower temperature components also exist, but with a lower intensity), yielding a broad band X-ray luminosity of $\sim 10^{32-33}$\,erg\,s$^{-1}$. The X-ray luminosity is thus intermediate between that of ``normal'' OB stars and of HMXBs. 

Our analyses confirm the properties of the known $\gamma$\,Cas analogs and further provide additional criteria that may constitute interesting alternatives or complements to existing criteria. The hardness ratio, defined as the ratio between the hard (2.0--10.0\,keV) and soft (0.5--2.0\,keV) ISM-corrected fluxes, is usually easy to compute with a few counts and is an excellent proxy for temperature: Table \ref{fits} shows that $HR>1.6$ correspond to kT $\geq$5\,keV. In addition, the \loglxlb\ ratio is also high for $\gamma$\,Cas analogs, usually reaching at least --6, a value that is extreme even for magnetic or colliding-wind O stars. Finally, Fig. \ref{selection.fig} shows that $\gamma$\,Cas analogs are also remarkable in terms of hard X-ray luminosity, with $L_{\rm X}(hard)\geq 10^{31}$\,erg\,s$^{-1}$. It may be noted that such boundaries exclude the possible contributions of even the most extreme PMS companion stars.

All previously known $\gamma$\,Cas analogs dwell well inside this parameter region. However, they are not alone: these criteria enable us to discover eight new analogs (Table \ref{candidates.tab}, Fig. \ref{findingchart}). Spectra are available for seven of them (Fig. \ref{spec}):
\begin{itemize}
\item V782\,Cas: this star displays a high peculiar space velocity of 102 km\,s$^{-1}$ \citep{ber01}, suggesting that binary evolution affected it. It might either be a kicked binary system, or more classically, a disrupted runaway star. 
\item V767\,Cen: this star displays a very low projected rotational velocity $v \sin(i)$, suggesting that it is seen nearly pole-on, as was subsequently shown by \citet{fre02}. A spectropolarimetric FORS observation was reported by \citet{sch2017}, with a formal 2--3$\sigma$ measurement of a weak dipolar magnetic field but larger, 5$\sigma$ levels are generally required for FORS (see \citealt{bag15}), hence confirmation is awaited. Based on a preliminary assessment of its X-ray properties, \citet{sch2017} proposed V767\,Cen as a possible $\gamma$\,Cas analog. 
\item CQ\,Cir: this star is often referred to as a Herbig Ae/Be PMS star, but it was found to be in the classical IR Be region by \citet{mat08}, who explained this oddity as due to a peculiar evolution scenario in which an HAeBe star looses its disk and therefore dereddens to a classical Be location. However, the observed X-ray luminosity is at least two orders of magnitudes above that emitted by the most extreme Herbig Ae/Be stars \citep{stel06}.
\item V771\,Sgr
\item HD\,316568: this star shows a high temperature - hardness ratio, and hard X-ray luminosity, but a slightly lower broad-band X-ray luminosity than the other analogs.
\item V2156\,Cyg
\item V810\,Cas
\end{itemize}
Even if none of those stars has an X-ray luminosity typical of HMXBs (which are brighter than $10^{34}$\,erg\,s$^{-1}$), we have also attempted a power-law fit: the power-law photon exponent is typical of what is found in the known $\gamma$\,Cas objects ($\Gamma\sim1.5-1.8$) for all but V782\,Cas (which has a small $\Gamma\sim0.8$) and V810\,Cas ($\Gamma\sim1.2$, a value typical of HMXBs). It is certain that these seven objects display a very hard spectrum and intermediate X-ray luminosities, both reminiscent of what is seen in $\gamma$\,Cas analogs.

For simple detections (see section \ref{sd}), two sources appear in the range of $\gamma$\,Cas luminosities: HD\,90563 and BQ\,Cru. Doubts have already been detailed in section 3 about the detection of BQ\,Cru, but the case of HD\,90563 is different. Not only does it have an X-ray luminosity in the range of $\gamma$\,Cas analogs, but most of the counts were collected in the hard band: regardless of the EPIC camera, the count rate in 2.0--10.0\,keV is at least twice that in the 0.5--2.0\,keV energy band. This star must therefore be added to the list of new $\gamma$\,Cas analogs.

\begin{figure}
\includegraphics[width=9cm]{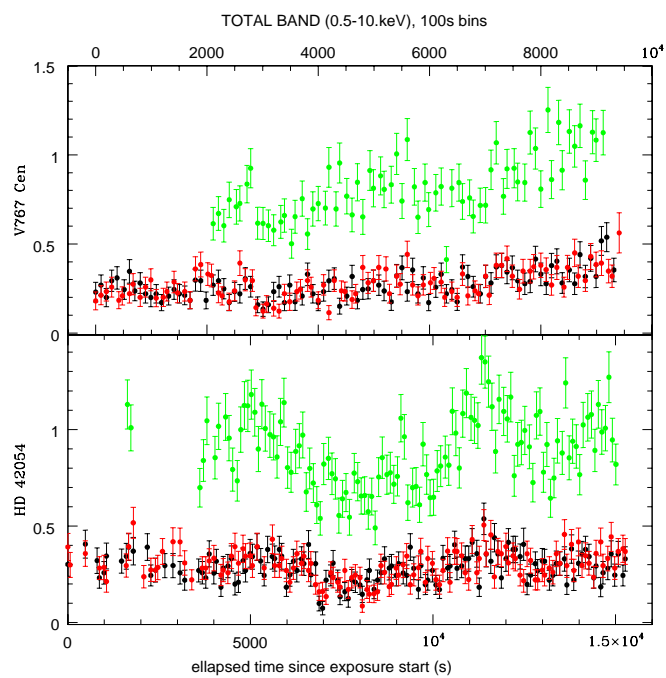}
\caption{Background-corrected \xmm\ light curves of V767\,Cen and HD\,42054, derived from the same extraction regions as the spectra and corrected for the loss of photons due to vignetting, off-axis angle, or other problems such as bad pixels using {\it epiclccorr}. EPIC-MOS1, MOS2, and pn data are shown in black, red, and green, respectively. Bins with effective exposure times shorter than the total bin length (100\,s) are not shown.}
\label{variab}
\end{figure}

$\gamma$\,Cas analogs share several other characteristics that may be taken into account when assessing the likelihood that a star belong to that class. For instance, the presence of a fluorescence iron line next to the thermal lines from ionized iron at 6.7\,keV is also obvious in $\gamma$\,Cas analog spectra, although it requires a well-exposed spectrum to be detected \citep{gim15}. For our new detections, the iron line complex clearly appears in the spectrum of V782\,Cas and there may be hints of its presence for V767\,Cen and V810\,Cas (the spectral bin around 6\,keV appears slighly above the best-fit spectrum). However, there are not enough counts to test for its presence in the other cases with available spectra.

\begin{figure*}
\includegraphics[width=8cm]{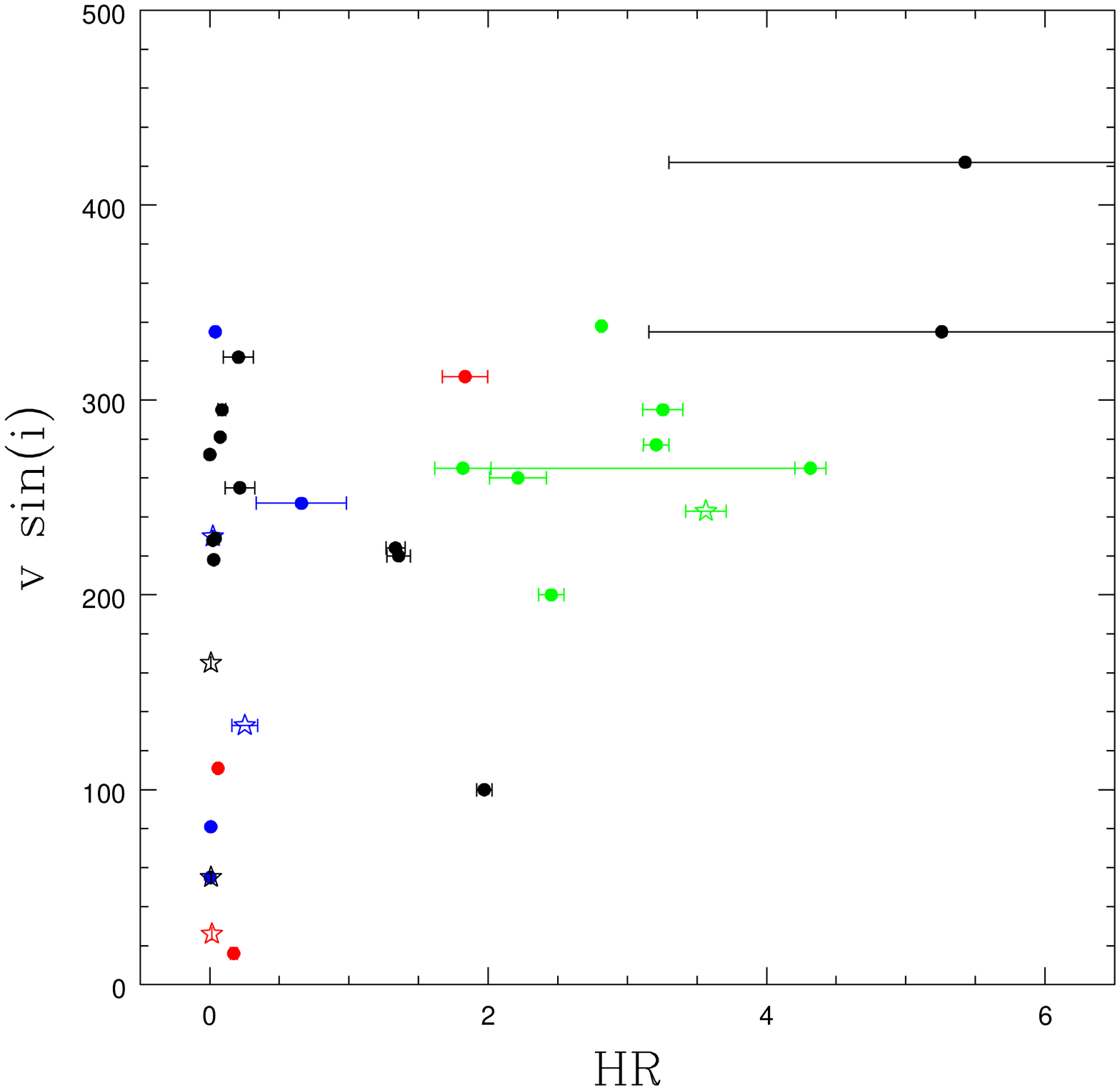}
\includegraphics[width=8cm]{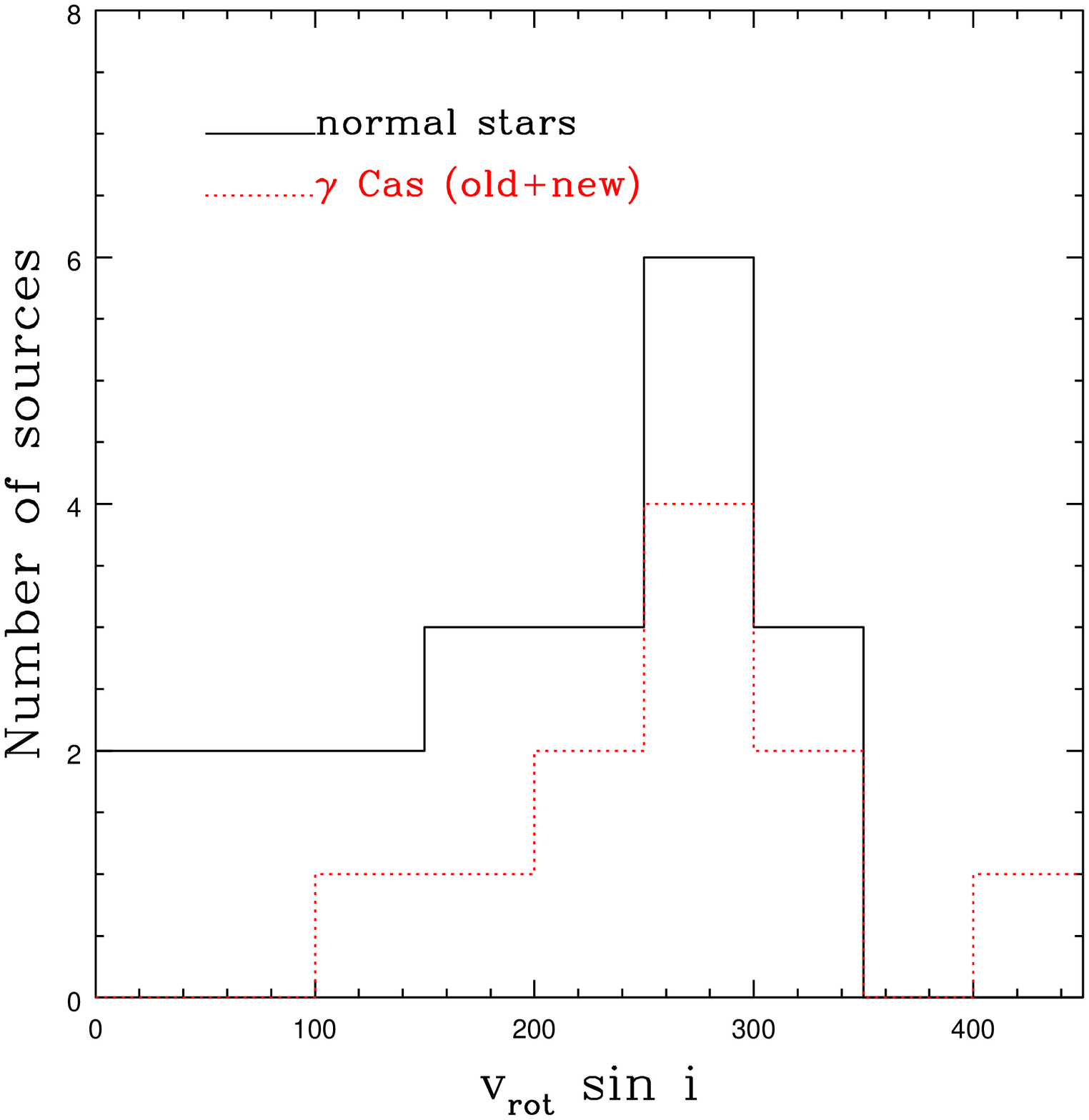}
\caption{{\it Left:} Comparisons of projected rotational velocities (when known, see Table \ref{journal2}) to hardness ratios. Symbols are the same as in Fig. \ref{plot2d}.  {\it Right:} Histogram of projected rotational velocities for all $\gamma$\,Cas analogs (previously known ones and new ones) and stars without any $\gamma$\,Cas characteristics but with similar effective temperatures ($\log(T_{eff})$= 4.28 to 4.55).}
\label{vsini}
\end{figure*}

In addition, the spectral fitting of $\gamma$\,Cas analogs often reveals an absorbing column larger than the interstellar one, and this may vary from one observation to the next \citep[e.g.,][]{smi12}. Additional absorption is detected for our new detections, except for HD\,316568, but other Oe and Be stars also display local absorption: this criterion is thus a (very) loose one. 

All $\gamma$\,Cas analogs also exhibit X-ray variability on various timescales, down to timescales that are limited by the count rate (i.e., about 10 seconds for the X-ray brightest ones; \citealt{smi98,lop07,smi12,ham16}). In addition to fast X-ray shot-noise like variability, large amplitude modulations on timescales of one hour or more are ubiquitous \citep{smith2016} and long-term variations linked to the disappearance and building-up of the disk have also been detected \citep{rau18}. A correlation between X-ray luminosity and disk density was indeed found by \citet{mot15} and \citet{rau18}. The presence of short-term variability was assessed when enough counts were present (see the variability flag in Table \ref{journal}). V767\,Cen appears significantly variable, with a strong increasing trend during its exposure (Fig. \ref{variab}), while V782\,Cas, HD\,316568, and V810\,Cas are not statistically variable. Longer-term variability can be assessed if several exposures separated by weeks, months, or years are available. When this was the case, fluxes (Table \ref{fits}) or count rates were compared and changes were detected in all cases: the flux of HD\,316568 changes from 4.0$\pm$1.3 to 6.5$\pm$0.6$\times 10^{-14}$\,erg\,cm$^{-2}$\,s$^{-1}$, the \sw\ count rate$^{10}$ of CQ\,Cir varies between 0.045$\pm$0.011 and 0.116$\pm$0.008\,cts\,s$^{-1}$, and the \sw\ count rate$^{10}$ of V771\,Sgr is found in the interval 0.025$\pm$0.010 -- 0.098$\pm$0.013\,cts\,s$^{-1}$.

Optically, a dense decretion disk seems also required for a star to display a $\gamma$\,Cas behavior. Unfortunately, except for $\gamma$\,Cas and $\pi$\,Aqr, no optical spectra at the exact dates of the X-ray observations are available in the BeSS database for $\gamma$\,Cas analogs. However, available data at other times seem to confirm this tendency (Table \ref{candidates.tab}). 

Known $\gamma$\,Cas analogs also display a limited range in spectral type, from late-O to early-B. This is also the case of our new cases, although V782\,Cas and V771\,Sgr may be of slightly later type than the bulk of $\gamma$\,Cas analogs. We note, however, that the spectral types are quite uncertain for these two stars. 

Finally, we examined the rotational velocities. In our sample, the harder emissions are usually reached only for the fastest rotators (left panel of Fig. \ref{vsini}), but there is no formal, significant correlation between rotation velocity and X-ray luminosity overall. For $\gamma$\,Cas objects, the origin of the magnetic field in the star-disk interaction paradigm is essentially unconstrained, but \cite{mot15}  proposed that the phenomenon appears when the star rotates very close to the critical velocity so that subsurface convecting layers may eventually generate an equatorially condensed magnetic field with no large-scale structure. Measuring the V$_{\mathrm{rot}}$/V$_{\mathrm{crit}}$ quantity requires an estimate of the rotation axis inclination, however. Assuming that disk and star rotation axis are coaligned, inclination can be obtained by optical or infrared interferometry measuring the oblateness of the circumstellar disk \cite[see][e.g.]{ste12}. Only few data are available, but in the scantily populated group of Be stars with interferometric measurements, the two most critically rotating Be stars are $\gamma$\,Cas and BZ\,Cru, and they exhibit the $\gamma$\,Cas phenomenon. Conversely, none of the other Be stars with available X-ray data in this sample exhibits $\gamma$\,Cas-like properties, thus apparently supporting the proposed mechanism.

For our sample, without the knowledge of the inclination, we are left with the comparison of projected rotational velocities. Since the critical velocity depends sensitively on the spectral type \cite[see, e.g.,][and references therein]{zorec2016} and $\gamma$\,Cas analogs display spectral types between B2.5 and O9, we compared the $v_{rot} sini$ distribution of the $\gamma$\,Cas objects with known rotational velocities to that of the Be stars in our survey in the same range of effective temperatures ($\log(T_{eff})$= 4.28 to 4.55) but do not show any sign of $\gamma$\,Cas activity. Comparing these two samples originating from the same input list and processed in the same manner is probably the best manner to minimize selection effects. The right panel of Fig. \ref{vsini} shows that $\gamma$\,Cas stars tend to have higher projected rotational velocities than non-$\gamma$\,Cas stars of similar spectral types, thus supporting the idea that high (or critical) rotation may be a key ingredient of the $\gamma$\,Cas phenomenon. However, owing to the small number statistics, the difference is not really statistically significant when formally using a Kolgomorov-Smirnov test. 
  
\medskip

Although they do not formally fulfill all criteria, some other objects also appear quite remarkable in our sample because of their bright/hard X-ray emission. The first two cases, HD\,42054 and V1230\,Ori, appear close to the $\gamma$\,Cas domain (Fig. \ref{selection.fig}), although they display slightly lower hardness ratios ($1<HR<1.6$) that reflect plasma temperatures of 2--5\,keV. They are also variable, both on short (Table \ref{journal}) and long (Table \ref{fits}) timescales. While V1230\,Ori is not bright enough to compete with $\gamma$\,Cas objects and some confusion with a nearby active star cannot be totally excluded for it, HD\,42054 appears particularly remarkable (with $<L_{\rm X}(hard)>=1.8\times 10^{31}$\,erg\,s$^{-1}$) and can be considered as a $\gamma$\,Cas candidate. This star, which was the subject of several studies, was not detected to have a sdO companion \citep{wan18} and is not a runaway star \citep{tet11}. Based on its irregular optical variations measured by {\it Hipparcos}, \citet{rim12} found it to be similar to $\gamma$\,Cas, with a probability of 42--86\% (depending on the method used). Compared to the bulk of the established $\gamma$\,Cas, it displays somewhat softer X-rays, slightly less intense hard X-ray luminosities, a later spectral type, and smaller $EW(H\alpha)$ (Table \ref{candidates.tab}, Fig. \ref{findingchart}), but its spectra clearly show a strong iron line complex (Fig. \ref{spec}) and remarkable short-term variations (oscillating behavior detected during the \xmm\ exposure, see Fig. \ref{variab}). It may represent the first case of a low-intensity $\gamma$\,Cas phenomenon. 
  
Two other sources, HD\,17505 and EM*\,MWC\,659, have X-ray luminosities in the range $10^{33}-10^{34}$\,erg\,s$^{-1}$ (Table \ref{det}), probably too high for $\gamma$\,Cas analogs and more similar to HMXB luminosities. HD\,17505 is a multiple O-star system, and it remains to be seen how this influences the recorded level of X-ray emission. In any case, the data are scarce for both objects (only a simple detection), therefore more information is required before we definitively conclude on the nature of these two objects.

\section{Summary and conclusion}

We have performed a survey of Oe and Be stars in the X-ray range using \xmm, \ch, and \sw\ data. To this aim, we have cross-correlated the BeSS catalog of Be stars with the 3XMM-DR7, \ch-CXOGSG, and XMM-SL2 catalogs. We also searched for more recent public observations of such stars (not yet included in these catalogs). Some shortcomings are inherent to this approach: the BeSS database may be incomplete, Be phases may be missed for some objects (which would thus not appear in the list of Be stars), some X-ray datasets may not yet be public, and upper limits were not investigated (because of the large variation in quality of such limits between objects and the absence of reliable exposure maps in some cases, such as the XMM slew survey). However, focusing on homogeneous catalogs and secure detections ensures reliable results for the targets under investigation, which is why we preferred this approach.

Overall, 84 matches were found. Nine of these are dubious associations mostly because of the relatively large distance between the Be star and its X-ray counterpart. In 51 cases, enough counts could be extracted to perform a spectral analysis. Using absorbed optically thin thermal models, we derived the X-ray luminosities in several bands, the hardness ratios, the plasma temperature(s), and the local absorption. For the remaining cases, the count rates were converted into broad-band X-ray luminosities. Compared to other surveys of OB stars, our targets show a similar range of X-ray luminosities, but with a substantial contribution of bright and hard (defined as high hardness ratios and/or high temperatures) sources. This is caused by the presence of $\gamma$\,Cas objects. We note, however, that our targets are not split into clear-cut groups in X-ray luminosity versus hardness graphs.

Splitting the targets in categories, we found that O-type stars and magnetic OB stars display the typical characteristics of their classes, while confusion with nearby active stars may become important for the latest stars of our sample, however. On the other hand, the properties of the known $\gamma$\,Cas analogs led us to identify eight new cases. They display high X-ray luminosities ($L_{\rm X}^{ISM\,cor}(0.5-10. keV)>4\times10^{31}$\,erg\,s$^{-1}$, $L_{\rm X}^{ISM\,cor}(2.-10. keV)>10^{31}$\,erg\,s$^{-1}$, \loglxlb $>-6$), high hardness ratios ($HR>1.6$ or $kT>5$\,keV), and early spectral types. When the information is available, most objects also present the iron line complex in their spectra, evidence of local absorption, short- and/or long-term variations of their X-ray emission, strong emission in H$\alpha$, and/or relatively high projected rotational velocities. In addition, one star appears to be in an intermediate stage between ``normal'' OB stars and  $\gamma$\,Cas objects, which could be linked to a low-intensity $\gamma$\,Cas phenomenon, for example, because of a later spectral type and a less dense circumstellar disk. We therefore consider this star as a $\gamma$\,Cas candidate. Finally, two Be stars may be even more luminous, but the lack of detailed information prevents us from drawing secure conclusions regarding their nature.

Further investigation is required to study these new $\gamma$\,Cas analogs and candidate in more detail, as well as the peculiar objects found in our survey. In particular, optical and X-ray monitoring would provide invaluable information to constrain the so-called $\gamma$\,Cas phenomenon.

\setcounter{table}{2}
\begin{sidewaystable*}
\centering
\caption{Results of the spectral fitting.}
\label{fits}
\tiny 
\setlength{\tabcolsep}{1pt}
\begin{tabular}{cllccccccccccccc}
\hline\hline
\#& Name                    & ObsID     & $N_{\rm H}^{ISM}$ &$N_{\rm H}$ & $kT_1$ & $norm_1$ &$ kT_2$ & $norm_2$ & $kT_3$ & $norm_3$ & $\chi^2_{\nu}$(dof) & $F_{\rm X}^{obs}$(tot) & $L_{\rm X}^{ISM\,cor}$(tot) & $\log(L_{\rm X}/L_{\rm BOL})$ & $HR$ \\
&                        &           & \multicolumn{2}{c}{($10^{22}$\,cm$^{-2}$)} & (keV) & (cm$^{-5}$) & (keV) & (cm$^{-5}$) & (keV) & (cm$^{-5}$) & & (erg\,cm$^{-2}$\,s$^{-1}$) & (erg\,s$^{-1}$) & & \\
\hline
1 & $\gamma$\,Cas           &0651670201 &0.027 &0.104$\pm$0.001 &15.55$\pm$0.15 &0.10491$\pm$0.00016 &               &                    &               &                   &3.63(909)&(1.816$\pm$0.003)e-10&(7.94$\pm$1.67)e32 &--5.4136$\pm$0.0007&3.246$\pm$0.008  \\
1 & $\gamma$\,Cas           &0651670301 &0.027 &0.101$\pm$0.001 &14.34$\pm$0.17 &0.09293$\pm$0.00016 &               &                    &               &                   &2.82(897)&(1.609$\pm$0.003)e-10&(7.03$\pm$1.48)e32 &--5.4660$\pm$0.0008&3.164$\pm$0.010  \\
1 & $\gamma$\,Cas           &0651670401 &0.027 &0.050$\pm$0.001 &15.57$\pm$0.13 &0.12750$\pm$0.00018 &               &                    &               &                   &4.80(921)&(2.258$\pm$0.003)e-10&(9.88$\pm$2.08)e32 &--5.3184$\pm$0.0006&2.978$\pm$0.008  \\
1 & $\gamma$\,Cas           &0651670501 &0.027 &0.099$\pm$0.001 &14.20$\pm$0.14 &0.09272$\pm$0.00013 &               &                    &               &                   &3.54(915)&(1.607$\pm$0.003)e-10&(7.02$\pm$1.48)e32 &--5.4668$\pm$0.0008&3.145$\pm$0.009  \\
1 & $\gamma$\,Cas           &0743600101 &0.027 &0.125$\pm$0.001 &25.26$\pm$0.23 &0.14451$\pm$0.00022 &               &                    &               &                   &8.64(936)&(2.442$\pm$0.003)e-10&(1.07$\pm$0.22)e33 &--5.2857$\pm$0.0005&3.740$\pm$0.007  \\
2 & Achernar                &           &0.0006&0.024$\pm$0.013 &0.21$\pm$0.02  &(2.15$\pm$0.29)e-4  &0.76$\pm$0.04  &(1.28$\pm$0.15)e-4  &               &                   &1.03(103)&(4.32$\pm$0.16)e-13  &(9.51$\pm$0.57)e28 &--8.025$\pm$0.016   &0.029$\pm$0.003  \\
3 & V782\,Cas               &           &0.30  &6.05$\pm$1.02   &6.85$\pm$1.55  &(3.23$\pm$0.57)e-3  &               &                    &               &                   &1.52(34) &(2.74$\pm$0.44)e-12  &(3.03$\pm$0.78)e32 &--5.25$\pm$0.07     &63.1$\pm$16.2    \\
9 & HD\,19818               &           &0.010 &0.72$\pm$0.27   &0.54$\pm$0.28  &(8.55$\pm$9.08)e-4  &               &                    &               &                   &2.38(11) &(3.70$\pm$3.70)e-13  &(4.31$\pm$4.31)e30 &--4.42$\pm$0.43     &0.07$\pm$0.10    \\
10& Merope                  &13         &0.013 &0.03$\pm$0.04   &0.765$\pm$0.019&(1.18$\pm$0.11)e-4  &               &                    &               &                   &1.62(54) &(2.36$\pm$0.09)e-13  &(3.27$\pm$0.33)e29 &--6.869$\pm$0.017   &0.049$\pm$0.004  \\
10& Merope                  &366        &0.013 &0.03$\pm$0.06   &0.63$\pm$0.03  &(9.97$\pm$1.72)e-5  &               &                    &               &                   &0.79(28) &(1.95$\pm$0.17)e-13  &(2.72$\pm$0.35)e29 &--6.95$\pm$0.04     &0.028$\pm$0.004  \\
10& Merope                  &17250      &0.013 &0.29$\pm$0.06   &0.60$\pm$0.04  &(1.88$\pm$0.28)e-4  &               &                    &               &                   &1.33(29) &(1.99$\pm$0.13)e-13  &(2.76$\pm$0.32)e29 &--6.94$\pm$0.03     &0.043$\pm$0.007  \\
10& Merope                  &17251      &0.013 &0.06$\pm$0.07   &0.76$\pm$0.04  &(1.14$\pm$0.16)e-4  &               &                    &               &                   &1.23(26) &(2.13$\pm$0.18)e-13  &(2.96$\pm$0.37)e29 &--6.91$\pm$0.04     &0.051$\pm$0.007  \\
10& Merope                  &17252      &0.013 &0.18$\pm$0.04   &0.61$\pm$0.03  &(1.49$\pm$0.16)e-4  &               &                    &               &                   &1.26(56) &(2.06$\pm$0.10)e-13  &(2.85$\pm$0.30)e29 &--6.93$\pm$0.02     &0.035$\pm$0.004  \\
11& Menkhib                 &0770990101 &0.12  &0.055$\pm$0.005 &0.159$\pm$0.002&(2.95$\pm$0.14)e-3  &0.656$\pm$0.004&(1.058$\pm$0.014)e-3&               &                   &7.96(309)&(2.17$\pm$0.02)e-12  &(6.05$\pm$2.38)e31 &--6.831$\pm$0.004   &0.0208$\pm$0.0005\\
11& Menkhib                 &0770990201 &0.12  &0.004$\pm$0.004 &0.193$\pm$0.002&(1.81$\pm$0.05)e-3  &0.739$\pm$0.003&(8.92$\pm$0.08)e-4  &               &                   &7.53(318)&(2.26$\pm$0.02))e-12 &(6.21$\pm$2.44)e31 &--6.820$\pm$0.004   &0.0243$\pm$0.0006\\
12& $\lambda$\,Eri          &           &0.013 &0.07$\pm$0.05   &0.241$\pm$0.019&(4.95$\pm$1.89)e-5  &1.23$\pm$0.13  &(2.34$\pm$0.44)e-5  &               &                   &1.36(36) &(6.85$\pm$0.67)e-14  &(5.27$\pm$0.69)e29 &--7.73$\pm$0.04     &0.09$\pm$0.03    \\
14& V1230\,Ori              &0212480301 &0.023 &0.64$\pm$0.07   &1.05$\pm$0.23  &(2.13$\pm$0.66)e-4  &3.33$\pm$1.28  &(1.18$\pm$0.22)e-3  &               &                   &1.13(64) &(1.32$\pm$0.09)e-12  &(3.03$\pm$0.37)e31 &--3.76$\pm$0.03     &1.93$\pm$0.20    \\
14& V1230\,Ori              &0403200101 &0.023 &0.99$\pm$0.09   &1.05$\pm$0.11  &(5.31$\pm$1.36)e-4  &3.53$\pm$0.40  &(1.05$\pm$0.08)e-3  &               &                   &1.54(95) &(1.25$\pm$0.06)e-12  &(2.87$\pm$0.32)e31 &--3.79$\pm$0.02     &2.07$\pm$0.15    \\
14& V1230\,Ori              &3          &0.023 &0.45$\pm$0.20   &1.90$\pm$0.27  &(4.84$\pm$0.64)e-4  &               &                    &               &                   &1.04(11) &(3.94$\pm$0.30)e-13  &(9.10$\pm$1.15)e30 &--4.28$\pm$0.03     &1.01$\pm$0.15    \\
14& V1230\,Ori              &4          &0.023 &0.25$\pm$0.20   &2.28$\pm$0.42  &(5.05$\pm$0.85)e-4  &               &                    &               &                   &1.17(7)  &(4.96$\pm$0.47)e-13  &(1.15$\pm$0.16)e31 &--4.18$\pm$0.04     &1.07$\pm$0.23    \\
14& V1230\,Ori              &18         &0.023 &0.64$\pm$0.04   &2.50$\pm$0.10  &(7.81$\pm$0.25)e-4  &               &                    &               &                   &1.18(139)&(6.48$\pm$0.16)e-13  &(1.49$\pm$0.16)e31 &--4.070$\pm$0.011   &1.86$\pm$0.07    \\
14& V1230\,Ori              &1522       &0.023 &0.80$\pm$0.08   &0.96$\pm$0.07  &(1.44$\pm$0.43)e-4  &2.65$\pm$0.18  &(5.75$\pm$0.36)e-4  &               &                   &1.14(111)&(5.54$\pm$0.20)e-13  &(1.27$\pm$0.14)e31 &--4.138$\pm$0.016   &1.65$\pm$0.09    \\
14& V1230\,Ori              &2567       &0.023 &0.36$\pm$0.08   &2.49$\pm$0.18  &(5.80$\pm$0.42)e-4  &               &                    &               &                   &1.16(34) &(5.50$\pm$0.28)e-13  &(1.27$\pm$0.14)e31 &--4.14$\pm$0.02     &1.38$\pm$0.13    \\
14& V1230\,Ori              &2568       &0.023 &0.38$\pm$0.09   &2.80$\pm$0.30  &(5.01$\pm$0.40)e-4  &               &                    &               &                   &0.81(29) &(5.00$\pm$0.28)e-13  &(1.15$\pm$0.13)e31 &--4.18$\pm$0.02     &1.64$\pm$0.19    \\
14& V1230\,Ori              &3498       &0.023 &0.82$\pm$0.10   &0.93$\pm$0.05  &(2.15$\pm$0.71)e-4  &2.46$\pm$0.13  &(5.63$\pm$0.34)e-4  &               &                   &1.08(147)&(5.57$\pm$0.13)e-13  &(1.28$\pm$0.13)e31 &--4.135$\pm$0.010   &1.35$\pm$0.06    \\
14& V1230\,Ori              &3744       &0.023 &0.96$\pm$0.05   &0.95$\pm$0.03  &(2.14$\pm$0.40)e-4  &2.44$\pm$0.08  &(7.03$\pm$0.24)e-4  &               &                   &1.53(244)&(6.22$\pm$0.10)e-13  &(1.43$\pm$0.15)e31 &--4.088$\pm$0.007   &1.64$\pm$0.04    \\
14& V1230\,Ori              &4373       &0.023 &0.75$\pm$0.04   &0.93$\pm$0.05  &(1.23$\pm$0.23)e-4  &2.71$\pm$0.07  &(9.64$\pm$0.23)e-4  &               &                   &1.46(286)&(8.86$\pm$0.12)e-13  &(2.03$\pm$0.21)e31 &--3.935$\pm$0.006   &1.87$\pm$0.04    \\
14& V1230\,Ori              &4374       &0.023 &0.87$\pm$0.05   &0.91$\pm$0.06  &(1.77$\pm$0.34)e-4  &2.24$\pm$0.07  &(6.34$\pm$0.21)e-4  &               &                   &1.58(225)&(5.48$\pm$0.08)e-13  &(1.26$\pm$0.13)e31 &--4.142$\pm$0.006   &1.39$\pm$0.04    \\
14& V1230\,Ori              &4395       &0.023 &0.93$\pm$0.07   &0.99$\pm$0.04  &(2.07$\pm$0.52)e-4  &2.71$\pm$0.10  &(7.69$\pm$0.30)e-4  &               &                   &1.24(202)&(7.17$\pm$0.15)e-13  &(1.65$\pm$0.17)e31 &--4.026$\pm$0.009   &1.90$\pm$0.06    \\
14& V1230\,Ori              &4396       &0.023 &0.75$\pm$0.05   &0.96$\pm$0.05  &(1.08$\pm$0.21)e-4  &2.56$\pm$0.08  &(6.81$\pm$0.20)e-4  &               &                   &1.31(243)&(6.19$\pm$0.09)e-13  &(1.42$\pm$0.15)e31 &--4.090$\pm$0.006   &1.68$\pm$0.04    \\
14& V1230\,Ori              &4473       &0.023 &0.24$\pm$0.13   &2.48$\pm$0.24  &(5.12$\pm$0.56)e-4  &               &                    &               &                   &1.11(26) &(5.25$\pm$0.31)e-13  &(1.21$\pm$0.14)e31 &--4.16$\pm$0.03     &1.17$\pm$0.14    \\
14& V1230\,Ori              &4474       &0.023 &0.26$\pm$0.12   &2.95$\pm$0.35  &(5.15$\pm$0.53)e-4  &               &                    &               &                   &1.06(29) &(5.60$\pm$0.32)e-13  &(1.29$\pm$0.15)e31 &--4.13$\pm$0.02     &1.51$\pm$0.20    \\
14& V1230\,Ori              &7407       &0.023 &0.84$\pm$0.54   &1.34$\pm$0.35  &(7.16$\pm$2.36)e-4  &               &                    &               &                   &0.88(4)  &(4.28$\pm$0.75)e-13  &(9.90$\pm$2.00)e30 &--4.25$\pm$0.08     &0.71$\pm$0.25    \\
14& V1230\,Ori              &7408       &0.023 &0.23$\pm$0.13   &2.22$\pm$0.29  &(4.08$\pm$0.48)e-4  &               &                    &               &                   &1.66(12) &(4.03$\pm$0.37)e-13  &(9.35$\pm$1.28)e30 &--4.27$\pm$0.04     &1.00$\pm$0.19    \\
14& V1230\,Ori              &7409       &0.023 &0.21$\pm$0.11   &2.29$\pm$0.29  &(3.57$\pm$0.41)e-4  &               &                    &               &                   &2.12(12) &(3.51$\pm$0.28)e-13  &(8.37$\pm$1.07)e30 &--4.32$\pm$0.03     &1.02$\pm$0.19    \\
14& V1230\,Ori              &7410       &0.023 &0.44$\pm$0.31   &2.03$\pm$0.44  &(4.72$\pm$1.04)e-4  &               &                    &               &                   &2.87(6)  &(3.93$\pm$0.50)e-13  &(9.08$\pm$1.47)e30 &--4.29$\pm$0.06     &1.12$\pm$0.33    \\
14& V1230\,Ori              &7411       &0.023 &0.14$\pm$0.10   &2.03$\pm$0.33  &(3.75$\pm$0.40)e-4  &               &                    &               &                   &0.73(11) &(3.89$\pm$0.32)e-13  &(9.05$\pm$1.18)e30 &--4.29$\pm$0.04     &0.78$\pm$0.15    \\
14& V1230\,Ori              &7412       &0.023 &0.31$\pm$0.13   &2.50$\pm$0.34  &(4.65$\pm$0.55)e-4  &               &                    &               &                   &0.72(13) &(4.57$\pm$0.33)e-13  &(1.06$\pm$0.13)e31 &--4.22$\pm$0.03     &1.29$\pm$0.18    \\
14& V1230\,Ori              &8568       &0.023 &0.25$\pm$0.13   &2.27$\pm$0.26  &(3.81$\pm$0.44)e-4  &               &                    &               &                   &1.18(17) &(3.74$\pm$0.26)e-13  &(8.64$\pm$1.06)e30 &--4.31$\pm$0.03     &1.06$\pm$0.16    \\
14& V1230\,Ori              &8589       &0.023 &1.17$\pm$0.27   &0.95$\pm$0.21  &(4.34$\pm$1.95)e-4  &3.81$\pm$15.52 &(1.97$\pm$1.03)e-4  &               &                   &0.84(21) &(3.66$\pm$0.77)e-13  &(8.42$\pm$1.96)e30 &--4.32$\pm$0.09     &1.26$\pm$0.56    \\
14& V1230\,Ori              &8895       &0.023 &0.35$\pm$0.18   &2.33$\pm$0.34  &(6.19$\pm$0.90)e-4  &               &                    &               &                   &1.32(11) &(5.75$\pm$0.48)e-13  &(1.33$\pm$0.17)e31 &--4.12$\pm$0.04     &1.24$\pm$0.24    \\
14& V1230\,Ori              &8896       &0.023 &0.81$\pm$0.36   &0.85$\pm$0.16  &(2.45$\pm$2.57)e-4  &2.50$\pm$1.73  &(3.36$\pm$1.48)e-4  &               &                   &0.73(9)  &(4.01$\pm$0.71)e-13  &(9.28$\pm$1.89)e30 &--4.28$\pm$0.08     &1.03$\pm$0.35    \\
14& V1230\,Ori              &8897       &0.023 &0.46$\pm$0.30   &2.31$\pm$0.41  &(4.32$\pm$0.98)e-4  &               &                    &               &                   &1.01(10) &(3.76$\pm$0.31)e-13  &(8.64$\pm$1.13)e30 &--4.31$\pm$0.04     &1.40$\pm$0.26    \\
14& V1230\,Ori              &13637      &0.023 &1.34$\pm$0.14   &0.96$\pm$0.08  &(4.97$\pm$1.46)e-4  &2.41$\pm$0.94  &(2.56$\pm$0.96)e-4  &               &                   &1.03(41) &(3.47$\pm$0.35)e-13  &(8.01$\pm$1.14)e30 &--4.34$\pm$0.04     &1.16$\pm$0.21    \\
14& V1230\,Ori              &14334      &0.023 &0.26$\pm$0.11   &2.90$\pm$0.43  &(5.11$\pm$0.51)e-4  &               &                    &               &                   &0.74(19) &(5.52$\pm$0.47)e-13  &(1.27$\pm$0.17)e31 &--4.14$\pm$0.04     &1.48$\pm$0.22    \\
14& V1230\,Ori              &14335      &0.023 &1.18$\pm$0.14   &0.86$\pm$0.07  &(3.64$\pm$1.27)e-4  &2.23$\pm$0.23  &(5.27$\pm$0.69)e-4  &               &                   &0.79(54) &(4.79$\pm$0.30)e-13  &(1.10$\pm$0.13)e31 &--4.20$\pm$0.03     &1.37$\pm$0.15    \\
14& V1230\,Ori              &15546      &0.023 &1.06$\pm$0.26   &1.18$\pm$0.20  &(3.65$\pm$2.47)e-4  &3.08$\pm$10.54 &(2.95$\pm$1.74)e-4  &               &                   &1.05(31) &(4.26$\pm$0.77)e-13  &(9.78$\pm$2.03)e30 &--4.25$\pm$0.08     &1.53$\pm$0.59    \\
14& V1230\,Ori              &17735      &0.023 &0.53$\pm$0.05   &2.58$\pm$0.11  &(7.58$\pm$0.29)e-4  &               &                    &               &                   &1.28(147)&(6.71$\pm$0.15)e-13  &(1.54$\pm$0.15)e31 &--4.055$\pm$0.010   &1.76$\pm$0.07    \\
\hline
\end{tabular}
\end{sidewaystable*}
\setcounter{table}{2}
\begin{sidewaystable*}
\centering
\caption{Continued.  }
\tiny 
\setlength{\tabcolsep}{1pt}
\begin{tabular}{cllccccccccccccc}
\hline\hline
\#& Name                    & ObsID     & $N_{\rm H}^{ISM}$ &$N_{\rm H}$ & $kT_1$ & $norm_1$ &$ kT_2$ & $norm_2$ & $kT_3$ & $norm_3$ & $\chi^2_{\nu}$(dof) & $F_{\rm X}^{obs}$(tot) & $L_{\rm X}^{ISM\,cor}$(tot) & $\log(L_{\rm X}/L_{\rm BOL})$ & $HR$ \\
&                        &           & \multicolumn{2}{c}{($10^{22}$\,cm$^{-2}$)} & (keV) & (cm$^{-5}$) & (keV) & (cm$^{-5}$) & (keV) & (cm$^{-5}$) & & (erg\,cm$^{-2}$\,s$^{-1}$) & (erg\,s$^{-1}$) & & \\
\hline
15& 43\,Ori                 &0212480301 &0.032 &0.22$\pm$0.04   &0.248$\pm$0.014&(8.29$\pm$2.55)e-4  &0.77$\pm$0.05  &(2.76$\pm$0.42)e-4  &2.34$\pm$0.34  &(3.64$\pm$0.32)e-4 &1.29(101)&(1.07$\pm$0.05)e-12  &(2.90$\pm$0.78)e31 &--6.07$\pm$0.02     &0.24$\pm$0.04    \\
15& 43\,Ori                 &0403200101 &0.032 &0.22$\pm$0.02   &0.245$\pm$0.007&(8.18$\pm$1.07)e-4  &0.77$\pm$0.02  &(2.62$\pm$0.19)e-4  &1.88$\pm$0.30  &(1.31$\pm$0.14)e-4 &1.80(182)&(7.92$\pm$0.12)e-13  &(2.16$\pm$0.58)e31 &--6.199$\pm$0.007   &0.103$\pm$0.014  \\
15& 43\,Ori                 &3          &0.032 &0.000$\pm$0.017 &0.60$\pm$0.04  &(2.08$\pm$0.16)e-4  &1.58$\pm$0.12  &(1.80$\pm$0.20)e-4  &               &                   &1.66(43) &(6.31$\pm$0.50)e-13  &(1.70$\pm$0.47)e31 &--6.30$\pm$0.03     &0.12$\pm$0.02    \\
15& 43\,Ori                 &4          &0.032 &0.000$\pm$0.019 &0.72$\pm$0.06  &(2.18$\pm$0.19)e-4  &2.61$\pm$0.34  &(3.95$\pm$0.39)e-4  &               &                   &0.53(35) &(9.47$\pm$0.50)e-13  &(2.52$\pm$0.68)e31 &--6.13$\pm$0.02     &0.35$\pm$0.04    \\
15& 43\,Ori                 &2567       &0.032 &0(fixed)        &0.71$\pm$0.06  &(2.43$\pm$0.26)e-4  &3.73$\pm$1.70  &(6.75$\pm$0.46)e-5  &               &                   &0.59(12) &(5.99$\pm$0.98)e-13  &(1.62$\pm$0.50)e31 &--6.32$\pm$0.07     &0.13$\pm$0.12    \\
15& 43\,Ori                 &2568       &0.032 &0.00$\pm$0.05   &0.62$\pm$0.04  &(2.61$\pm$0.49)e-4  &2.99$\pm$2.73  &(9.91$\pm$3.20)e-5  &               &                   &1.34(22) &(6.47$\pm$1.17)e-13  &(1.75$\pm$0.56)e31 &--6.29$\pm$0.08     &0.13$\pm$0.05    \\
15& 43\,Ori                 &4473       &0.032 &0.40$\pm$0.17   &0.25$\pm$0.04  &(2.01$\pm$1.55)e-3  &1.57$\pm$0.33  &(1.38$\pm$0.24)e-4  &               &                   &1.23(28) &(5.93$\pm$1.35)e-13  &(1.63$\pm$0.57)e31 &--6.32$\pm$0.10     &0.08$\pm$0.04    \\
15& 43\,Ori                 &4474       &0.032 &0.09$\pm$0.02   &1.01$\pm$0.04  &(4.18$\pm$0.50)e-4  &7.73$\pm$0.50  &(3.19$\pm$0.07)e-3  &               &                   &1.17(243)&(5.98$\pm$0.11)e-12  &(1.54$\pm$0.41)e32 &--5.346$\pm$0.007   &1.84$\pm$0.05    \\
15& 43\,Ori                 &7407       &0.032 &0(fixed)        &0.31$\pm$0.04  &(2.92$\pm$0.51)e-4  &1.14$\pm$0.10  &(1.64$\pm$0.22)e-4  &               &                   &0.61(14) &(6.00$\pm$0.39)e-13  &(1.65$\pm$0.45)e31 &--6.32$\pm$0.03     &0.060$\pm$0.013  \\
15& 43\,Ori                 &7408       &0.032 &0.67$\pm$0.36   &0.21$\pm$0.04  &(8.36$\pm$1.38)e-3  &1.59$\pm$1.75  &(1.49$\pm$2.07)e-4  &               &                   &0.60(6)  &(6.55$\pm$0.60)e-13  &(1.80$\pm$0.50)e31 &--6.28$\pm$0.04     &0.076$\pm$0.013  \\
15& 43\,Ori                 &7409       &0.032 &0.00$\pm$0.04   &0.60$\pm$0.05  &(2.57$\pm$0.40)e-4  &2.81$\pm$1.87  &(1.05$\pm$0.32)e-4  &               &                   &0.69(14) &(6.39$\pm$1.33)e-13  &(1.73$\pm$0.58)e31 &--6.29$\pm$0.09     &0.13$\pm$0.03    \\
15& 43\,Ori                 &7410       &0.032 &0(fixed)        &0.81$\pm$0.13  &(1.96$\pm$0.35)e-4  &2.01$\pm$7.48  &(1.33$\pm$0.92)e-4  &               &                   &1.17(6)  &(5.57$\pm$0.75)e-13  &(1.50$\pm$0.45)e31 &--6.36$\pm$0.06     &0.16$\pm$0.07    \\
15& 43\,Ori                 &7411       &0.032 &0(fixed)        &0.63$\pm$0.16  &(2.20$\pm$0.22)e-4  &2.38$\pm$0.61  &(2.50$\pm$0.37)e-4  &               &                   &1.54(20) &(7.46$\pm$0.45)e-13  &(2.00$\pm$0.55)e31 &--6.23$\pm$0.03     &0.23$\pm$0.05    \\
15& 43\,Ori                 &7412       &0.032 &0.52$\pm$0.14   &0.26$\pm$0.04  &(3.33$\pm$3.64)e-3  &4.59$\pm$9.44  &(1.28$\pm$0.39)e-4  &               &                   &1.34(17) &(7.48$\pm$4.00)e-13  &(2.03$\pm$1.21)e31 &--6.23$\pm$0.23     &0.18$\pm$0.15    \\
15& 43\,Ori                 &8568       &0.032 &0.00$\pm$0.05   &0.64$\pm$0.04  &(2.51$\pm$0.44)e-4  &3.75$\pm$2.01  &(1.42$\pm$0.30)e-4  &               &                   &1.45(24) &(7.08$\pm$1.50)e-13  &(1.91$\pm$0.65)e31 &--6.25$\pm$0.09     &0.21$\pm$0.07    \\
15& 43\,Ori                 &8589       &0.032 &0.69$\pm$0.14   &0.23$\pm$0.04  &(6.42$\pm$0.10)e-3  &2.10$\pm$0.45  &(2.36$\pm$0.38)e-4  &               &                   &0.70(31) &(7.00$\pm$1.84)e-13  &(1.90$\pm$0.71)e31 &--6.26$\pm$0.11     &0.17$\pm$0.06    \\
15& 43\,Ori                 &8895       &0.032 &0.00$\pm$0.03   &0.75$\pm$0.04  &(2.33$\pm$0.24)e-4  &2.63$\pm$0.55  &(1.34$\pm$0.35)e-4  &               &                   &1.96(16) &(6.49$\pm$1.40)e-13  &(1.74$\pm$0.60)e31 &--6.29$\pm$0.09     &0.17$\pm$0.07    \\
15& 43\,Ori                 &8896       &0.032 &0(fixed)        &0.73$\pm$0.09  &(2.62$\pm$0.25)e-4  &2.54$\pm$0.84  &(1.94$\pm$0.44)e-4  &               &                   &0.63(19) &(7.82$\pm$0.45)e-13  &(2.10$\pm$0.57)e31 &--6.21$\pm$0.02     &0.20$\pm$0.04    \\
15& 43\,Ori                 &8897       &0.032 &0(fixed)        &0.64$\pm$0.08  &(2.46$\pm$0.25)e-4  &2.63$\pm$0.71  &(1.91$\pm$0.38)e-4  &               &                   &1.00(17) &(7.38$\pm$0.50)e-13  &(1.99$\pm$0.54)e31 &--6.23$\pm$0.03     &0.20$\pm$0.04    \\
16& HD\,42054               &0402121401 &0.006 &0.001$\pm$0.003 &0.93$\pm$0.03  &(1.33$\pm$0.11)e-4  &4.53$\pm$0.18  &(1.58$\pm$0.03)e-3  &               &                   &1.21(448)&(2.74$\pm$0.04)e-12  &(3.75$\pm$0.34)e31 &--5.319$\pm$0.006   &1.23$\pm$0.03    \\
16& HD\,42054               &11021      &0.006 &0.000$\pm$0.012 &1.02$\pm$0.06  &(1.23$\pm$0.12)e-4  &5.51$\pm$0.51  &(1.15$\pm$0.05)e-3  &               &                   &1.04(213)&(2.15$\pm$0.06)e-12  &(2.94$\pm$0.27)e31 &--5.425$\pm$0.012   &1.39$\pm$0.07    \\
16& HD\,42054               &12226      &0.006 &0.00$\pm$0.02   &1.59$\pm$0.23  &(3.56$\pm$1.49)e-4  &8.58$\pm$2.10  &(8.00$\pm$1.33)e-4  &               &                   &1.10(51) &(1.91$\pm$0.18)e-12  &(2.61$\pm$0.34)e31 &--5.48$\pm$0.04     &1.46$\pm$0.20    \\
17& PZ\,Gem                 &0670080301 &0.085 &0.31$\pm$0.02   &16.18$\pm$4.18 &(7.12$\pm$0.12)e-4  &               &                    &               &                   &1.00(394)&(1.14$\pm$0.02)e-12  &(9.66$\pm$1.77)e31 &--6.138$\pm$0.008   &4.32$\pm$0.11    \\
17& PZ\,Gem                 &0760220601 &0.085 &0(fixed)        &5.79$\pm$0.89  &(7.80$\pm$0.29)e-5  &               &                    &               &                   &1.03(81) &(1.25$\pm$0.09)e-13  &(1.10$\pm$0.22)e31 &--7.08$\pm$0.03     &1.82$\pm$0.20    \\
20& 15\,Mon                 &0011420101 &0.007 &0.188$\pm$0.011 &0.055$\pm$0.004&0.37$\pm$0.28       &0.229$\pm$0.003&(2.75$\pm$0.20)e-3  &0.86$\pm$0.06  &(8.01$\pm$1.43)e-5 &1.79(191)&(1.52$\pm$0.04)e-12  &(1.55$\pm$0.44)e31 &--6.363$\pm$0.011   &0.0071$\pm$0.0009\\
20& 15\,Mon                 &5401       &0.007 &0(fixed)        &0.201$\pm$0.016&(1.21$\pm$0.09)e-3  &0.66$\pm$0.11  &(1.62$\pm$0.40)e-4  &               &                   &1.22(43) &(1.29$\pm$0.07)e-12  &(1.32$\pm$0.38)e31 &--6.43$\pm$0.02     &0.0079$\pm$0.0016\\
20& 15\,Mon                 &6247       &0.007 &0(fixed)        &0.197$\pm$0.018&(1.18$\pm$0.11)e-3  &0.57$\pm$0.10  &(2.21$\pm$1.05)e-4  &               &                   &1.10(36) &(1.35$\pm$0.07)e-12  &(1.38$\pm$0.40)e31 &--6.41$\pm$0.02     &0.0067$\pm$0.0014\\
20& 15\,Mon                 &6248       &0.007 &0(fixed)        &0.22$\pm$0.03  &(1.15$\pm$0.24)e-3  &0.86$\pm$0.14  &(1.70$\pm$0.79)e-4  &               &                   &1.11(8)  &(1.38$\pm$0.19)e-12  &(1.41$\pm$0.45)e31 &--6.40$\pm$0.06     &0.016$\pm$0.009  \\
21& 19\,Mon                 &           &0.020 &0.000$\pm$0.009 &0.25$\pm$0.02  &(3.10$\pm$0.37)e-5  &               &                    &               &                   &1.51(10) &(2.96$\pm$0.58)e-14  &(5.43$\pm$1.39)e29 &--7.94$\pm$0.09     &0.0009$\pm$0.0005\\
22& HD\,57682               &           &0.088 &0.000$\pm$0.002 &0.78$\pm$0.02  &(1.56$\pm$0.08)e-4  &1.77$\pm$0.19  &(1.60$\pm$0.14)e-4  &               &                   &1.39(134)&(4.60$\pm$0.22)e-13  &(1.04$\pm$0.29)e32 &--6.56$\pm$0.02     &0.17$\pm$0.02    \\
23& BN\,Gem                 &           &0.019 &0.008$\pm$0.033 &0.25$\pm$0.03  &(2.17$\pm$0.71)e-5  &5.69$\pm$3.25  &(2.00$\pm$0.41)e-5  &               &                   &1.25(13) &(5.25$\pm$0.60)e-14  &(2.76$\pm$1.53)e30 &--7.25$\pm$0.05     &0.66$\pm$0.32    \\
24& V392\,Pup               &0694730301 &0.006 &0.07$\pm$0.06   &0.07$\pm$0.02  &(3.30$\pm$6.33)e-4  &0.49$\pm$0.05  &(2.15$\pm$0.50)e-5  &               &                   &0.92(46) &(3.46$\pm$0.77)e-14  &(1.75$\pm$0.40)e29 &--6.87$\pm$0.10     &0.014$\pm$0.005  \\
24& V392\,Pup               &0694730401 &0.006 &0.53$\pm$0.19   &0.28$\pm$0.07  &(1.30$\pm$1.04)e-4  &               &                    &               &                   &0.91(16) &(2.84$\pm$1.20)e-14  &(1.43$\pm$0.61)e29 &--6.96$\pm$0.18     &0.007$\pm$0.005  \\
25& V374\,Car               &0113890601 &0.054 &0.05$\pm$0.04   &1.03$\pm$0.17  &(2.09$\pm$0.57)e-5  &12.81$\pm$25.36&(1.05$\pm$0.13)e-4  &               &                   &0.88(34) &(2.18$\pm$0.69)e-13  &(2.86$\pm$1.01)e30 &--6.70$\pm$0.14     &1.76$\pm$1.13    \\
25& V374\,Car               &0113891001 &0.054 &0.009$\pm$0.018 &3.71$\pm$0.47  &(1.19$\pm$0.05)e-4  &               &                    &               &                   &0.93(78) &(1.66$\pm$0.12)e-13  &(2.20$\pm$0.39)e30 &--6.81$\pm$0.03     &1.29$\pm$0.17    \\
25& V374\,Car               &0113891101 &0.054 &0.011$\pm$0.033 &4.26$\pm$1.01  &(7.95$\pm$0.55)e-5  &               &                    &               &                   &1.18(34) &(1.17$\pm$0.12)e-13  &(1.54$\pm$0.29)e30 &--6.97$\pm$0.04     &1.47$\pm$0.23    \\
25& V374\,Car               &0126511201 &0.054 &0.000$\pm$0.013 &3.83$\pm$0.34  &(1.31$\pm$0.05)e-4  &               &                    &               &                   &0.93(103)&(1.86$\pm$0.12)e-13  &(2.45$\pm$0.42)e30 &--6.76$\pm$0.03     &1.31$\pm$0.06    \\
25& V374\,Car               &0134531201 &0.054 &0.000$\pm$0.011 &4.09$\pm$0.50  &(8.92$\pm$0.34)e-5  &               &                    &               &                   &0.87(81) &(1.30$\pm$0.07)e-13  &(1.72$\pm$0.29)e30 &--6.92$\pm$0.02     &1.39$\pm$0.13    \\
25& V374\,Car               &0134531301 &0.054 &0.02$\pm$0.04   &4.22$\pm$1.09  &(9.26$\pm$0.72)e-5  &               &                    &               &                   &1.35(32) &(1.35$\pm$0.13)e-13  &(1.77$\pm$0.33)e30 &--6.91$\pm$0.04     &1.49$\pm$0.24    \\
25& V374\,Car               &0134531501 &0.054 &0.000$\pm$0.014 &3.61$\pm$0.42  &(1.01$\pm$0.04)e-4  &               &                    &               &                   &1.02(74) &(1.40$\pm$0.10)e-13  &(1.86$\pm$0.33)e30 &--6.88$\pm$0.03     &1.24$\pm$0.15    \\
25& V374\,Car               &65         &0.054 &0.00$\pm$0.04   &5.08$\pm$2.88  &(1.46$\pm$0.13)e-4  &               &                    &               &                   &1.47(7)  &(2.29$\pm$0.39)e-13  &(3.02$\pm$0.71)e30 &--6.67$\pm$0.07     &1.64$\pm$0.49    \\
25& V374\,Car               &66         &0.054 &0.00$\pm$0.04   &3.04$\pm$0.72  &(1.03$\pm$0.10)e-4  &               &                    &               &                   &0.54(7)  &(1.34$\pm$0.21)e-13  &(1.78$\pm$0.40)e30 &--6.90$\pm$0.07     &1.04$\pm$0.32    \\
25& V374\,Car               &1229       &0.054 &0.00$\pm$0.08   &3.36$\pm$0.85  &(8.71$\pm$1.23)e-5  &               &                    &               &                   &3.84(6)  &(1.18$\pm$0.24)e-13  &(1.55$\pm$0.40)e30 &--6.96$\pm$0.09     &1.16$\pm$0.35    \\
25& V374\,Car               &1232       &0.054 &0.00$\pm$0.06   &3.37$\pm$1.03  &(1.33$\pm$0.16)e-4  &               &                    &               &                   &1.21(6)  &(1.79$\pm$0.30)e-13  &(2.38$\pm$0.55)e30 &--6.78$\pm$0.07     &1.16$\pm$0.19    \\
25& V374\,Car               &1458       &0.054 &0.05$\pm$0.07   &2.92$\pm$0.91  &(1.21$\pm$0.13)e-4  &               &                    &               &                   &0.88(10) &(1.49$\pm$0.20)e-13  &(1.98$\pm$0.41)e30 &--6.86$\pm$0.06     &1.08$\pm$0.31    \\
28& HD\,93190               &9484       &0.92  &0.18$\pm$0.21   &0.76$\pm$0.18  &(3.42$\pm$1.28)e-5  &               &                    &               &                   &1.90(5)  &(1.27$\pm$0.35)e-14  &(6.64$\pm$4.54)e31 &--8.18$\pm$0.12     &0.07$\pm$0.04    \\
29& HD\,93843               &9508       &0.15  &0.50$\pm$0.15   &0.24$\pm$0.03  &(9.04$\pm$7.19)e-4  &1.05$\pm$0.16  &(4.05$\pm$1.54)e-5  &               &                   &0.92(26) &(1.27$\pm$0.60)e-13  &(1.81$\pm$1.32)e32 &--6.84$\pm$0.21     &0.043$\pm$0.017  \\
29& HD\,93843               &9857       &0.15  &0.32$\pm$0.18   &0.15$\pm$0.05  &(0.73$\pm$1.60)e-3  &0.61$\pm$0.08  &(1.30$\pm$0.47)e-4  &               &                   &1.32(15) &(1.34$\pm$0.39)e-13  &(1.98$\pm$1.24)e32 &--6.82$\pm$0.13     &0.031$\pm$0.011  \\
30& HD\,305891              &0152570101 &0.92  &0.55$\pm$0.15   &0.71$\pm$0.13  &(1.35$\pm$0.60)e-4  &               &                    &               &                   &1.35(16) &(3.22$\pm$0.56)e-14  &(9.05$\pm$5.31)e31 &--5.37$\pm$0.08     &0.11$\pm$0.03    \\
32& V863\,Cen               &           &0.006 &0.03$\pm$0.02   &0.23$\pm$0.03  &(3.52$\pm$0.82)e-5  &1.00$\pm$0.07  &(2.42$\pm$0.30)e-5  &               &                   &0.88(34) &(7.27$\pm$0.53)e-14  &(9.21$\pm$0.86)e28 &--7.20$\pm$0.03     &0.060$\pm$0.012  \\
33& $\delta$\,Cen           &           &0.007 &0.001$\pm$0.014 &0.19$\pm$0.09  &(0.35$\pm$1.42)e-4  &0.70$\pm$0.07  &(2.25$\pm$1.07)e-5  &               &                   &0.89(26) &(7.43$\pm$0.95)e-14  &(1.49$\pm$0.27)e29 &--8.35$\pm$0.06     &0.022$\pm$0.006  \\
34& BZ\,Cru                 &           &0.20  &0.079$\pm$0.003 &0.97$\pm$0.03  &(1.40$\pm$0.09)e-4  &13.33$\pm$0.22 &(7.21$\pm$0.02)e-3  &               &                   &1.54(815)&(1.199$\pm$0.004)e-11&(2.76$\pm$0.37)e32 &--5.6930$\pm$0.0014 &2.814$\pm$0.014  \\
37& HD\,119682              &0087940201 &0.20  &0.014$\pm$0.008 &8.12$\pm$0.34  &(1.169$\pm$0.013)e-3&               &                    &               &                   &1.21(471)&(1.89$\pm$0.03)e-12  &(7.68$\pm$2.85)e32 &--5.558$\pm$0.007   &2.26$\pm$0.04    \\
37& HD\,119682              &0551000201 &0.20  &0.025$\pm$0.008 &8.41$\pm$0.53  &(7.05$\pm$0.09)e-4  &               &                    &               &                   &1.10(443)&(1.14$\pm$0.02)e-12  &(4.63$\pm$1.72)e32 &--5.778$\pm$0.008   &2.34$\pm$0.05    \\
37& HD\,119682              &8929       &0.20  &0.00$\pm$0.03   &9.92$\pm$3.03  &(1.13$\pm$0.04)e-3  &               &                    &               &                   &0.89(56) &(1.90$\pm$0.13)e-12  &(7.68$\pm$2.90)e32 &--5.56$\pm$0.03     &2.39$\pm$0.15    \\
37& HD\,119682              &10834      &0.20  &0.00$\pm$0.03   &12.01$\pm$2.92 &(8.64$\pm$0.28)e-4  &               &                    &               &                   &1.20(89) &(1.46$\pm$0.08)e-12  &(5.88$\pm$2.20)e32 &--5.67$\pm$0.02     &2.56$\pm$0.19    \\
37& HD\,119682              &10835      &0.20  &0.00$\pm$0.04   &9.47$\pm$2.49  &(8.44$\pm$0.33)e-4  &               &                    &               &                   &1.14(44) &(1.41$\pm$0.60)e-12  &(5.69$\pm$3.21)e32 &--5.69$\pm$0.18     &2.35$\pm$1.05    \\
37& HD\,119682              &10836      &0.20  &0(fixed)        &17.01$\pm$7.56 &(1.02$\pm$0.05)e-3  &               &                    &               &                   &0.78(52) &(1.72$\pm$0.08)e-12  &(6.91$\pm$2.58)e32 &--5.60$\pm$0.02     &2.81$\pm$0.23    \\
\hline
\end{tabular}
\end{sidewaystable*}
\setcounter{table}{2}
\begin{sidewaystable*}
\centering
\caption{Continued.  }
\tiny 
\setlength{\tabcolsep}{1pt}
\begin{tabular}{cllccccccccccccc}
\hline\hline
\# & Name                    & ObsID     & $N_{\rm H}^{ISM}$ &$N_{\rm H}$ & $kT_1$ & $norm_1$ &$ kT_2$ & $norm_2$ & $kT_3$ & $norm_3$ & $\chi^2_{\nu}$(dof) & $F_{\rm X}^{obs}$(tot) & $L_{\rm X}^{ISM\,cor}$(tot) & $\log(L_{\rm X}/L_{\rm BOL})$ & $HR$ \\
&                        &           & \multicolumn{2}{c}{($10^{22}$\,cm$^{-2}$)} & (keV) & (cm$^{-5}$) & (keV) & (cm$^{-5}$) & (keV) & (cm$^{-5}$) & & (erg\,cm$^{-2}$\,s$^{-1}$) & (erg\,s$^{-1}$) & & \\
\hline
38& $\mu$\,Cen              &           &0.008 &0.063$\pm$0.013 &0.257$\pm$0.009&(1.50$\pm$0.16)e-4  &1.27$\pm$0.04  &(8.29$\pm$0.56)e-5  &     &     &1.53(95) &(2.35$\pm$0.09)e-13  &(6.87$\pm$0.44)e29 &--7.496$\pm$0.017   &0.099$\pm$0.011  \\
39& V767\,Cen               &           &0.043 &0.067$\pm$0.011 &0.25$\pm$0.02  &(1.40$\pm$0.36)e-4  &6.44$\pm$0.26  &(1.85$\pm$0.03)e-3  &     &     &1.11(432)&(3.09$\pm$0.05)e-12  &(2.62$\pm$0.63)e32 &--5.365$\pm$0.007   &1.97$\pm$0.05    \\
40& CQ\,Cir                 &           &0.28  &0.70$\pm$0.32   &8.66$\pm$7.95  &(3.11$\pm$0.53)e-3  &               &                    &     &     &1.03(50) &(4.27$\pm$1.63)e-12  &(1.75$\pm$0.93)e33 &--4.30$\pm$0.17     &5.26$\pm$2.10    \\
41& V1040\,Sco              &           &0.067 &0.07$\pm$0.06   &0.46$\pm$0.37  &(1.99$\pm$1.85)e-5  &6.40$\pm$1.12  &(3.02$\pm$0.13)e-4  &     &     &1.13(78) &(5.07$\pm$0.29)e-13  &(1.26$\pm$0.10)e30 &--6.48$\pm$0.02     &1.83$\pm$0.16    \\
42& $\delta$\,Sco           &           &0.087 &0.000$\pm$0.004 &0.227$\pm$0.003&(7.09$\pm$0.21)e-4  &0.683$\pm$0.013&(1.13$\pm$0.07)e-4  &     &     &1.56(167)&(6.62$\pm$0.13)e-13  &(2.59$\pm$0.71)e30 &--7.860$\pm$0.009   &0.0090$\pm$0.0007\\
43& $\zeta$\,Oph            &2571       &0.12  &0.02$\pm$0.06   &0.16$\pm$0.03  &(2.54$\pm$1.57)e-3  &0.62$\pm$0.03  &(1.52$\pm$0.27)e-3  &$^a$ &$^b$ &1.05(81) &(3.79$\pm$0.47)e-12  &(8.16$\pm$1.10)e30 &--7.18$\pm$0.05     &0.043$\pm$0.005  \\
43& $\zeta$\,Oph            &4367       &0.12  &0.18$\pm$0.05   &0.151$\pm$0.011&(8.44$\pm$3.79)e-3  &0.59$\pm$0.02  &(2.04$\pm$0.29)e-3  &$^c$ &$^d$ &1.27(87) &(3.46$\pm$0.17)e-12  &(7.55$\pm$0.55)e30 &--7.22$\pm$0.02     &0.037$\pm$0.003  \\
45& V1075\,Sco              &           &0.16  &0(fixed)        &0.197$\pm$0.003&(8.11$\pm$0.16)e-4  &0.605$\pm$0.018&(1.37$\pm$0.08)e-4  &     &     &1.47(165)&(4.96$\pm$0.08)e-13  &(1.15$\pm$0.37)e32 &--6.874$\pm$0.007   &0.0076$\pm$0.0006\\
46& $\gamma$\,Ara           &           &0.061 &0.001$\pm$0.011 &0.185$\pm$0.008&(2.03$\pm$0.22)e-4  &0.741$\pm$0.016&(1.74$\pm$0.08)e-4  &$^e$ &$^f$ &1.46(211) &(5.40$\pm$0.17)e-13  &(8.95$\pm$1.03)e30 &--7.261$\pm$0.014   &0.075$\pm$0.007  \\
47& V750\,Ara               &           &0.087 &0.193$\pm$0.011 &9.73$\pm$0.73  &(1.65$\pm$0.02)e-3  &               &                    &     &     &1.05(408)&(2.68$\pm$0.05)e-12  &(4.19$\pm$1.14)e32 &--5.291$\pm$0.008   &3.21$\pm$0.09    \\
50& Cl*\,NGC\,6383\,FJL\,24 &           &0.28  &0.04$\pm$0.41   &0.26$\pm$0.08  &(0.21$\pm$2.44)e-4  &1.67$\pm$0.41  &(2.28$\pm$0.58)e-5  &     &     &0.37(17) &(2.77$\pm$1.00)e-14  &(1.30$\pm$0.65)e31 &--4.93$\pm$0.16     &0.22$\pm$0.11    \\
51& V3892\,Sgr              &0201200101 &0.48  &0(fixed)        &6.92$\pm$0.40  &(1.205$\pm$0.014)e-3&               &                    &     &     &1.1(400) &(1.73$\pm$0.04)e-12  &(3.86$\pm$1.10)e32 &--5.677$\pm$0.010   &2.03$\pm$0.07    \\
51& V3892\,Sgr              &0691760101 &0.48  &0(fixed)        &7.17$\pm$0.41  &(7.81$\pm$0.10)e-4  &               &                    &     &     &0.99(378)&(1.13$\pm$0.02)e-12  &(2.51$\pm$0.71)e32 &--5.863$\pm$0.008   &2.07$\pm$0.05    \\
51& V3892\,Sgr              &8647       &0.48  &0.00$\pm$0.14   &12.18$\pm$12.14&(8.08$\pm$0.83)e-4  &               &                    &     &     &1.37(8)  &(1.26$\pm$0.13)e-12  &(2.71$\pm$0.82)e32 &--5.83$\pm$0.04     &2.55$\pm$1.11    \\
53& V771\,Sgr               &           &0.19  &0.97$\pm$0.29   &8.17$\pm$23.2  &(2.48$\pm$0.37)e-3  &               &                    &     &     &1.22(37) &(3.26$\pm$1.37)e-12  &(2.45$\pm$1.19)e32 &--4.64$\pm$0.18     &6.29$\pm$3.37    \\
54& HD\,316568              &0206590201 &0.33  &0.00$\pm$0.06   &4.87$\pm$1.94  &(3.66$\pm$0.33)e-5  &               &                    &     &     &1.31(11) &(4.89$\pm$1.18)e-14  &(3.83$\pm$2.17)e31 &--6.27$\pm$0.10     &1.61$\pm$0.58    \\
54& HD\,316568              &0402280101 &0.33  &0(fixed)        &3.72$\pm$0.58  &(5.47$\pm$0.28)e-5  &               &                    &     &     &0.87(41) &(6.48$\pm$0.58)e-14  &(5.21$\pm$2.71)e31 &--6.14$\pm$0.04     &1.27$\pm$0.20    \\
54& HD\,316568              &4547       &0.33  &0(fixed)        &5.73$\pm$7.41  &(2.57$\pm$0.37)e-5  &               &                    &     &     &1.15(3)  &(3.64$\pm$1.27)e-14  &(2.81$\pm$1.74)e31 &--6.40$\pm$0.15     &1.81$\pm$1.28    \\
55& ALS\,4570               &           &1.00  &0.01$\pm$0.16   &0.31$\pm$0.04  &(6.01$\pm$4.80)e-4  &2.61$\pm$0.21  &(2.69$\pm$0.16)e-4  &     &     &1.10(83) &(2.57$\pm$0.36)e-13  &(1.12$\pm$0.38)e33 &--8.01$\pm$0.06     &0.18$\pm$0.06    \\
56& Cl*\,NGC\,6530\,ZCW\,175&0008820101 &0.24  &0(fixed)        &2.27$\pm$0.41  &(2.93$\pm$0.29)e-5  &               &                    &     &     &1.36(15) &(2.93$\pm$0.40)e-14  &(5.09$\pm$1.42)e30 &--5.89$\pm$0.06     &0.72$\pm$0.21    \\
56& Cl*\,NGC\,6530\,ZCW\,175&3754       &0.24  &0.53$\pm$0.19   &0.96$\pm$0.11  &(1.43$\pm$0.40)e-5  &               &                    &     &     &1.16(11) &(9.07$\pm$1.22)e-15  &(1.68$\pm$0.47)e30 &--6.37$\pm$0.06     &0.21$\pm$0.05    \\
57& HD\,164906              &0008820101 &0.26  &0.58$\pm$0.14   &0.78$\pm$0.08  &(6.06$\pm$1.82)e-5  &               &                    &     &     &1.97(18) &(3.16$\pm$0.40)e-14  &(8.66$\pm$2.68)e30 &--7.48$\pm$0.05     &0.14$\pm$0.04    \\
57& HD\,164906              &0720540401 &0.26  &0.79$\pm$0.12   &0.73$\pm$0.13  &(9.09$\pm$3.20)e-5  &               &                    &     &     &1.60(20) &(3.45$\pm$0.62)e-14  &(9.16$\pm$3.05)e30 &--7.45$\pm$0.08     &0.16$\pm$0.05    \\
57& HD\,164906              &0720540501 &0.26  &0.70$\pm$0.11   &0.78$\pm$0.07  &(8.23$\pm$1.67)e-5  &               &                    &     &     &1.64(26) &(3.77$\pm$0.38)e-14  &(9.84$\pm$2.94)e30 &--7.42$\pm$0.04     &0.16$\pm$0.04    \\
57& HD\,164906              &0720540601 &0.26  &0.000$\pm$0.009 &2.00$\pm$0.18  &(5.26$\pm$0.39)e-5  &               &                    &     &     &1.72(32) &(5.01$\pm$0.68)e-14  &(1.24$\pm$0.39)e31 &--7.32$\pm$0.06     &0.58$\pm$0.13    \\
57& HD\,164906              &977        &0.26  &0.00$\pm$0.21   &0.72$\pm$0.15  &(5.13$\pm$3.13)e-6  &               &                    &     &     &0.49(1)  &(6.41$\pm$4.00)e-15  &(2.15$\pm$1.47)e30 &--8.08$\pm$0.27     &0.04$\pm$0.02    \\
58& Cl*\,NGC\,6530\,ZCW\,221&0008820101 &0.21  &0.06$\pm$0.09   &2.57$\pm$0.75  &(4.61$\pm$0.58)e-5  &               &                    &     &     &1.13(14) &(4.76$\pm$0.68)e-14  &(1.19$\pm$0.38)e31 &--5.58$\pm$0.06     &0.93$\pm$0.22    \\
58& Cl*\,NGC\,6530\,ZCW\,221&0720540401 &0.21  &0.05$\pm$0.12   &2.20$\pm$0.52  &(5.61$\pm$0.75)e-5  &               &                    &     &     &0.77(14) &(5.47$\pm$0.61)e-14  &(1.40$\pm$0.43)e31 &--5.51$\pm$0.05     &0.74$\pm$0.25    \\
58& Cl*\,NGC\,6530\,ZCW\,221&0720540501 &0.21  &0.00$\pm$0.02   &2.84$\pm$0.31  &(1.22$\pm$0.06)e-4  &               &                    &     &     &1.15(57) &(1.37$\pm$0.11)e-13  &(3.47$\pm$1.03)e31 &--5.11$\pm$0.03     &0.96$\pm$0.14    \\
58& Cl*\,NGC\,6530\,ZCW\,221&0720540601 &0.21  &0(fixed)        &2.09$\pm$0.25  &(5.64$\pm$0.43)e-5  &               &                    &     &     &1.48(27) &(5.64$\pm$0.47)e-14  &(1.48$\pm$0.44)e31 &--5.48$\pm$0.04     &0.63$\pm$0.10    \\
59& Cl*\,NGC\,6530\,ZCW\,228&0008820101 &0.23  &0.00$\pm$0.02   &2.07$\pm$0.33  &(7.28$\pm$0.55)e-5  &               &                    &     &     &0.90(25) &(7.15$\pm$0.67)e-14  &(1.25$\pm$0.33)e31 &--5.51$\pm$0.04     &0.62$\pm$0.11    \\
59& Cl*\,NGC\,6530\,ZCW\,228&0720540401 &0.23  &0.00$\pm$0.18   &1.27$\pm$0.11  &(2.67$\pm$0.58)e-5  &               &                    &     &     &1.20(13) &(2.91$\pm$0.59)e-14  &(5.63$\pm$1.78)e30 &--5.85$\pm$0.09     &0.21$\pm$0.07    \\
59& Cl*\,NGC\,6530\,ZCW\,228&0720540601 &0.23  &0(fixed)        &1.95$\pm$0.14  &(7.32$\pm$0.45)e-5  &               &                    &     &     &1.24(41) &(7.12$\pm$0.50)e-14  &(1.26$\pm$0.32)e31 &--5.50$\pm$0.03     &0.56$\pm$0.08    \\
59& Cl*\,NGC\,6530\,ZCW\,228&977        &0.23  &0.00$\pm$0.04   &1.60$\pm$0.08  &(2.88$\pm$0.23)e-5  &               &                    &     &     &1.73(15) &(2.82$\pm$0.46)e-14  &(5.15$\pm$1.51)e30 &--5.89$\pm$0.07     &0.39$\pm$0.09    \\
59& Cl*\,NGC\,6530\,ZCW\,228&3754       &0.23  &0.00$\pm$0.08   &1.64$\pm$0.43  &(1.06$\pm$0.24)e-5  &               &                    &     &     &0.72(31) &(1.03$\pm$0.20)e-14  &(1.87$\pm$0.58)e30 &--6.33$\pm$0.08     &0.41$\pm$0.13    \\
59& Cl*\,NGC\,6530\,ZCW\,228&4444       &0.23  &0.00$\pm$0.35   &1.87$\pm$0.54  &(1.41$\pm$0.33)e-5  &               &                    &     &     &0.93(9)  &(1.36$\pm$0.33)e-14  &(2.42$\pm$0.83)e30 &--6.22$\pm$0.11     &0.53$\pm$0.34    \\
60& V4379\,Sgr              &           &0.043 &0(fixed)        &2.30$\pm$0.49  &(2.59$\pm$0.27)e-3  &               &                    &     &     &1.16(66) &(3.12$\pm$0.40)e-12  &(5.82$\pm$4.36)e30 &--4.28$\pm$0.06     &0.74$\pm$0.23    \\
63& BD-13\,4928             &0605130101 &0.20  &0.82$\pm$0.16   &0.24$\pm$0.05  &(3.81$\pm$3.83)e-4  &2.13$\pm$0.45  &(4.87$\pm$0.67)e-5  &     &     &0.68(42) &(5.01$\pm$1.30)e-14  &(3.73$\pm$1.20)e31 &--6.22$\pm$0.11     &0.54$\pm$0.26    \\
69& HD\,190864              &           &0.37  &0.37$\pm$0.06   &0.25$\pm$0.02  &(7.08$\pm$2.61)e-4  &               &                    &     &     &1.32(64) &(7.32$\pm$1.03)e-14  &(1.14$\pm$0.51)e32 &--7.04$\pm$0.06     &0.0033$\pm$0.0009\\
74& HD\,198931              &           &0.25  &1.12$\pm$0.15   &0.69$\pm$0.11  &(1.02$\pm$0.33)e-4  &               &                    &     &     &1.19(14) &(2.57$\pm$0.75)e-14  &(2.74$\pm$1.40)e30 &--6.88$\pm$0.13     &0.21$\pm$0.11    \\
75& V2156\,Cyg              &           &0.19  &2.44$\pm$0.91   &2.63$\pm$2.14  &(1.25$\pm$0.61)e-3  &               &                    &     &     &1.11(6)  &(6.88$\pm$2.72)e-13  &(7.53$\pm$3.34)e31 &--5.30$\pm$0.17     &6.34$\pm$3.33    \\
76& Alfirk                  &0300490201 &0.004 &0.005$\pm$0.002 &0.207$\pm$0.003&(5.76$\pm$0.15)e-4  &0.700$\pm$0.008&(2.24$\pm$0.05)e-4  &     &     &2.14(208)&(9.49$\pm$0.12)e-13  &(1.27$\pm$0.27)e30 &--6.920$\pm$0.005   &0.0178$\pm$0.0008\\
76& Alfirk                  &0300490301 &0.004 &0.000$\pm$0.002 &0.197$\pm$0.003&(4.90$\pm$0.10)e-4  &0.648$\pm$0.010&(2.07$\pm$0.06)e-4  &     &     &1.96(200)&(8.20$\pm$0.12)e-13  &(1.10$\pm$0.23)e30 &--6.983$\pm$0.006   &0.0152$\pm$0.0007\\
76& Alfirk                  &0300490401 &0.004 &0.011$\pm$0.003 &0.197$\pm$0.003&(5.49$\pm$0.16)e-4  &0.643$\pm$0.010&(2.25$\pm$0.07)e-4  &     &     &1.79(200)&(8.67$\pm$0.12)e-13  &(1.16$\pm$0.24)e30 &--6.958$\pm$0.006   &0.0153$\pm$0.0007\\
76& Alfirk                  &0300490501 &0.004 &0.008$\pm$0.002 &0.201$\pm$0.003&(5.39$\pm$0.14)e-4  &0.651$\pm$0.011&(2.09$\pm$0.06)e-4  &     &     &1.71(198)&(8.52$\pm$0.11)e-13  &(1.14$\pm$0.24)e30 &--6.966$\pm$0.006   &0.0150$\pm$0.0006\\
79& $\pi$\,Aqr              &           &0.036 &0.226$\pm$0.003 &11.51$\pm$0.25 &(6.37$\pm$0.03)e-3  &               &                    &     &     &1.41(819)&(1.055$\pm$0.004)e-11&(7.44$\pm$0.99)e31 &--5.5918$\pm$0.0016 &3.56$\pm$0.14    \\
82& BD+61\,2355             &           &0.16  &0.64$\pm$2.63   &0.41$\pm$2.42  &(0.82$\pm$9.50)e-4  &               &                    &     &     &4.30(3)  &(2.01$\pm$2.01)e-14  &(1.10$\pm$1.11)e30 &--5.84$\pm$0.43     &0.03$\pm$0.04    \\
83& V810\,Cas               &           &0.18  &0.31$\pm$0.04   &62.97$\pm$31.30&(1.00$\pm$0.04)e-3  &               &                    &     &     &1.30(71) &(1.38$\pm$0.35)e-12  &(4.55$\pm$1.91)e32 &--5.17$\pm$0.11     &5.43$\pm$2.13    \\
\hline
\end{tabular}
\tablefoot{$^a$ 1.18$\pm$0.07, $^b$(6.02$\pm$0.89)e-4, $^c$1.13$\pm$0.24, $^d$(3.94$\pm$0.98)e-4, $^e$1.58$\pm$0.10, and $^f$(8.27$\pm$0.74)e-5\\
  Whenever several observations were fitted, the second column provides the corresponding ObsID. The spectral models are of the form $tbabs\times phabs \times \sum apec$. The interstellar columns are derived from the color excesses of Table \ref{journal2} using the formula of \citet{gud12}. The hardness ratios are defined by $HR=F_{\rm X}^{ISM\,cor}(hard)/F_{\rm X}^{ISM\,cor}(soft)$, with soft and hard energy bands being defined as 0.5--2.0\,keV and 2.0--10.0\,keV, respectively (the total band being 0.5--10.0\,keV). Errors correspond to 1$\sigma$ uncertainties. For the X-ray luminosities, they combine the distance errors (see Table \ref{journal2}) with errors on X-ray fluxes (derived from the ``flux err'' command in Xspec), but do not integrate the impact of model choices; errors on \loglxlb, however, do not depend on distance and reflect only X-ray flux uncertainties.
}
\end{sidewaystable*}
\normalsize

\begin{table*}
\caption{X-ray luminosities in 0.5--10.0\,keV, corrected for interstellar absorption, for the simple detection cases (see Section 4.1.2 for details).}
\label{det}
\tiny 
\setlength{\tabcolsep}{2pt}
\begin{tabular}{clccc}
\hline\hline
\# & Name &  $N_{\rm H}^{ISM}$ & $L_{\rm X}^{ISM\,cor}$(min..max $\pm$ error) & $\log(L_{\rm X}/L_{\rm BOL})$ (min..max $\pm$ error)\\
   &      & ($10^{22}$\,cm$^{-2}$) & (erg\,s$^{-1}$) & \\
\hline
4 & Cl*\,NGC\,869\,LAV\,1039    &0.37   &(0.4..1.3$\pm$0.7)e30 &-7.1..-6.7$\pm$0.2   \\
5 & NGC\,869\,1164              &0.39   &(0.2..1.2$\pm$1.1)e31 &-4.9..-4.1$\pm$0.4   \\
6 & HD\,14162                   &0.39   &(0.6..2.0$\pm$1.5)e30 &-8.7..-8.2$\pm$0.2   \\
7 & Cl*\,NGC\,884\,LAV\,1703    &0.41   &(1.1..2.7$\pm$1.2)e31 &-6.0..-5.6$\pm$0.2   \\
8 & HD\,17505                   &0.47   &(0.7..1.0$\pm$0.8)e34 &-6.1..-5.9$\pm$0.2   \\
13& 25\,Ori                     &0.016  &(0.5..1.9$\pm$0.9)e29 &-8.7..-8.1$\pm$0.2   \\
18& Cl*\,NGC\,2244\,PS\,26      &0.16   &(4.1..6.1$\pm$1.9)e29 &-5.7..-5.6$\pm$0.1   \\
19& Cl*\,NGC\,2244\,JOHN\,33    &0.23   &(2.2..4.2$\pm$0.9)e30 &-5.52..-5.25$\pm$0.06\\
26& HD\,90563                   &0.42   &(2.0..4.4$\pm$3.3)e32 &-6.0..-5.7$\pm$0.1   \\
31& Phecda                      &0.0006 &(0.4..4.5$\pm$2.3)e29 &-6.9..-5.8$\pm$0.2   \\
35& BQ\,Cru                     &0.58   &(2.7..6.1$\pm$1.7)e32 &-4.2..-3.9$\pm$0.1   \\
36& HD\,117357                  &0.22   &(3.1..5.9$\pm$5.8)e30 &-7.2..-6.9$\pm$0.3   \\
44& HD\,153295                  &0.34   &<5.4e32              & <-5.0                \\
48& $\alpha$\,Ara               &0.004  &(1.2..6.5$\pm$2.9)e30 &-6.9..-6.1$\pm$0.2   \\
49& V864\,Ara                   &0.035  &(0.6..2.4$\pm$1.1)e31 &-5.1..-4.5$\pm$0.2   \\
58& Cl*\,NGC\,6530\,ZCW\,221    &0.21   &(1.0..1.8$\pm$0.7)e30 &-6.6..-6.4$\pm$0.1   \\
61& HD\,165783                  &0.39   &(0.4..1.5$\pm$1.3)e31 &-7.3..-6.6$\pm$0.3   \\
62& Cl*\,NGC\,6611\,PPM\,38     &0.55   &(0.2..1.1$\pm$0.2)e31 &-5.50..-4.80$\pm$0.08\\
64& Cl*\,NGC\,6611\,BKP\,29783  &0.59   &(1.4..4.4$\pm$0.7)e31 &-5.73..-5.24$\pm$0.07\\ 
65& BD-13\,4933                 &0.67   &(1.1..8.0$\pm$6.2)e30 &-8.3..-7.5$\pm$0.2   \\
66& EM*\,AS\,315                &0.80   &(1.8..8.0$\pm$7.7)e31 &-7.7..-7.1$\pm$0.2   \\
67& CX\,Dra                     &0.015  &(0.9..5.0$\pm$2.1)e31 &-6.1..-5.3$\pm$0.2   \\
68& HD\,344783                  &0.37   &(1.3..3.8$\pm$4.2)e31 &-6.8..-6.3$\pm$0.2   \\
70& HD\,228438                  &0.40   &(1.0..2.4$\pm$1.6)e31 &-7.8..-7.5$\pm$0.2   \\
71& HD\,228860                  &0.65   &(0.8..3.0$\pm$1.8)e31 &-7.7..-7.2$\pm$0.2   \\
73& W\,Del                      &0.031  &(1.0..2.9$\pm$0.9)e30 &-5.6..-5.1$\pm$0.1   \\
77& $\epsilon$\,Cap             &0.004  &(1.4..7.8$\pm$4.3)e30 &-6.6..-5.9$\pm$0.2   \\
78& EM\,Cep                     &0.13   &(2.3..3.0$\pm$1.7)e29 &-8.1..-8.0$\pm$0.1   \\
80& HD\,215227                  &0.15   &(0.06..1.1$\pm$0.8)e32&-6.9..-5.6$\pm$0.2   \\
81& EM*\,MWC\,659               &0.26   &(0.2..1.1$\pm$1.0)e34 &-5.0..-4.2$\pm$0.2   \\
84& $\beta$\,Scl                &0.0006 &(1.4..7.7$\pm$3.5)e29 &-6.3..-5.6$\pm$0.2   \\
\hline
\end{tabular}
\tablefoot{The interstellar columns are derived from the color excesses of Table \ref{journal2} using the formula of \citet{gud12}. Errors correspond to 1$\sigma$ uncertainties, combining the distance errors (see Table \ref{journal2}) with errors on maximum count rates or photon fluxes. When there were two detections (\xmm\ and \ch\ for 25\,Ori and \xmm\ slew survey and \sw\ for EM*\,MWC\,659 - for HD\,215227, only the \xmm\ detection was considered, not the new \ch\ one), the range reflects the minimum and maximum fluxes considering available data. For Cl*\,NGC\,6530\,ZCW\,221, the \xmm\ data (Table \ref{fits}) most probably refer to a brighter companion (see discussion in Sect. 3), therefore the \ch\ data reported here reflect the stellar X-ray properties better. }
\end{table*}
\normalsize

\begin{table*}
\caption{List of properties of known, new, and candidate $\gamma$\,Cas analogs, along with two other peculiar bright sources. }
\label{candidates.tab}
\setlength{\tabcolsep}{3pt}
\tiny 
\begin{tabular}{llccccccccc}
\hline
\# & Name     & Spectral &  \loglxlb &  $L_{\rm X}^{ISM\,cor}(tot)$ & $L_{\rm X}^{ISM\,cor}(hard)$ & $HR$ & $kT$ & var.? & --EW(H$\alpha$) & $v_{rot} sini$ \\
   &          & type     &           & \multicolumn{2}{c}{(erg\,s$^{-1}$)} &   & (keV) & short/long & (\AA) & (km\,s$^{-1}$) \\
\hline
\hline
\multicolumn{7}{l}{\it Known  $\gamma$\,Cas analogs} \\
1 & $\gamma$\,Cas & B0IV-Vpe & --5.39$\pm$0.04   & (8.50$\pm$0.84)e32 & (6.51$\pm$0.68)e32 & 3.25$\pm$0.14 & 14--25 & y/y& 34 & 295\\ 
17& PZ\,Gem (high)& O9pe     &--6.138$\pm$0.008  & (9.66$\pm$1.77)e31 & (7.87$\pm$1.44)e31 & 4.32$\pm$0.11 & 16     & /y & 15 & 265\\
34& BZ\,Cru       & B0.5IVpe &--5.6930$\pm$0.0014& (2.76$\pm$0.37)e32 & (2.03$\pm$0.27)e32 &2.814$\pm$0.014& 13     & y/ & 32 & 338\\
37& HD\,119682    & B0Ve     & --5.63$\pm$0.04   & (6.67$\pm$0.58)e32 & (4.79$\pm$0.43)e32 & 2.55$\pm$0.13 & 8--17  &y/y &  5 & 200\\
47& V750\,Ara     & B2Vne    & --5.291$\pm$0.008 & (4.19$\pm$1.14)e32 & (3.19$\pm$0.87)e32 & 3.21$\pm$0.09 & 10     &y/  & 31 & 277\\
51& V3892\,Sgr    & Oe       & --5.78$\pm$0.07   & (3.08$\pm$0.49)e32 & (2.12$\pm$0.32)e32 & 2.24$\pm$0.24 & 7--14  &y/y &    & 260\\
79& $\pi$\,Aqr    & B1Ve     &--5.5918$\pm$0.0016& (7.44$\pm$0.99)e31 & (5.80$\pm$0.77)e31 & 3.56$\pm$0.14 & 12     &y/  &  3 & 243\\
\hline
\multicolumn{7}{l}{\it New $\gamma$\,Cas analogs}\\
3 & V782\,Cas  & B2.5III:[n]e+ & --5.25$\pm$0.07 & (3.03$\pm$0.78)e32 & (2.99$\pm$0.75)e32 & 63.1$\pm$16.2 & 7  &n/  & 22 &    \\
26& HD\,90563  & B2Ve          & --6.0..--5.7    & (2.0..4.4)e32      &                    &               &    &    & 65 &    \\
39& V767\,Cen  & B2Ve          &--5.365$\pm$0.007& (2.62$\pm$0.63)e32 & (1.74$\pm$0.42)e32 & 1.97$\pm$0.05 & 6  &y/  &  6 & 100\\
40& CQ\,Cir    & B1Ve          & --4.30$\pm$0.17 & (1.75$\pm$0.93)e33 & (1.47$\pm$0.79)e33 & 5.26$\pm$2.10 & 9  & /y &    & 335\\
53& V771\,Sgr  & B3/5ne        & --4.64$\pm$0.18 & (2.45$\pm$1.19)e32 & (2.11$\pm$1.23)e32 & 6.29$\pm$3.37 & 8  & /y & 53 &    \\
54& HD\,316568 & B2IVpe        & --6.26$\pm$0.08 & (4.04$\pm$0.77)e31 & (2.43$\pm$0.32)e31 & 1.60$\pm$0.23 &4--6&n/y &    &    \\
75& V2156\,Cyg & B1.5Vnnpe     & --5.30$\pm$0.17 & (7.53$\pm$3.34)e31 & (6.51$\pm$3.38)e31 & 6.34$\pm$3.33 & 3  &    & 34 &    \\
83& V810\,Cas  & B1npe         &--5.139$\pm$0.015& (4.84$\pm$1.62)e32 & (4.11$\pm$1.38)e32 & 5.58$\pm$0.30 & 64 &n/  & 33 & 422\\
\hline
\multicolumn{7}{l}{\it Candidate $\gamma$\,Cas analog}   \\
16& HD 42054  & B4IVe   & --5.39$\pm$0.05 & (3.24$\pm$0.34)e31 & (1.88$\pm$0.13)e31 & 1.40$\pm$0.11 & 5--9 &y/y & 12 & 220\\   
\hline
\multicolumn{7}{l}{\it Others}   \\
8 & HD\,17505    & O6.5III((f))n+O8V & --6.1..--5.9 & (0.7..1.0)e34 & & & &   & &126\\
81& EM*\,MWC\,659& B0IIIpe           & --5.0..--4.2 & (0.2..1.1)e34 & & & &   & &   \\ 
\hline
\end{tabular}
\tablefoot{Spectral type and projected rotational velocities are reproduced from Table \ref{journal2}. X-ray properties come from the fits of Table \ref{fits} or from Table \ref{det}, except when several observations are available, in which case the average values and the dispersions around them are provided. When several thermal components were fit, only the temperature of the hottest one is reported here. For PZ\,Gem, only the ``high'' state (i.e., the observation with the highest X-ray luminosity) is reported here. The short-term variability flag comes from Table \ref{journal}, and the long-term variability flag was set when significant variations of the observed fluxes (see Table \ref{fits}) or \sw\ count rates are detected. Equivalent widths of the H$\alpha$ line are estimated from the spectra in the BeSS database taken at the closest time with respect to the X-ray observation(s) (see also \citealt{smith2016} for typical values for the previously known $\gamma$\,Cas analogs). These BeSS observations are rarely simultaneous. In this context, we note that \citet{rau13} reported EW(H$\alpha$)=--22\AA\ at the time of the \xmm\ observation in ``high'' state, while the emission reported here is lower because of its later date.
}
\end{table*}

\begin{figure*}
\includegraphics[width=15cm]{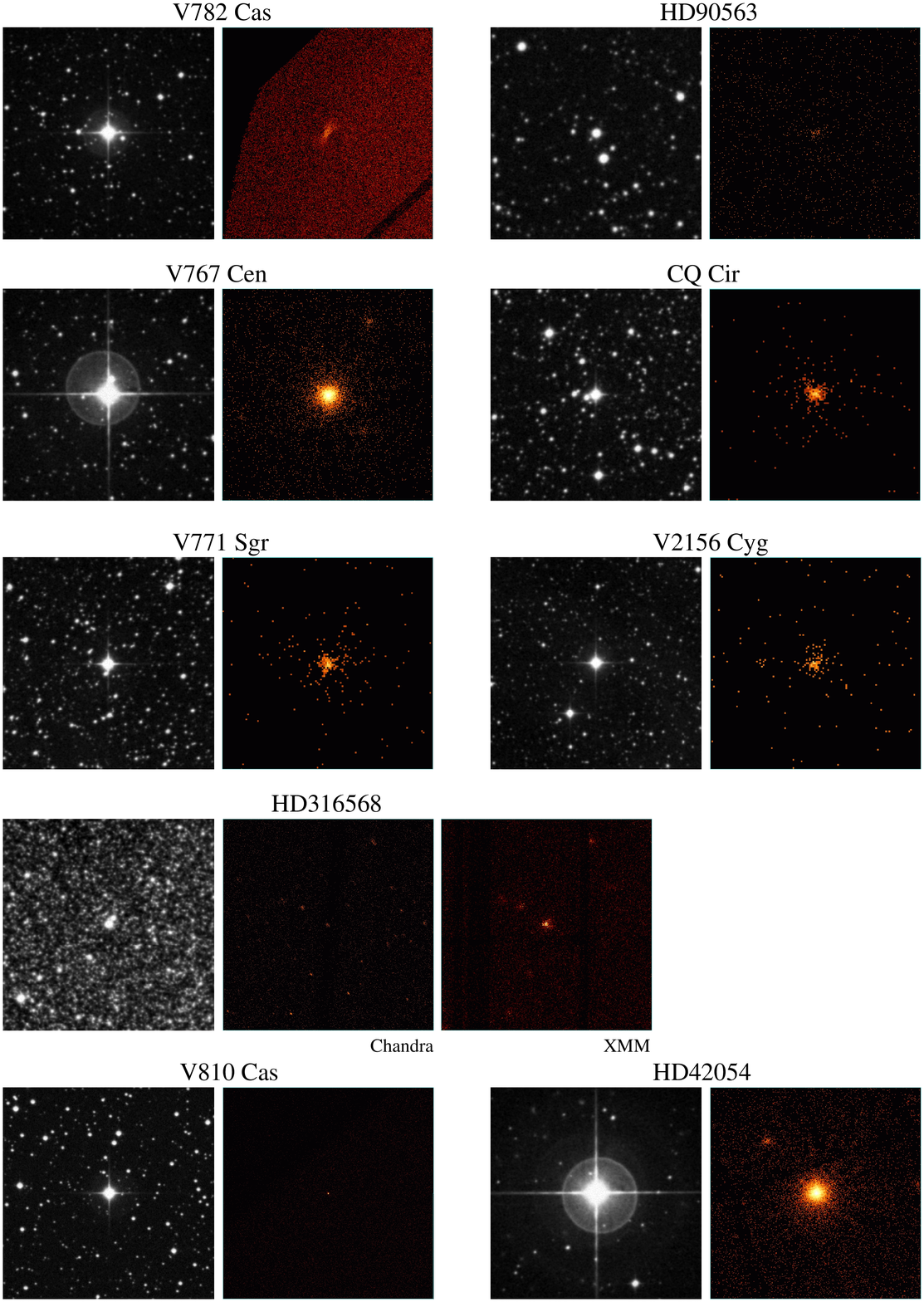}
\caption{Optical and X-ray images centered on the eight new $\gamma$\,Cas analogs and the $\gamma$\,Cas candidate. All images are centered on the targets, 5'$\times$5' in size, and oriented with north up and east to the left. The X-ray images correspond to the 0.5--10.0\,keV energy band (after merging all EPIC data when \xmm\ is used), while the optical images are from the DSS survey (POSS2/UKSTU Red). For HD\,316568, the \ch\ data with ObsID 4547 and the \xmm\ data with ObsID 0402280101 are both shown; for HD\,42054, the \xmm\ dataset is shown rather than the zeroth-order \ch\ data; for CQ\,Cir, V771\,Sgr, and V2156\,Cyg, images merging all \sw\ data were created with the online tool (see Sect. 4.2.1). }
\label{findingchart}
\end{figure*}

\begin{acknowledgements}
YN acknowledges support from the Fonds National de la Recherche Scientifique (Belgium), the Communaut\'e Fran\c caise de Belgique, and the PRODEX \xmm\ contract. CM acknowledges financial support from the French Centre National d'Etudes Spatiales (CNES). We thank R. Lallement for providing us with automated distance-reddening curves from the STILISM database. This research has made use of the ADS as well as the SIMBAD database, operated at CDS (Strasbourg, France). It also used the VizieR catalogue access tool (CDS,  Strasbourg, France - initially described in A\&AS 143, 23) and the facilities of the XCat-DB developped by the Survey Science Center of the \xmm\ satellite at Strasbourg Observatory. In addition to \xmm, this work has made use of data from the European Space Agency (ESA) mission {\it Gaia} (\url{https://www.cosmos.esa.int/gaia}), processed by the {\it Gaia} Data Processing and Analysis Consortium (DPAC, \url{https://www.cosmos.esa.int/web/gaia/dpac/consortium}). Funding for the DPAC has been provided by national institutions, in particular the institutions participating in the {\it Gaia} Multilateral Agreement. This work has also made use of the BeSS database, operated at LESIA (Observatoire de Meudon, France) and available on http://basebe.obspm.fr, as well as data and/or software provided by the High Energy Astrophysics Science Archive Research Center (HEASARC), which is a service of the Astrophysics Science Division at NASA/GSFC and the High Energy Astrophysics Division of the Smithsonian Astrophysical Observatory.
 
\end{acknowledgements}


\begin{thebibliography}{00}
\bibitem[Abt et al.(2002)]{abt2002} Abt, H.~A., Levato, H., \& Grosso, M.\ 2002, \apj, 573, 359 
\bibitem[Antokhin et al.(2008)]{ant08} Antokhin, I.~I., Rauw, G., Vreux, J.-M., van der Hucht, K.~A., \& Brown, J.~C.\ 2008, \aap, 477, 593
\bibitem[Asplund et al.(2009)]{asp09} Asplund, M., Grevesse, N., Sauval, A.J., \& Scott, P.\ 2009, \araa, 47, 481 
\bibitem[Bagnulo et al.(2006)]{bag06} Bagnulo, S., Landstreet, J.~D., Mason, E., et al.\ 2006, \aap, 450, 777
\bibitem[Bagnulo et al.(2015)]{bag15} Bagnulo, S., Fossati, L., Landstreet, J.~D., \& Izzo, C.\ 2015, \aap, 583, A115
\bibitem[Baume et al.(2003)]{baume2003} Baume, G., V{\'a}zquez, R.~A., Carraro, G., \& Feinstein, A.\ 2003, \aap, 402, 549 
\bibitem[Berger \& Gies(2001)]{ber01} Berger, D.~H., \& Gies, D.~R.\ 2001, \apj, 555, 364
\bibitem[Berghoefer et al.(1997)]{ber97} Berghoefer, T.~W., Schmitt, J.~H.~M.~M., Danner, R., \& Cassinelli, J.~P.\ 1997, \aap, 322, 167 
\bibitem[Bragg \& Kenyon(2002)]{bragg2002} Bragg, A.~E. \& Kenyon, S.~J.\ 2002, \aj, 124, 3289.
\bibitem[Capitanio et al.(2017)]{capitanio2017} Capitanio, L., Lallement, R., Vergely, J.~L., Elyajouri, M., \& Monreal-Ibero, A.\ 2017, \aap, 606, A65 
\bibitem[Chauville et al.(2001)]{chauville2001} Chauville, J., Zorec, J., Ballereau, D., et al.\ 2001, \aap, 378, 861 
\bibitem[Claeskens et al.(2011)]{cla11} Claeskens, J.-F., Gosset, E., Naz{\'e}, Y., Rauw, G., \& Vreux, J.-M.\ 2011, \aap, 525, A142
\bibitem[Cohen et al.(1997)]{coh97} Cohen, D.~H., Cassinelli, J.~P., \& MacFarlane, J.~J.\ 1997, \apj, 487, 867 
\bibitem[Currie et al.(2010)]{currie2010} Currie, T., Hernandez, J., Irwin, J., et al.\ 2010, \apjs, 186, 191 
\bibitem[David \& Hillenbrand(2015)]{david2015} David, T.~J., \& Hillenbrand, L.~A.\ 2015, \apj, 804, 146 
\bibitem[de Jager \& Nieuwenhuijzen(1987)]{dejager1987} de Jager, C., \& Nieuwenhuijzen, H.\ 1987, \aap, 177, 217 
\bibitem[Evans(1967)]{evans1967} Evans, D.~S.\ 1967, Determination of Radial Velocities and their Applications, 30, 57 
\bibitem[Fr{\'e}mat et al.(2002)]{fre02} Fr{\'e}mat, Y., Zorec, J., Hubert, A.-M., et al.\ 2002, \aap, 385, 986
\bibitem[Gaia Collaboration et al.(2016)]{gaiadr1} Gaia Collaboration, Brown, A.~G.~A., Vallenari, A., et al.\ 2016, \aap, 595, A2 
\bibitem[Gaia Collaboration et al.(2018)]{gaiadr2g} Gaia Collaboration, Brown, A.~G.~A., Vallenari, A., et al.\ 2018, arXiv:1804.09365 
\bibitem[Gagn\'e et al.(2008)]{gag08} Gagn\'e, M., Cohen, D., Owocki, S., Townsley, L., \& Broos, P., poster presented at ``Contifest, a lifetime of influence'', available on http://www2.lowell.edu/workshops/Contifest/talks/Gagne.pdf
\bibitem[Gim{\'e}nez-Garc{\'{\i}}a et al.(2015)]{gim15} Gim{\'e}nez-Garc{\'{\i}}a, A., Torrej{\'o}n, J.~M., Eikmann, W., et al.\ 2015, \aap, 576, A108
\bibitem[Gkouvelis et al.(2016)]{gko16} Gkouvelis, L., Fabregat, J., Zorec, J., et al.\ 2016, \aap, 591, A140 
\bibitem[G{\l}{\c e}bocki \& Gnaci{\'n}ski(2005)]{gle05} G{\l}{\c e}bocki, R., \& Gnaci{\'n}ski, P.\ 2005, 13th Cambridge Workshop on Cool Stars, Stellar Systems and the Sun, 560, 571
\bibitem[Gontcharov \& Mosenkov(2018)]{gon2018} Gontcharov, G.~A., \& Mosenkov, A.~V.\ 2018, \mnras, 475, 1121 
\bibitem[Grunhut et al.(2012)]{gru12} Grunhut, J.~H., Wade, G.~A., \& MiMeS Collaboration 2012, American Institute of Physics Conference Series, 1429, 67 
\bibitem[Grunhut et al.(2017)]{gru17} Grunhut, J.~H., Wade, G.~A., Neiner, C., et al.\ 2017, \mnras, 465, 2432 
\bibitem[Guarcello et al.(2007)]{guar2007} Guarcello, M.~G., Prisinzano, L., Micela, G., et al.\ 2007, \aap, 462, 245 
\bibitem[Gudennavar et al.(2012)]{gud12} Gudennavar, S.~B., Bubbly, S.~G., Preethi, K., \& Murthy, J.\ 2012, \apjs, 199, 8 
\bibitem[Hamaguchi et al.(2016)]{ham16} Hamaguchi, K., Oskinova, L., Russell, C.~M.~P., et al.\ 2016, \apj, 832, 140
\bibitem[Huang et al.(2010)]{huang2010} Huang, W., Gies, D.~R., \& McSwain, M.~V.\ 2010, \apj, 722, 605 
\bibitem[Hensberge et al.(2000)]{hens2000} Hensberge, H., Pavlovski, K., \& Verschueren, W.\ 2000, Star Formation from the Small to the Large Scale, 445, 395 
\bibitem[Jaschek \& Egret(1982)]{jas82} Jaschek, M., \& Egret, D.\ 1982, Be Stars, 98, 261
\bibitem[Kaltcheva \& Golev(2012)]{kal2012} Kaltcheva, N.~T., \& Golev, V.~K.\ 2012, \pasp, 124, 128 
\bibitem[Li et al.(2018)]{li18} Li, G.-W., Shi, J.-R., Yanny, B., et al.\ 2018, \apj, 863, 70
\bibitem[Lopes de Oliveira et al.(2007)]{lop07} Lopes de Oliveira, R., Motch, C., Smith, M.~A., Negueruela, I., \& Torrej{\'o}n, J.~M.\ 2007, \aap, 474, 983
\bibitem[Lopes de Oliveira et al.(2010)]{lop10} Lopes de Oliveira, R., Smith, M.~A., \& Motch, C.\ 2010, \aap, 512, A22
\bibitem[Luri et al.(2018)]{luri2018} Luri, X., Brown, A.~G.~A., Sarro, L.~M., et al.\ 2018, arXiv:1804.09376 
\bibitem[Maryeva et al.(2017)]{mary2017} Maryeva, O.~V., Chentsov, E.~L., Goranskij, V.~P., Dyachenko, V.~V., \& Karpov, S.~V.\ 2017, Stars: From Collapse to Collapse, 510, 187 
\bibitem[Marshall et al.(2006)]{marshall2006} Marshall, D.~J., Robin, A.~C., Reyl{\'e}, C., Schultheis, M., \& Picaud, S.\ 2006, \aap, 453, 635
\bibitem[Mathew et al.(2008)]{mat08} Mathew, B., Subramaniam, A., \& Bhatt, B.~C.\ 2008, \mnras, 388, 1879
\bibitem[Mennickent et al.(2002)]{men02} Mennickent, R.~E., Pietrzy{\'n}ski, G., Gieren, W., \& Szewczyk, O.\ 2002, \aap, 393, 887 
\bibitem[Meurs et al.(1992)]{meu92} Meurs, E.~J.~A., Piters, A.~J.~M., Pols, O.~R., et al.\ 1992, \aap, 265, L41
\bibitem[Motch et al.(2015)]{mot15} Motch, C., Lopes de Oliveira, R., \& Smith, M.~A.\ 2015, \apj, 806, 177 
\bibitem[Naz{\'e}(2009)]{naz09} Naz{\'e}, Y.\ 2009, \aap, 506, 1055 
\bibitem[Naz{\'e} et al.(2011)]{naz11} Naz{\'e}, Y., Broos, P.~S., Oskinova, L., et al.\ 2011, \apjs, 194, 7 
\bibitem[Naz{\'e} et al.(2014a)]{naz14} Naz{\'e}, Y., Petit, V., Rinbrand, M., et al.\ 2014a, \apjs, 215, 10 (+ erratum \apjs, 224, 13)
\bibitem[Naz{\'e} et al.(2014b)]{naz14n11} Naz{\'e}, Y., Wang, Q.~D., Chu, Y.-H., Gruendl, R., \& Oskinova, L.\ 2014b, \apjs, 213, 23 
\bibitem[Naz{\'e} et al.(2017)]{naz17} Naz{\'e}, Y., Rauw, G., \& Cazorla, C.\ 2017, \aap, 602, L5 
\bibitem[Nebot G{\'o}mez-Mor{\'a}n et al.(2013)]{nebot2013} Nebot G{\'o}mez-Mor{\'a}n, A., Motch, C., Barcons, X., et al.\ 2013, \aap, 553, A12 
\bibitem[Nebot G{\'o}mez-Mor{\'a}n et al.(2015)]{nebot2015} Nebot G{\'o}mez-Mor{\'a}n, A., Motch, C., Pineau, F.-X., et al.\ 2015, \mnras, 452, 884 
\bibitem[Neiner et al.(2011)]{nei11} Neiner, C., de Batz, B., Cochard, F., et al.\ 2011, \aj, 142, 149 
\bibitem[Nieva(2013)]{nieva2013} Nieva, M.-F.\ 2013, \aap, 550, A26 
\bibitem[Ochsenbein \& Halbwachs(1987)]{ochs1987} Ochsenbein, F. \& Halbwachs, J.~L.\ 1987, Bulletin d'Information du Centre de Donnees Stellaires, 32, 83.
\bibitem[Oskinova et al.(2013)]{osk13} Oskinova, L.~M., Sun, W., Evans, C.~J., et al.\ 2013, \apj, 765, 73 
\bibitem[Pallavicini et al.(1981)]{pal1981} Pallavicini, R., Golub, L., Rosner, R., et al.\ 1981, \apj, 248, 279 
\bibitem[Piatti \& Clari{\'a}(2001)]{piatti2001} Piatti, A.~E., \& Clari{\'a}, J.~J.\ 2001, \aap, 370, 931 
\bibitem[Rauw et al.(2013)]{rau13} Rauw, G., Naz{\'e}, Y., Spano, M., Morel, T., \& ud-Doula, A.\ 2013, \aap, 555, L9
\bibitem[Rauw et al.(2015)]{rau15} Rauw, G., Naz{\'e}, Y., Wright, N.~J., et al.\ 2015, \apjs, 221, 1 
\bibitem[Rauw \& Naz{\'e}(2016)]{rau16} Rauw, G., \& Naz{\'e}, Y.\ 2016, Advances in Space Research, 58, 761 
\bibitem[Rauw et al.(2018)]{rau18} Rauw, G., Naz{\'e}, Y., Smith, M.~A., et al.\ 2018, arXiv:1802.05512 
\bibitem[Rimoldini et al.(2012)]{rim12} Rimoldini, L., Dubath, P., S{\"u}veges, M., et al.\ 2012, \mnras, 427, 2917
\bibitem[{{Robinson} {et~al.}(2002){Robinson}, {Smith}, \&  {Henry}}]{robinson2002} {Robinson}, R.~D., {Smith}, M.~A., \& {Henry}, G.~W. 2002, \apj, 575, 435
\bibitem[Rosen et al.(2016)]{ros16} Rosen, S.~R., Webb, N.~A., Watson, M.~G., et al.\ 2016, \aap, 590, A1 
\bibitem[Royer et al.(2007)]{royer2007} Royer, F., Zorec, J., \& G{\'o}mez, A.~E.\ 2007, \aap, 463, 671 
\bibitem[Savage et al.(1985)]{savage1985} Savage, B.~D., Massa, D., Meade, M., \& Wesselius, P.~R.\ 1985, \apjs, 59, 397 
\bibitem[Saxton et al.(2008)]{sax08} Saxton, R.~D., Read, A.~M., Esquej, P., et al.\ 2008, \aap, 480, 611 
\bibitem[Schulz et al.(2006)]{sch06} Schulz, N.~S., Testa, P., Huenemoerder, D.~P., Ishibashi, K., \& Canizares, C.~R.\ 2006, \apj, 653, 636 
\bibitem[Selim et al.(2016)]{selim2016} Selim, I.~M., Essam, A., Hendy, Y.~H.~M., \& Bendary, R.\ 2016, NRIAG Journal of Astronomy and Geophysics, 5, 16 
\bibitem[Silaj \& Landstreet(2014)]{sil14} Silaj, J., \& Landstreet, J.~D.\ 2014, \aap, 566, A132
\bibitem[Sim{\'o}n-D{\'{\i}}az \& Herrero(2014)]{sim2014} Sim{\'o}n-D{\'{\i}}az, S., \& Herrero, A.\ 2014, \aap, 562, A135 
\bibitem[Sim{\'o}n-D{\'{\i}}az et al.(2017)]{sim2017} Sim{\'o}n-D{\'{\i}}az, S., Godart, M., Castro, N., et al.\ 2017, \aap, 597, A22 
\bibitem[Sch{\"o}ller et al.(2017)]{sch2017} Sch{\"o}ller, M., Hubrig, S., Fossati, L., et al.\ 2017, \aap, 599, A66 
\bibitem[Smith et al.(1998)]{smi98} Smith, M.~A., Robinson, R.~D., \& Corbet, R.~H.~D.\ 1998, \apj, 503, 877
\bibitem[{{Smith} \& {Robinson}(1999)}]{smith1999}{Smith}, M.~A., \& {Robinson}, R.~D. 1999, \apj, 517, 866
\bibitem[Smith et al.(2012a)]{smi12} Smith, M.~A., Lopes de Oliveira, R., \& Motch, C.\ 2012a, \apj, 755, 64
\bibitem[Smith et al.(2012b)]{smi12b} Smith, M.~A., Lopes de Oliveira, R., Motch, C., et al.\ 2012b, \aap, 540, A53   
\bibitem[Smith et al.(2016)]{smith2016} Smith, M.~A., Lopes de Oliveira, R., \& Motch, C.\ 2016, Advances in Space Research, 58, 782
\bibitem[Stee et al.(2012)]{ste12} Stee, P., Delaa, O., Monnier, J.~D., et al.\ 2012, \aap, 545, A59 
\bibitem[Stelzer et al.(2006)]{stel06} Stelzer, B., Micela, G., Hamaguchi, K., \& Schmitt, J.~H.~M.~M.\ 2006, \aap, 457, 223 
\bibitem[Sung et al.(1997)]{sung1997} Sung, H., Bessell, M.~S., \& Lee, S.-W.\ 1997, \aj, 114, 2644 
\bibitem[Tetzlaff et al.(2011)]{tet11} Tetzlaff, N., Neuh{\"a}user, R., \& Hohle, M.~M.\ 2011, \mnras, 410, 190
\bibitem[ud-Doula \& Naz{\'e}(2016)]{udd16} ud-Doula, A., \& Naz{\'e}, Y.\ 2016, Advances in Space Research, 58, 680 
\bibitem[ud-Doula et al.(2018)]{udd17} ud-Doula, A., Owocki, S.~P., \& Kee, N.~D.\ 2018, \mnras, 478, 3049
\bibitem[Uesugi \& Fukuda(1970)]{uesugi1970} Uesugi, A., \& Fukuda, I.\ 1970, Contributions from the Institute of Astrophysics and Kwasan Observatory, University of Kyoto, Kyoto: University, Kwasan Observatory, Institute of Astrophysics, 1970,  
\bibitem[van Leeuwen(2007)]{vanl2007} van Leeuwen, F.\ 2007, \aap, 474, 653 
\bibitem[Wang et al.(2016)]{wan16} Wang, S., Liu, J., Qiu, Y., et al.\ 2016, \apjs, 224, 40 
\bibitem[Wang et al.(2018)]{wan18} Wang, L., Gies, D.~R., \& Peters, G.~J.\ 2018, \apj, 853, 156
\bibitem[Wegner(1993)]{wegner1993} Wegner, W.\ 1993, \actaa, 43, 209 
\bibitem[Wegner(2007)]{wegner2007} Wegner, W.\ 2007, \mnras, 374, 1549 
\bibitem[Witham et al.(2008)]{wit08} Witham, A.~R., Knigge, C., Drew, J.~E., et al.\ 2008, \mnras, 384, 1277 
\bibitem[Wright et al.(2010)]{wrihgt2010} Wright, N.~J., Drake, J.~J., Drew, J.~E., \& Vink, J.~S.\ 2010, \apj, 713, 871 
\bibitem[Yudin(2001)]{yudin2001} Yudin, R.~V.\ 2001, \aap, 368, 912 
\bibitem[Zorec et al.(2016)]{zorec2016} Zorec, J., Fr{\'e}mat, Y., Domiciano de Souza, A., et al.\ 2016, \aap, 595, A132 

\end{thebibliography}
\end{document}